\documentclass[aps,twocolumn,prc,superscriptaddress,showpacs,nofootinbib,amssymb,amsfonts,amsmath]{revtex4-1}

\usepackage{graphicx}
\usepackage{dcolumn}
\usepackage{bm}

\usepackage{amsmath}    
\usepackage{amsfonts}   
\usepackage{amssymb}
\usepackage{graphicx}   

\begin{document}
\title{Pairing and rotational properties of actinides 
and superheavy nuclei in covariant density functional theory.}

\author{A.\ V.\ Afanasjev}

\affiliation{Department of Physics and Astronomy, Mississippi 
State University, MS 39762}

\author{O.\ Abdurazakov}
\affiliation{Department of Physics and Astronomy, Mississippi 
State University, MS 39762}

\date{\today}

\begin{abstract}

  The cranked relativistic Hartree-Bogoliubov (CRHB) theory has been applied 
for a systematic study of pairing and rotational properties of actinides and 
light superheavy nuclei. Pairing correlations are taken into account by 
the Brink-Booker part of finite range Gogny D1S force.
For the first time in the covariant
density functional theory (CDFT) 
framework the pairing properties are studied via the quantities (such as 
three-point $\Delta^{(3)}$ indicators) related to odd-even mass staggerings. 
The investigation of the moments of inertia at low spin 
and the $\Delta^{(3)}$ indicators shows the need for an attenuation of the strength of 
the Brink-Booker part of the Gogny D1S force
in pairing channel. The investigation of rotational properties
of even-even and odd-mass nuclei at normal deformation, performed 
in the density functional theory framework in such a systematic way
for the first time, reveals that in the majority of the cases the 
experimental data are well described. These include the evolution of the moments 
of inertia with spin, band crossings in the $A\geq 242$ nuclei, the impact of the 
particle in specific orbital on the moments of inertia in odd-mass nuclei. The analysis 
of the discrepancies between theory and experiment in the band crossing region of 
$A\leq 240$ nuclei suggests the stabilization of octupole deformation at high 
spin, not included in the present calculations. The evolution of pairing with 
deformation, which is important for the fission barriers, has been investigated 
via the analysis of the moments of inertia in the superdeformed minimum. The 
dependence of the results on the CDFT parametrization has been studied by
comparing the results of the calculations obtained with the NL1 and NL3* 
parametrizations.
\end{abstract}

\maketitle

\section{Introduction}

Starting from the dawn of the 21st century, there is an increased interest 
to a detailed spectroscopic study of the heaviest actinides and light 
superheavy nuclei. Rotational, single-particle and other properties of 
these nuclei were and are studied both in experiment and in theory 
(see Refs.\ \cite{A250,HG.08} and references quoted therein). 
There is a hope that detailed spectroscopic information 
on such nuclei will allow to better test and constrain theoretical models 
so that the location of the island of enhanced stability of spherical 
superheavy nuclei can be predicted with higher level of confidence.

 A continuing experimental effort to study rotational properties of such 
nuclei is driven in part by the fact that in odd-mass nuclei they provide an 
important additional fingerprint for the Nilsson configuration assignment for 
the bandheads on which the rotational structures are built \cite{INPC2010}. 
The investigation of high spin structures also provides important information 
on stability of nuclei against fission \cite{No254-exp1}. Among recent 
surprises is the observation of rotational band in the $Z=104$ $^{256}$Rf 
nucleus up to very high spin of $I=20^+$ \cite{Rf256}; this is the highest-$Z$ 
nucleus is which such structures were observed. In addition, there is a 
revival 
of theoretical interest to the description of such structures. This is 
illustrated by recent systematic investigations of rotational properties 
in heavy actinides and light superheavy nuclei performed within total routhian 
surface (TRS) approach \cite{LXW.12} and particle-number conserving method 
based on a cranked shell model (PNC+CSM) \cite{ZZZZ.11,ZHZZZ.12}. These 
approaches are based on phenomenological Woods-Saxon and Nilsson potentials, 
respectively.

 Alternative and more microscopic approaches are based on non-relativistic 
and relativistic density functional theories (DFT) \cite{BHP.03,VALR.05}. 
Unfortunately, these approaches\footnote{We consider here only the 
calculations which include approximate particle number projection 
since it is needed for a proper description of rotational properties
\cite{GBDFH.94,VER.00,CRHB}.} were only occasionally used for the 
description of rotational structures in the pairing regime and no 
systematic assesment of their errors 
and the sources of these errors are available.
 It turns out that 
within these approaches more efforts were dedicated to the 
investigation of superdeformed (SD) rotational bands in different mass regions 
(see Refs.\ \cite{BHP.03,VALR.05} and references quoted therein) 
than to the study of rotational bands at normal deformation.
However, in contrast with normal-deformed (ND) bands, neither spin nor
parity are known for absolute majority of the SD bands. 
The studies of the ND bands over observable 
frequency range have been performed only in a few nuclei within the cranked 
Hartree(+Fock)+Bogoliubov [HB or HFB] frameworks based on DFT. These are 
$^{72,74,76}$Kr \cite{AF.05}, $^{74}$Rb\cite{Rb74}, $^{76}$Sr \cite{Sr76}, 
$^{80}$Zr \cite{AF.05} studied in covariant DFT [further CDFT]) as well as
$^{48,50}$Cr \cite{AER.01b}, a few even-even Er and Yb nuclei 
\cite{ER.94,AER.01,AER.01b}, and $^{240}$Pu \cite{AER.01} studied 
in Gogny DFT [further GDFT]. 
Somewhat more attention has been paid to rotational structures of 
actinides within Skyrme DFT [further SDFT] \cite{DBH.01,BBDH.03}, but even 
these investigations are away from being systematic.

  The situation is even worse in odd-mass nuclei where only few
rotational bands across the nuclear chart have been studied in 
the DFT framework so far (see Sect.\ \ref{Rot-odd} for a detailed 
overview). However, rotational properties of one-quasiparticle 
configurations give an important information on 
their underlying structure, thus providing an extra tool for a 
configuration assignment. This is especially 
important for light superheavy nuclei at the edge of the region 
where spectroscopic studies are still feasible (the nuclei with 
masses $A\sim 255$  and proton  number $Z\geq 102$)  \cite{INPC2010}
since alternative methods of configuration assignment are either
unreliable/questionable or cannot be employed because of the 
limitations of the experimental measurements.

 Covariant density functional theory  \cite{VALR.05} is 
well suited for the description of rotational structures. It exploits 
basic properties of QCD at low energies, in particular symmetries and 
the separation of scales \cite{EDF}. Built on the Dirac equation, it 
provides a consistent treatment of the spin degrees of freedom 
\cite{EDF,CFG.92} and spin-orbit splittings \cite{BRRMG.99,LA.11}; 
the latter has an essential influence on the underlying shell structure. 
It also includes the complicated interplay between the 
large Lorentz scalar and vector self-energies induced on the QCD 
level by the in-medium changes of the scalar and vector quark 
condensates \cite{CFG.92}. Lorentz covariance of CDFT leads to the fact 
that time-odd mean fields of this theory are determined as spatial 
components of Lorentz vectors and therefore coupled with the same 
constants as the time-like components \cite{AA.10} which are fitted
to ground-state properties of finite nuclei. This is important for
the description of odd-mass nuclei \cite{AA.10}, the excitations with 
unsaturated spins, magnetic moments [25], and nuclear rotations 
\cite{VALR.05,AR.00,TO-rot}. The successes of the CDFT in the description
of rotating nuclei both in paired and unpaired regimes and at 
different extremes (superdeformation [see Ref.\ \cite{VALR.05} and
references therein], ultrahigh spins \cite{ASN.12} and the limits 
of angular momentum in nuclear configurations 
\cite{VALR.05,Kr74-no-term}) are well documented.

  One should note that our understanding of the pairing properties in 
the CDFT framework is far from being satisfactory. Although it has been 
shown in Ref.\ \cite{SRR.01} that a relativistic bare potential (Bonn 
potential) reproduces pairing correlations at the Fermi surface in the 
CDFT application to infinite nuclear matter, its mathematical properties 
make a numerical solution of the relativistic Hartree+Bogoliubov (RHB)
equations with this potential in pairing channel extremely difficult 
task. This task has not been 
solved so far. On the other side, the relativity does not affect
pairing significantly \cite{SR.02}. As a consequence, simpler versions 
of phenomenological non-relativistic pairing such as constant gap pairing, 
monopole pairing, zero-range $\delta$-pairing (see Ref.\ \cite{KALR.10} 
and references quoted therein), separable pairing \cite{TMR.09} and 
the pairing based on the Brink-Booker part of finite range Gogny force
\cite{GELR.96,CRHB} are used in the CDFT 
calculations. 

  In all CDFT applications of the first three types of pairing the 
selection of the pairing strength has been guided either by 
non-relativistic results (see, for example, Refs.\ \cite{R.96,KALR.10}) 
or by the local fits to experimental/empirical pairing gaps (see, for 
example, Ref.\ \cite{AAR.10}). The strengths of separable pairing 
have been fitted to the properties of the Brink-Booker parts of 
finite range Gogny D1S and D1 forces
in nuclear matter \cite{TMR.09}.  However, to our knowledge 
the results of the calculations with these types of pairing have 
not been directly confronted with experimental observables sensitive 
to pairing such as the moments of inertia and/or the indicators 
related to odd-even mass staggerings (such as three-point $\Delta^{(3)}$ 
indicators, see Sect.\ \ref{Sub-indicat} below for details). Thus, at 
present it is not clear how accurately these types of pairing perform. 
The global investigations of the pairing in the CDFT framework similar to 
those  performed in non-relativistic SDFT framework (see Refs.\ 
\cite{BBNSS.09,MAB.11}) are not available yet.

  Somewhat more is known in the CDFT about the properties of the pairing
force based on the Brink-Booker part of finite range Gogny D1S force
 via the studies of rotational structures. 
Available investigations within the cranked RHB theory with 
approximate particle number projection by means of Lipkin-Nogami 
method (further CRHB+LN) show that it performs rather well in nuclei 
with masses $A\leq 200$ \cite{CRHB,J1Rare,VALR.05,AF.05}, but its 
strength has to be decreased by approximately 10\% in actinides 
\cite{A250}.  On the contrary, available applications of the RHB 
theory with pairing force based on the Brink-Booker part of finite 
range Gogny D1S force
in pairing channel to the ground state 
properties across the nuclear chart follow the prescription of 
Ref.\ \cite{GELR.96} in which the strength of the Brink-Booker 
part
is increased by a factor 1.15. This difference in the 
selection of the pairing strength definitely requires the 
clarification.


  To our knowledge, the detailed analysis of pairing indicators
(such as the $\Delta^{(3)}$ indicators) has not been performed 
so far in either relativistic mean field+BCS, RHB or CRHB(+LN) 
frameworks because of the complexity of the definition of the 
ground states in odd-mass nuclei. In order to define the ground 
state in odd-mass 
nucleus, the binding energies have to be calculated for a number 
of one-quasiparticle configurations based on the orbitals active 
in the vicinity of the Fermi level and only then the lowest in 
energy state is assigned to the ground state. This non-trivial 
problem has only been solved first for few nuclei in Ref.\ 
\cite{A250} and then in the systematic studies of 
actinides and rare-earth nuclei in Ref.\ \cite{AS.11}.

  The current manuscript aims on detailed and systematic study of 
pairing properties of actinides in the RHB and CRHB(+LN) frameworks 
via simultaneous investigation of the moments of inertia and the
$\Delta^{(3)}$ indicators. Such an investigation covers not 
only normal-deformed but also superdeformed structures. 
The rotational structures in the SD minimum provide only available 
information on the evolution of pairing with deformation in 
actinides. This is important for an understanding of fission 
barriers which  according to Ref.\ \cite{KALR.10} sensitively 
depend on the pairing properties. In addition, the rotational 
properties of even-even and odd-mass actinides are studied in a 
systematic way up to high spin in order to see the typical 
accuracy of the description of rotational and band crossing 
features, the impact of blocked orbital on rotational properties 
and the feasibility of the use of rotational features in 
configuration assignment of light odd-mass superheavy 
nuclei. The systematic analysis is restricted to  reflection 
symmetric nuclei. As a consequence, light octupole  deformed 
actinides \cite{BN.96} are omitted. Based on the results obtained 
in actinides, the deformation  and rotational properties of 
superheavy nuclei are also studied.


  The manuscript is organized as follows. The CRHB(+LN)
theory and its details are discussed in Sec.\ \ref{CRHB-eq}.
Section \ref{Pairing-sect} is devoted to the pairing properties 
of actinides. In this section, the pairing strength is defined 
and the calculated deformation, low-spin rotational properties 
and the $\Delta^{(3)}$ indicators are compared with experiment. 
Rotational properties of even-even and odd-mass nuclei
are considered up to high spin in Secs.\ \ref{Rot-even-even} 
and \ref{Rot-odd}, respectively. Deformation, pairing and 
rotational properties of actinide fission isomers are discussed 
in Sec.\ \ref{Fis-def-rot}. We report the results for deformation
and rotational properties of even-even superheavy nuclei in Sec.\ 
\ref{SHE-def-rot}. Finally, Sec.\ \ref{Concl} summarizes the 
results of our work.

\begin{figure}[h]
\centering
\includegraphics[width=8.6cm]{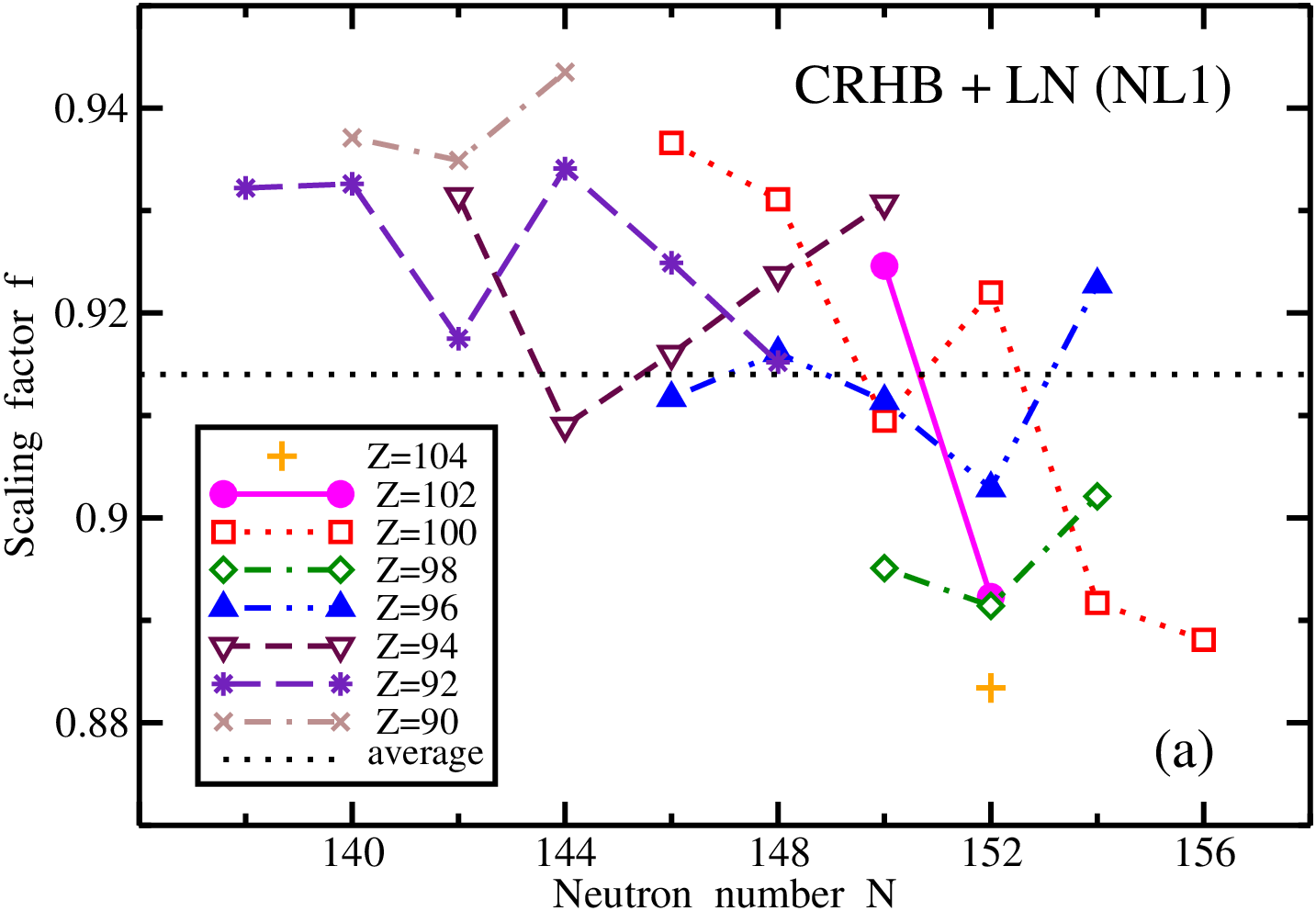}
\includegraphics[width=8.6cm]{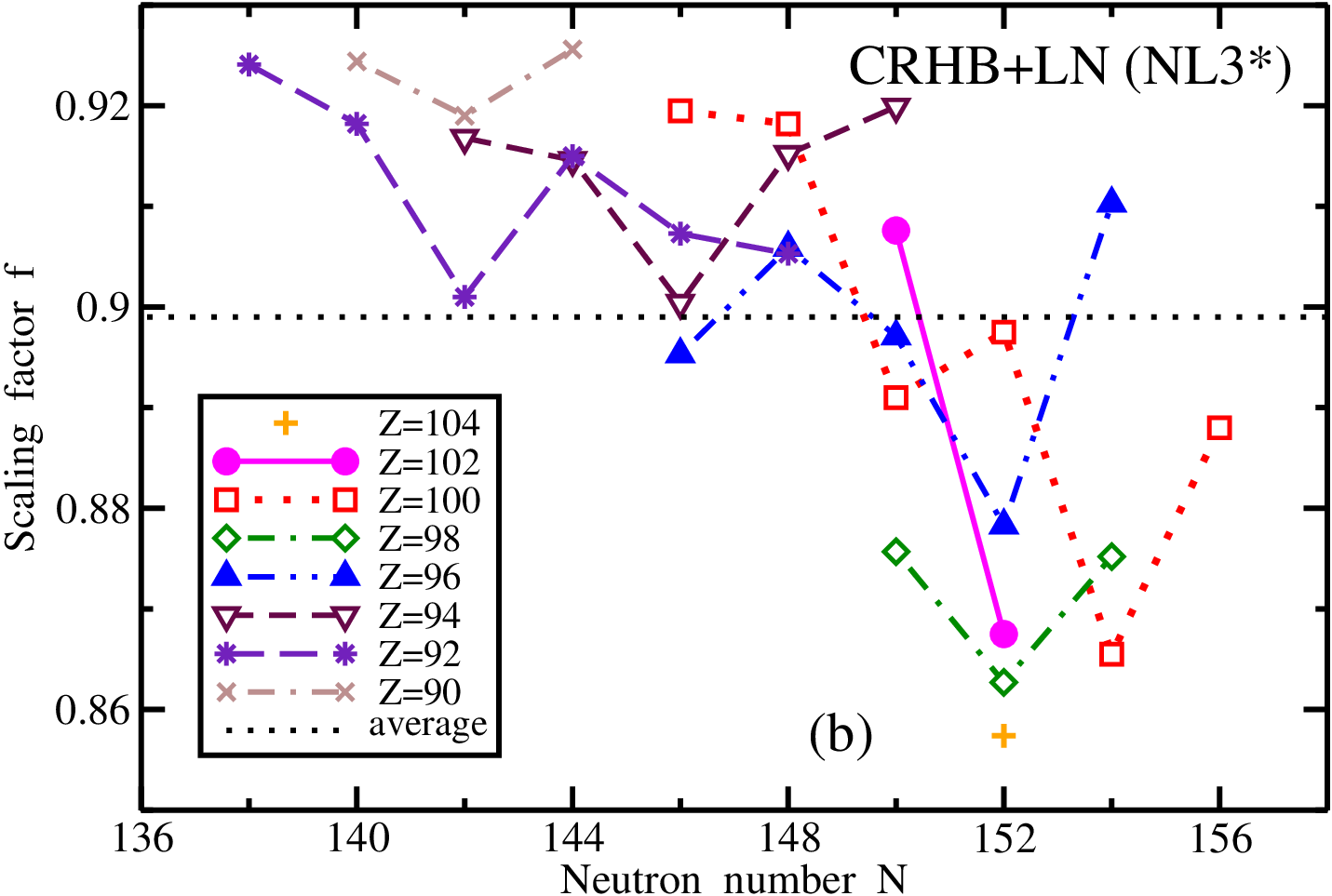}
\caption{(Color online) Scaling factors as a function of neutron number 
for individual nuclei of different isotope chains. The dotted 
line corresponds to the average scaling factor $f_{av}$ 
for a given parametrization of the CDFT. The results of the 
calculations with the NL1 and NL3* parametrizations are 
presented.}
\label{Scaling-f}
\end{figure}

\section{Theoretical formalism}
\label{CRHB-eq}

 The CRHB+LN equations for the fermions in the rotating 
frame (in one-dimensional cranking approximation) are 
given by \cite{CRHB}
\begin{eqnarray}
\begin{pmatrix}
  \hat{h}_D'-\lambda'-\Omega_x \hat{J_x} & \hat{\Delta} \\
 -\hat{\Delta}^*& -\hat{h}_D'^{\,*} +\lambda'+\Omega_x \hat{J_x}^* 
\end{pmatrix} 
\begin{pmatrix} 
U({\bm r}) \\ V({\bm r}) 
\end{pmatrix}_k \nonumber \\ 
= E_k' 
\begin{pmatrix} 
U({\bm r}) \\ V({\bm r}) 
\end{pmatrix}_k, \,\,\,\,\,\,\,\,\,
\end{eqnarray}
where
\begin{eqnarray}
  \hat{h}_D'&= & \hat{h}_D  + 4 \lambda_2 \rho - 2\lambda_2 Tr(\rho)
\,,  \\
\lambda' &=&  \lambda_1+ 2 \lambda_2 \,, \\
\label{equasi}
E_k' &=&  E_k - \lambda_2\,.
\end{eqnarray}
 Here, $\hat{h}_D$ is the Dirac Hamiltonian for the nucleon with mass $m$;
$\lambda_1$ is defined from the average particle number constraints for 
protons and neutrons; $\rho_{\tau} =V^{*}_{\tau} V^{T}_{\tau}$ is the 
density matrix;  $U_k ({\bm r})$ and $V_k ({\bm r})$ are quasiparticle 
Dirac spinors; $E_k$ denotes the quasiparticle energies; and $\hat{J_x}$
is the angular momentum component. The LN method corresponds to a 
restricted variation of $\lambda_2 \langle (\Delta N)^2 \rangle$ (see 
Ref.\ \cite{CRHB} for definitions of $\lambda_1$ and $\lambda_2$), where 
$\lambda_2$ is calculated self-consistently in each step of the iteration. 
The form of the CRHB+LN equations given above corresponds to the shift 
of the LN modification into the particle-hole channel.

The Dirac Hamiltonian $\hat{h}_D$ contains an attractive scalar potential 
$S({\bm r})$
\begin{eqnarray}
S(\bm r)=g_\sigma\sigma(\bm r),
\label{Spot}
\end{eqnarray}
a repulsive vector potential $V_0({\bm r})$
\begin{eqnarray}
V_0(\bm r)~=~g_\omega\omega_0(\bm r)+g_\rho\tau_3\rho_0(\bm r)
+e \frac{1-\tau_3} {2} A_0(\bm r),
\label{Vpot}
\end{eqnarray}
and a magnetic potential ${\bm V}({\bm r})$
\begin{eqnarray}
\bm V(\bm r)~=~g_\omega\bm\omega(\bm r)
+g_\rho\tau_3\bm\rho(\bm r)+
e\frac {1-\tau_3} {2} \bm A(\bm r).
\label{Vmag}
\end{eqnarray}
The last term breaks time-reversal symmetry and induces currents.
In rotating nuclei, the time-reversal symmetry is broken by the 
Coriolis field. Without rotation, it is broken when the time-reversal orbitals 
are not occupied pairwise. In the Dirac equation, the space-like
components of the vector mesons $\bm\omega(\bm r)$ and 
$\bm\rho(\bm r)$ have the same structure as the space-like
component $\bm A(\bm r)$ generated by the photons. 
Since $\bm A(\bm r)$ is the vector potential of the magnetic
field, by analogy the effect due to presence of the vector
field $\bm V(\bm r)$ is called {\it nuclear magnetism} \cite{KR.89}.
It has considerable influence on the magnetic moments \cite{HR.88}, 
the moments of inertia \cite{AR.00,TO-rot} and affects the properties
of odd- and odd-odd nuclei \cite{AA.10}. In the present calculations 
the spatial components of the vector  mesons are properly 
taken into account in a fully self-consistent way. The detailed 
description of the mesonic degrees of freedom in the CRHB+LN theory 
is presented in Ref.\ \cite{CRHB}.

  The CRHB(+LN) equations are solved in the basis of an anisotropic 
three-dimensional harmonic oscillator in Cartesian coordinates. 
The same basis deformation $\beta_0=0.3$, $\gamma=0^{\circ}$ and 
oscillator frequency $\hbar \omega_0=41$A$^{-1/3}$ MeV have been 
used in the calculations. All fermionic and bosonic states belonging 
to the shells up to $N_F=14$ and $N_B=20$ are taken into account 
in the normal-deformed minimum in the diagonalization of the Dirac 
equation and the matrix inversion of the Klein-Gordon equations, 
respectively. As follows from detailed analysis of Refs.\ 
\cite{A250,AS.11}, this truncation of basis provides sufficient
accuracy of the calculations. In order to have similar accuracy
in the superdeformed minimum, $N_F$ has been increased to 16 in
the calculations.

  The calculations are performed as a function of rotational 
frequency in the frequency range $\Omega_x=0.01-0.45$ MeV
in steps of 0.02 MeV outside the band crossing regions 
and 0.01 MeV in the band crossing regions and their vicinities. 
Note that not always full convergence is obtained at all 
frequencies; for these frequencies, no calculated curve 
is shown in the figures below. This typically happens in 
the regime of extremely 
week pairing at high rotational frequencies in the $Z=90-102$ 
nuclei (see Figs.\ \ref{sys-J1-NL1} and \ref{sys-J1-NL3s} below). 
Alternatively, no convergence takes place in the band crossing 
region or above of some Rf and Sg nuclei (see Figs.\ \ref{sys-J1-NL1} 
and \ref{sys-J1-NL3s} below) most likely because the solution 
jumps between two closely lying in energy minima.

The calculations have been performed with the NL1 \cite{NL1} and 
NL3* \cite{NL3*} parametrizations of the RMF Lagrangian. 
The selection of the parametrizations has been dictated by the 
following considerations:

\begin{itemize}

\item
{\it The accuracy of the description of single-particle properties.} 
The description of rotating  nuclei is more complicated as compared 
with the one of the ground states properties (such as binding energies, 
radii etc) of even-even nuclei. This is because it depends not only on 
the calculated deformations of nuclei, but also on the energies and 
alignment properties of the single-particle orbitals. For example, 
the alignments of proton or neutron angular momenta in the 
upbending/backbending region and whether it proceeds smoothly or
in a abrupt way strongly depend on the accuracy of  the description 
of the excitation energies of high-$j$ aligning orbitals 
with respect of quasiparticle vacuum \cite{F.priv}. So far, the accuracy of the 
description of deformed one-quasiparticle states has been systematically 
studied only with the NL1 and NL3* parametrization in Ref.\ 
\cite{AS.11}; this study covers all one-quasiparticle states
in actinide region. It is interesting that 
the overall accuracy of the description of the energies of deformed 
one-quasiparticle states in Ref.\ \cite{AS.11} is slightly better
in the NL1 parametrization, which was fitted 25 years ago
mostly to the nuclei at the $\beta$-stability line, than in the recent 
NL3* parametrization. This suggests that the inclusion of extra 
information on neutron rich nuclei into the fit of the NL3* 
parametrization may lead to some degradation of the description of 
single-particle states along the valley of $\beta$-stability.
\\
\\
So far the calculated alignment properties of single-particle 
orbitals have only been confronted with experiment in unpaired regime 
at normal deformation in the $A\sim 80$ \cite{AF.05} region and at 
superdeformation in the $A\sim 60$ \cite{A60} and 150 \cite{ALR.98} 
mass regions. These investigations have been mostly performed 
with the NL1 parametrization which describes well the alignment 
properties of the single-particle orbitals. 

\item

{\it The accuracy of the description of the moments of inertia 
in unpaired regime}. The pairing has a significant impact on 
the moments of inertia which is much stronger than its impact on 
other physical observables. As a consequence, it is very difficult 
to disentangle pairing and rotational alignment contributions to 
the moments of inertia. Fortunately, the pairing is very weak at 
high spin,  and, thus, can be neglected there \cite{ALR.98,SD-A150,VALR.05}.
As a result, it becomes possible to benchmark the performance of 
different CDFT parametrizations with respect of the description of 
the moments of inertia in unpaired regime.
\\
\\
 So far such detailed benchmark calculations across the nuclear 
chart are only available for the NL1 parametrization. They include 
the investigations of the moments of inertia of superdeformed bands
in the $A\sim 60$ \cite{A60} and 150 \cite{SD-A150,VALR.05} mass regions and
in $^{108}$Cd \cite{108Cd}. Moreover, the rotational properties 
of smooth terminating bands in the $A\sim 110$ mass region \cite{VALR.05}, 
triaxial
superdeformed bands both at ultra-high spin in $^{158}$Er \cite{ASN.12}
and at moderate/high spin in the $A\sim 170$ mass region \cite{Hf171-SD}
have been succesfully studied with this parametrization.
These detailed investigations showed that the NL1
parametrization provides very good description of rotational
and deformation properties of studied nuclei which in many
cases is similar but frequently better than the one obtained
with newer parametrizations such as NL3 and NLSH (for later
comparison see Refs.\ \cite{A60,ALR.98}). These results give us strong
confidence that the NL1 parametrization should perform
reasonably well also in actinides.

Limited benchmark
calculations in unpaired regime are available also for
the NL3* parametrization but only for a few nuclei ($^{58}$Cu,
$^{143}$Eu, $^{109}$Sb and $^{74}$Kr \cite{NL3*} and $^{158}$Er
\cite{ASN.12}) across nuclear chart.
However, these studies cover different types of bands such as
near-axial and triaxial superdeformed bands and smoothly 
terminating bands.

\item
{\it The accuracy of the description of pairing properties by 
the Brink-Booker part of finite range Gogny D1S force.}
 As discussed in the introduction, our knowledge of pairing 
properties of the Brink-Booker part of Gogny D1S 
force
comes mostly from the CRHB+LN
calculations. One interesting observation, born in the studies
of few rotational bands, is the need for an attenuation of this 
pairing force in the nobelium region \cite{A250}. To validitate 
this observation, the systematic calculations of the moments of
inertia and the $\Delta^{(3)}$ indicators in actinides have to be 
confronted with available systematic studies in lighter nuclei 
employing the same CDFT parametrization. Such studies of the 
moments of inertia are available only for normal deformed proton-rich 
$A \sim 70$ \cite{AF.05} and rare-earth \cite{J1Rare} nuclei and for 
superdeformed nuclei in the $A\sim 190$ region \cite{CRHB}. The later
two studies are performed with the NL1 parametrization, while 
the former one with NL3. However, it was verified that the results
for rotational structures in the $A\sim 70$ mass region with the
same pairing are similar for the NL1 and NL3 parametrizations.

\end{itemize}

 Thus, two different parametrizations, namely, NL1, fitted to the nuclei 
in the valley of beta-stability, and NL3*, tailored towards the description
of neutron-rich nuclei, are used in the current study. This selection
allows to study the dependence of the results on the CDFT parametrization.
In addition, the use of two parametrizations allows in many cases to 
circumvent the convergence problems for specific blocked orbitals in 
odd-mass nuclei which can show up in one parametrization but will not
affect the solution in another parametrization (see Sect.\ \ref{Rot-odd} 
for details). For example, the calculations for the moments of inertia
of the $\pi 5/2[523]$ and $\pi 3/2[521]$ rotational bands in $^{214}$Am
are possible only with the NL1 parametrization, while the ones for the
$\nu 9/2[734]$ bands in $^{247}$Cm and $^{249}$Cf only with NL3*
(see Figs.\ \ref{J1-241Am}, \ref{J1-247Cm} and \ref{J1-cf249} below). 
Other examples of complementarity of the calculations with NL1 and NL3*
can be found in Sect.\ \ref{Rot-odd}.

 It is clear that the NL3* parametrization is less tested than the NL1 one
in respect of the description of rotating nuclei. However, it has been 
successfully applied to the description of binding energies \cite{NL3*}, 
ground state properties of deformed nuclei \cite{SRH.10}, single-particle 
spectra of spherical odd-mass nuclei \cite{LA.11}, fission barriers  
\cite{AAR.10}, giant resonances \cite{NL3*}, and breathing mode 
\cite{GLLM.10}.

  The NL1 and NL3* parametrizations are representatives of the 
meson-exchange models with non-linear meson-nucleon couplings 
\cite{AAR.12}. This type of model can be supplemented, for example,
by isoscalar-isovector coupling as it is done in the FSUGold 
model \cite{TP.05}. There are two other classes of covariant 
density functional models such as density-dependent meson-exchange 
\cite{TW.99} (represented, for example, by the DD-ME2 \cite{DD-ME2} 
parametrization) and density-dependent point-coupling 
\cite{NHM.92,BMM.02} (represented, for example, by the DD-PC1 
\cite{NVR.08} and PC-PK1 \cite{PC-PK1} parametrizations) models.
However, these parametrizations have not been benchmarked with 
respect of the description of rotational structures in the unpaired 
regime and nothing is known about their accuracy of the description 
of one-quasiparticle deformed states in odd-mass nuclei. Thus, 
they are not employed in the current study.

  Nuclear configurations of deformed odd nuclei (further one-quasiparticle 
[1-qp] configurations) are labeled by means of the asymptotic quantum 
number $\Omega [Nn_z \Lambda]$ (Nilsson quantum number) of the dominant 
component of the wave function of blocked single-particle orbital at
low rotational frequency.

\begin{figure*}[ht]
\centering
\includegraphics[width=14.0cm]{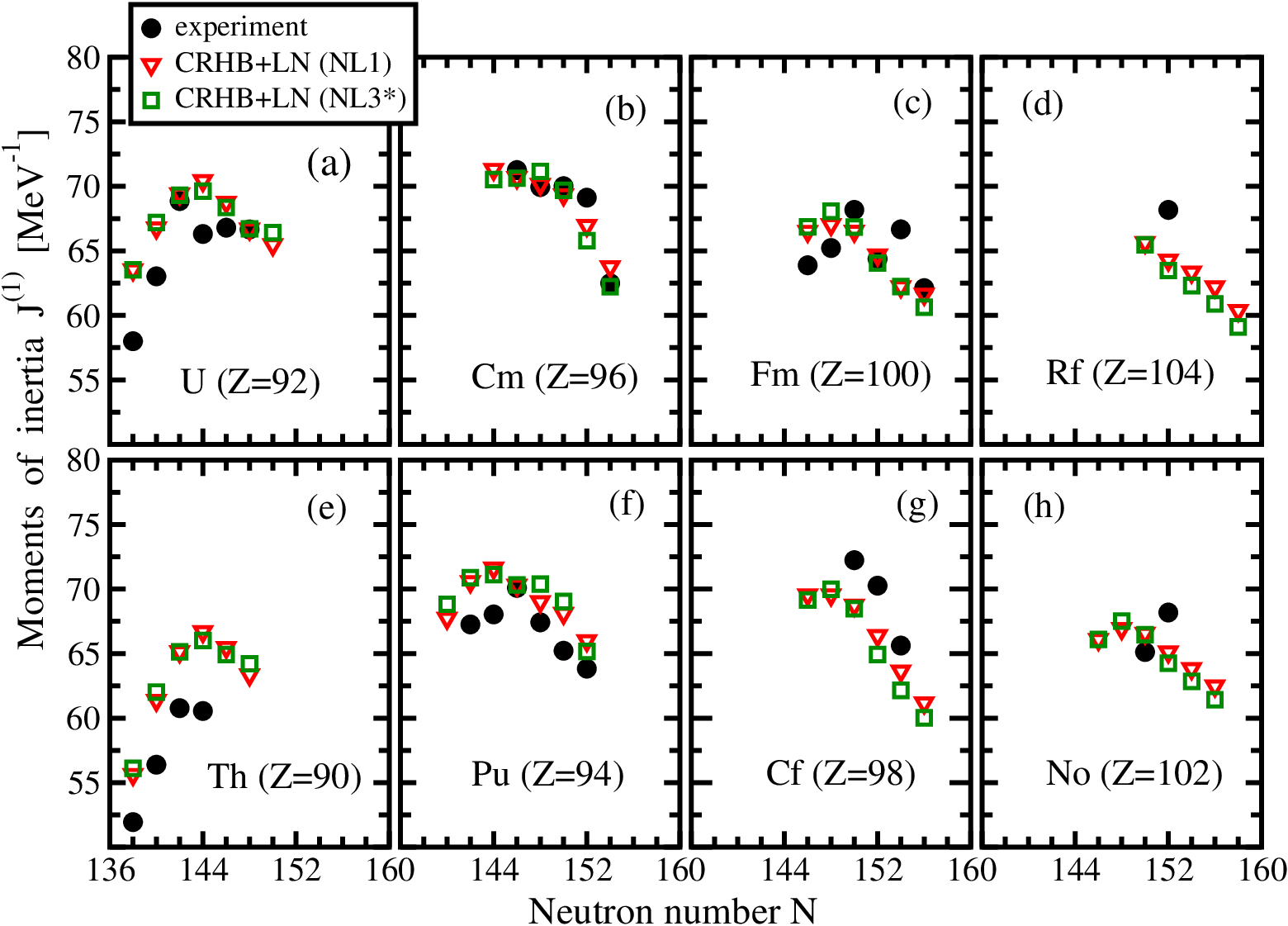}
\caption{(Color online) Calculated and experimental moments of inertia at 
low spin. Experimental moments of inertia are extracted from the 
energies of the $2^+$ states. Calculated values are obtained 
in the CRHB+LN calculations with $f_{av}$ specific to a given 
parametrization at the 
rotational frequency corresponding to experimental energy of the 
$2^+ \rightarrow 0^+$ transition. Experimental data are shown by 
filled black circles, while calculated values are given by  
red triangles (the NL1 parametrization) and green squares (the 
NL3* parametrization). Theoretical results are shown at 
$\Omega_x = 0.02$ MeV in the cases when experimental data are 
not available. 
}
\label{J1-low}
\end{figure*}

\section{Pairing properties}
\label{Pairing-sect}

\subsection{Formalism}

 The pair field $\hat{\Delta}$ in CRHB(+LN) theory is 
given by
\begin{eqnarray}
\hat{\Delta} \equiv \Delta_{ab}~=~\frac{1}{2}\sum_{cd} V^{pp}_{abcd}
\kappa_{cd}
\label{gap}
\end{eqnarray}
where the indices $a,b,\dots$ denote quantum numbers which specify
the single-particle states with the space coordinates $\bm r$, as 
well as the Dirac and isospin indices $s$ and
$\tau$. It contains the pairing tensor  $\kappa$
\begin{eqnarray}
\kappa = V^{*}U^{T}
\label{kappa}
\end{eqnarray}
and the matrix elements $V^{pp}_{abcd}$ of the effective interaction in 
the particle-particle ($pp$) channel, for which
the Brink-Booker part of phenomenological non-relativistic Gogny-type 
finite range interaction 
\begin{eqnarray}
V^{pp}(1,2) &  = & f \sum_{i=1,2} e^{-[({\bm r}_1-{\bm r} _2)/\mu_i]^2} \nonumber \\
&\times& (W_i+B_i P^{\sigma}- H_i P^{\tau} - M_i P^{\sigma} P^{\tau})
\label{Vpp}
\end{eqnarray}
is used.  The clear advantage of such a force is that it provides 
an automatic cutoff of high-momentum components. The motivation for 
such an approach to the description of pairing is given in 
Ref.\ \cite{CRHB}. In Eq.\ (\ref{Vpp}), $\mu_i$, $W_i$, $B_i$, $H_i$ 
and $M_i$ $(i=1,2)$ are the parameters of the force and $P^{\sigma}$ 
and $P^{\tau}$ are the exchange operators for the spin and isospin 
variables, respectively. The D1S parametrization of the Gogny force
\cite{D1S,D1S-a} is used here. Note  that a scaling  factor $f$ is 
introduced in Eq.\ (\ref{Vpp}), the role of which is discussed in 
Sect.\ \ref{Sel-f}.

  As a  measure for the size of the pairing correlations in 
Hartree-(Fock)-Bogoliubov calculations, we use the pairing 
energy 
\begin{eqnarray}
E_{pairing}~=~-\frac{1}{2}\mbox{Tr} (\Delta\kappa).
\label{Epair}
\end{eqnarray}
%

\subsection{The selection of the scaling factor $f$}
\label{Sel-f}

\begin{figure*}[ht]
\centering
\includegraphics[width=14.0cm]{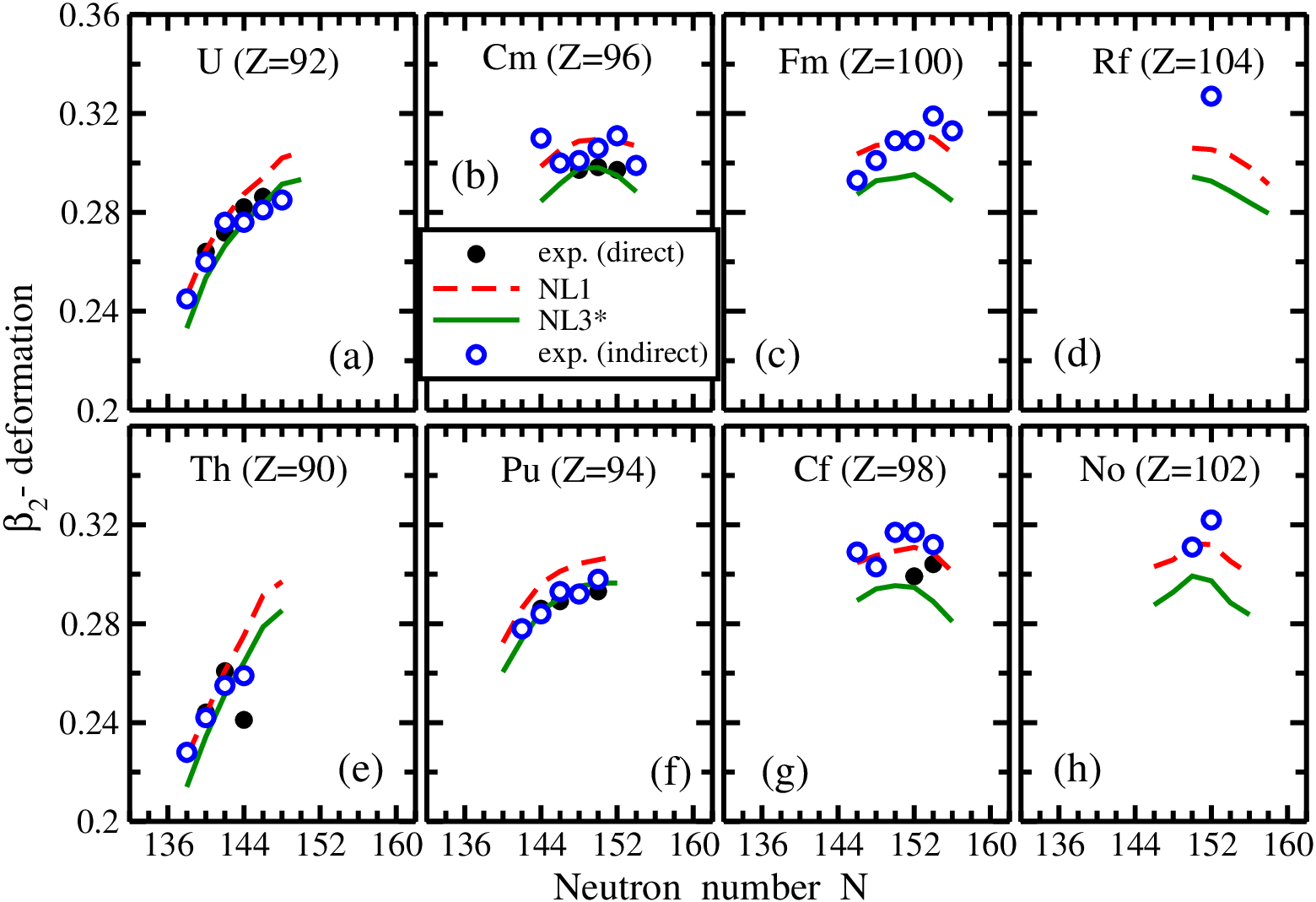}
\caption{(Color online) The calculated (lines) and experimental (circles) 
quadrupole deformation parameters $\beta_2$. The experimental 
values of $\beta_2$ obtained in the direct measurements 
\cite{RMMNS.87} are shown by solid circles, while those 
deduced from the $2^+\rightarrow 0^+$ transition energies, 
with the prescription of Ref.\ \cite{No252}, are given 
by open circles.  
The results of the calculations with the
NL1 and NL3* parametrizations are shown by red dashed and
green solid lines, respectively.}
\label{beta2-low}
\end{figure*}

  In the CRHB+LN framework, the original strength (scaling factor 
$f=1.0$ 
 in Eq.\ (\ref{Vpp})) of the Brink-Booker 
part of the Gogny D1S force provided a good description of the moments 
of inertia in the $A\sim 75$ \cite{AF.05}, $A\sim 160-170$ 
\cite{J1Rare} and $A\sim 190$ \cite{CRHB} mass regions. 
However, as discussed in detail in Ref.\ \cite{A250} it produces 
pairing correlations in the $A\sim 250$ mass region that are 
too strong in the CRHB+LN calculations, and, thus, it has to 
be attenuated ($f<1.0$) in order to reproduce the observed 
moments of inertia. The cranked HFB calculations in the Gogny 
DFT also show the same problem (see discussion in Sect.\ IIIA
of Ref.\ \cite{A250}). The need for attenuation of the strength
of the Brink-Booker part within the framework of the CRHB+LN theory is 
not surprising since its pairing properties were adjusted by 
fitting only the odd-even mass differences of the Sn isotopes
\cite{D1S,D1S-a} which are far away from actinides. In addition, 
this fit was done in the framework of the HFB theory completely 
based on the Gogny force, while only the Brink-Booker part of 
the Gogny force is used in the pairing channel of the CRHB+LN 
theory.

 In Ref.\ \cite{A250},  the scaling factor $f$ of the Brink-Booker 
part of finite range Gogny D1S force
(see Eq.\ (\ref{Vpp})) has been chosen to reproduce 
the experimental kinematic moment of inertia of the ground state 
rotational band in $^{254}$No at rotational frequency $\Omega_x=0.15$ 
MeV. For example, the value $f=0.893$ has been obtained for the 
CRHB+LN calculations with the NL1 parametrization (see Table 1 in 
Ref.\ \cite{A250}). It provided good description of the moments of 
inertia of rotational bands in $^{252,254}$No \cite{A250}, $^{250}$Fm 
\cite{Fm250}, $^{253}$No \cite{No253}, and $^{255}$Lr \cite{Lr255}.

 However, considering more systematic character of the current
investigation the scaling factor $f$ has been defined by the fit 
to the moments of inertia extracted from the $I^{\pi}=2^+$ states of
ground state rotational bands in all even-even actinides for which 
such experimental data were available by the end of June, 2012. 
If not explicitly specified the experimental data have been taken 
from Refs.\ \cite{ENSDF,240U-246Pu-250Cm}; the nuclei used in the 
fit are shown in Fig.\ \ref{Scaling-f}. The advantage of such 
approach, in which  the scaling factor $f$ is defined at very 
low frequency $\Omega_x \sim 0.02$ MeV, as compared with the 
one used in Ref.\ \cite{A250} is twofold. First, the definition 
of scaling factor $f$ in Ref.\
\cite{A250} at $\Omega_x=0.15$ MeV is affected by the 
accuracy of the description of the alignments of specific 
single-particle orbitals (in particular, the ones emerging 
from proton $i_{13/2}$ and neutron $j_{15/2}$ subshells). 
This factor affects the definition of $f$ at $\Omega_x=0.02$ 
MeV to a much smaller extent. As a result, the current
calculations test predictive power of the model with respect
of rotational response in a more straightforward way.
Second, such a definition of $f$ allows to verify whether the 
fit of pairing strength to the moments of inertia leads to a 
consistent description of three-point indicators $\Delta^{(3)}$ 
(extracted from experimental odd-even mass staggerings) which 
are defined at no rotation.

\begin{figure*}[ht]
\centering
\includegraphics[width=14.0cm]{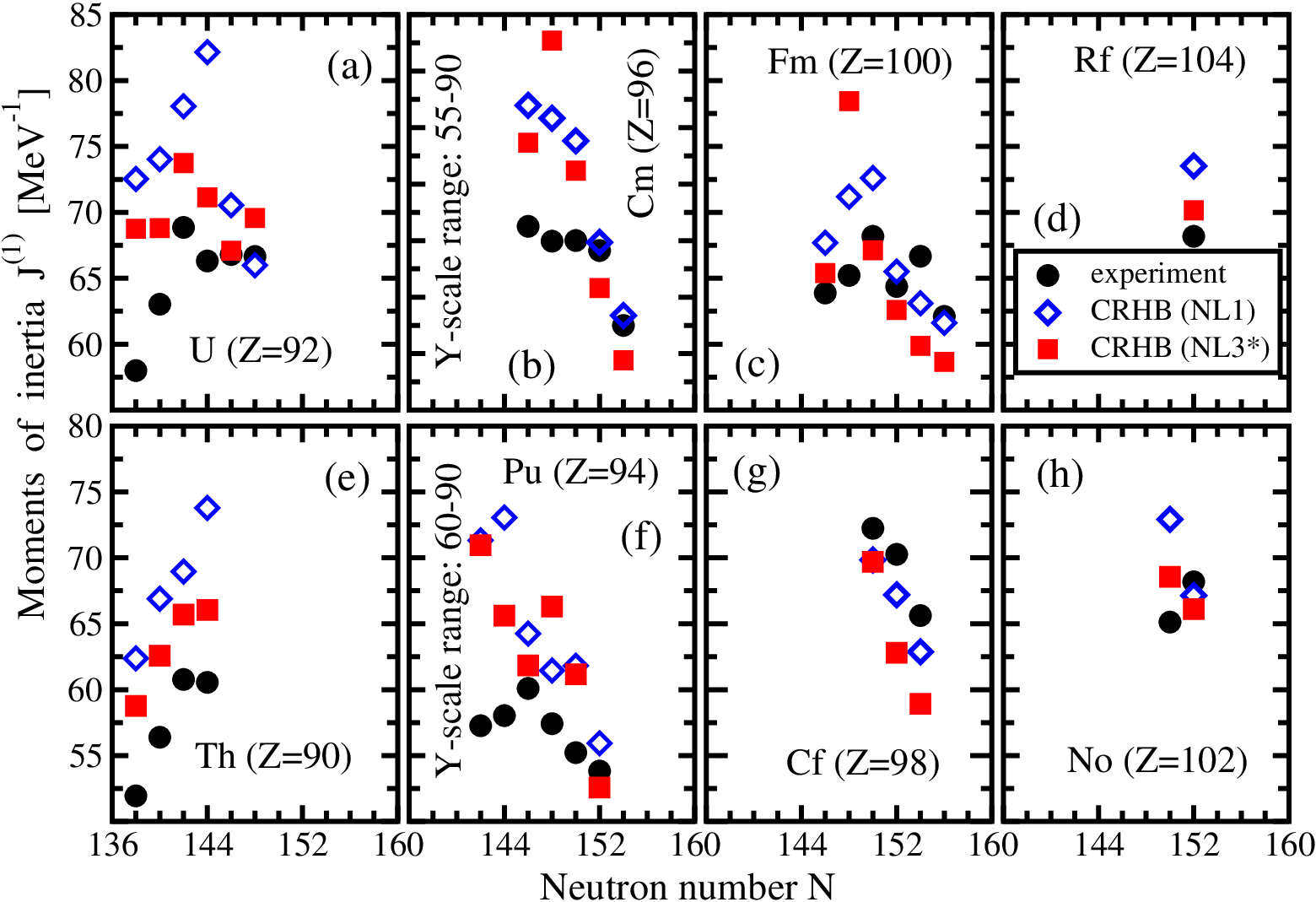}
\caption{(Color online) The same as Fig.\ \ref{J1-low} but for the CRHB 
calculations with scaling factor $f=1.0$ and only for the nuclei 
in which experimental data are available. To make the comparison 
with Fig.\ \ref{J1-low} easier, the difference between the highest 
and lowest values on the vertical axis is kept the same as in 
Fig.\ \ref{J1-low}; the only exception is the panel with the Cm 
isotopes. As a consequence, the different ranges for vertical 
axis are used for top and bottom panels. Moreover, the 
different ranges are used for vertical axis in the case 
of the Cm and Pu isotopes.}
\label{J1-low-CRHB}
\end{figure*}

Fig.\ \ref{Scaling-f} shows individual scaling factors $f_i$
for the nuclei used in the fit; these factors exactly reproduce 
the moments of inertia extracted from the $2^+$ states of the 
ground state rotational bands. In addition, the average scaling
factor
\begin{eqnarray}
f_{av}=\frac{\sum\limits_{i=1}^K f_i}{K}, 
\end{eqnarray}
where $K$ is the number of nuclei used in the fit, is shown by 
dotted lines. The $f_{av}$ is equal to 0.9147 and 0.899 in the NL1 
and NL3* parametrizations, respectively. Similar to Ref.\ \cite{A250}
these scaling factors only weakly depend on the CDFT parametrization.
For the NL1 parametrization, the obtained value of $f_{av}=0.9147$ 
is reasonably close to $f=0.893$ obtained for $^{254}$No in Ref.\ 
\cite{A250}. These average scaling factors $f_{av}$ will be used in 
systematic calculations of rotational bands in the actinides and 
superheavy elements.

\begin{figure}[h]
\centering
\includegraphics[width=8.6cm]{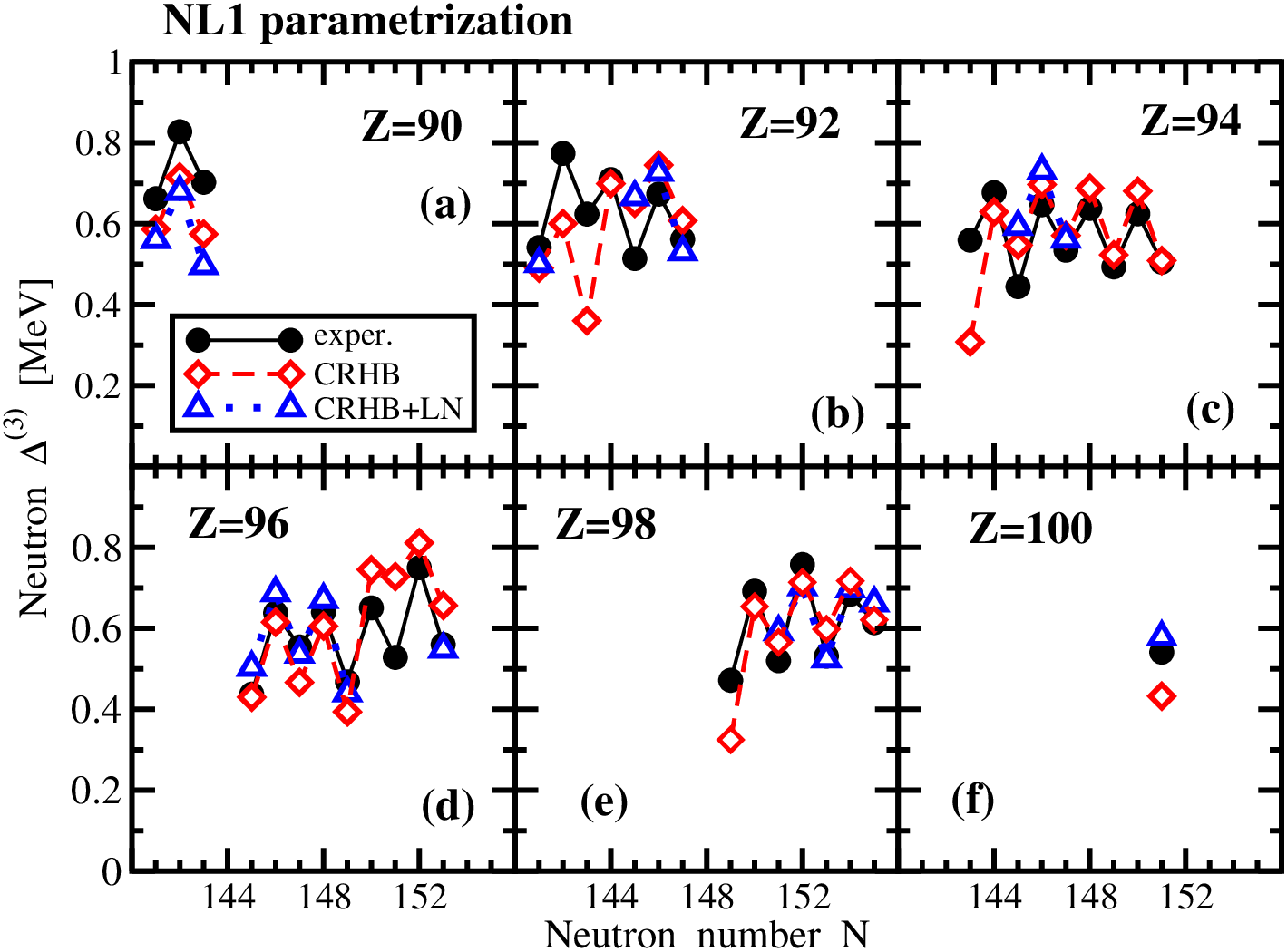}
\caption{(Color online) Experimental and calculated neutron three-point 
indicators $\Delta^{(3)}_{\nu}(N)$ as a function of neutron number $N$. The 
results of the CRHB and CRHB+LN calculations with the NL1 
parametrization are shown.}
\label{Delta3N-NL1}
\end{figure}

\begin{figure}[h]
\centering
\includegraphics[width=8.6cm]{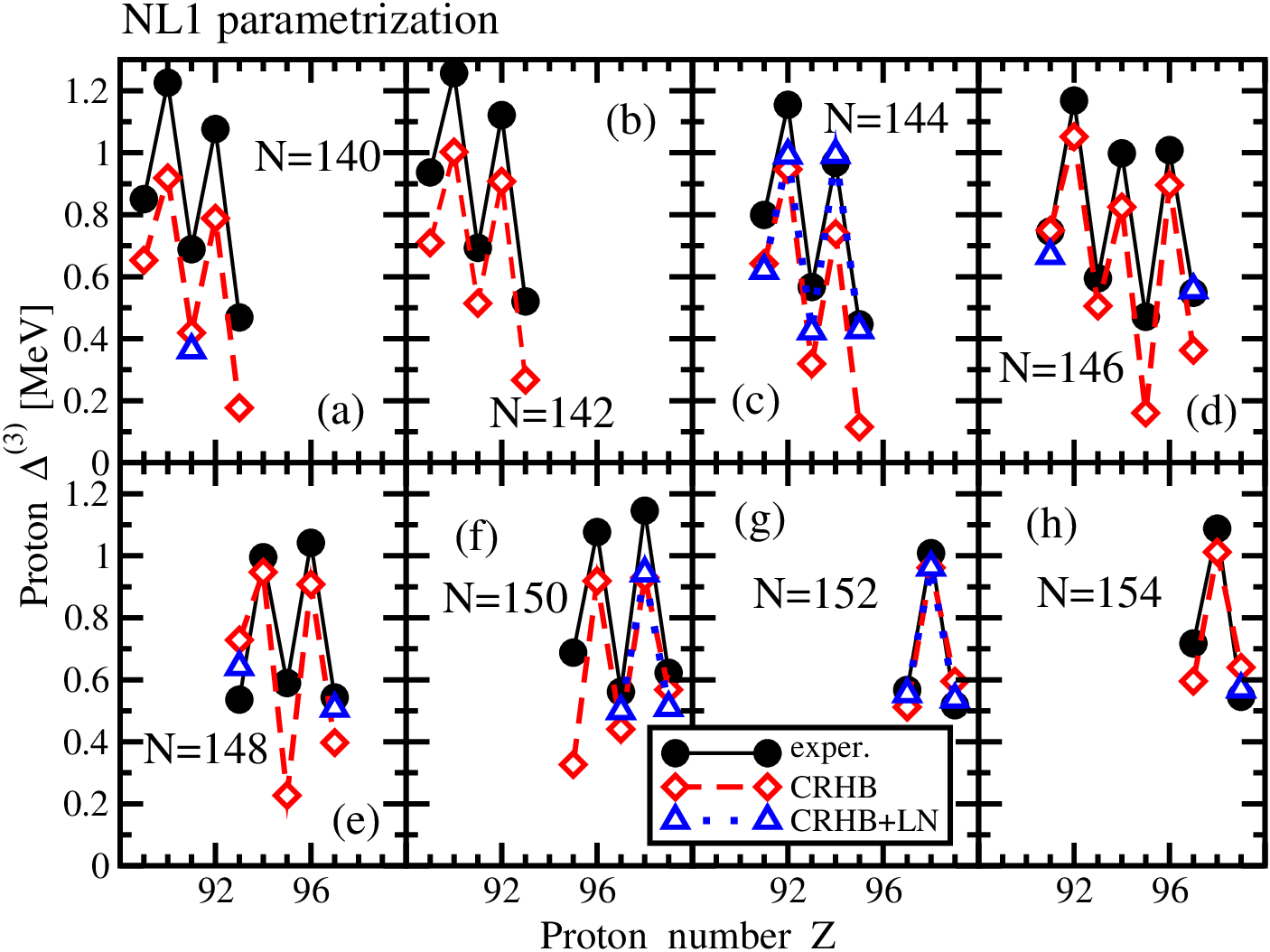}
\caption{(Color online) The same as Fig.\ \ref{Delta3N-NL1} but for proton 
three-point indicators $\Delta^{(3)}_{\pi}(Z)$ as a function of 
proton number $Z$.}
\label{Delta3P-NL1}
\end{figure}

\begin{figure}[h]
\centering
\includegraphics[width=8.6cm]{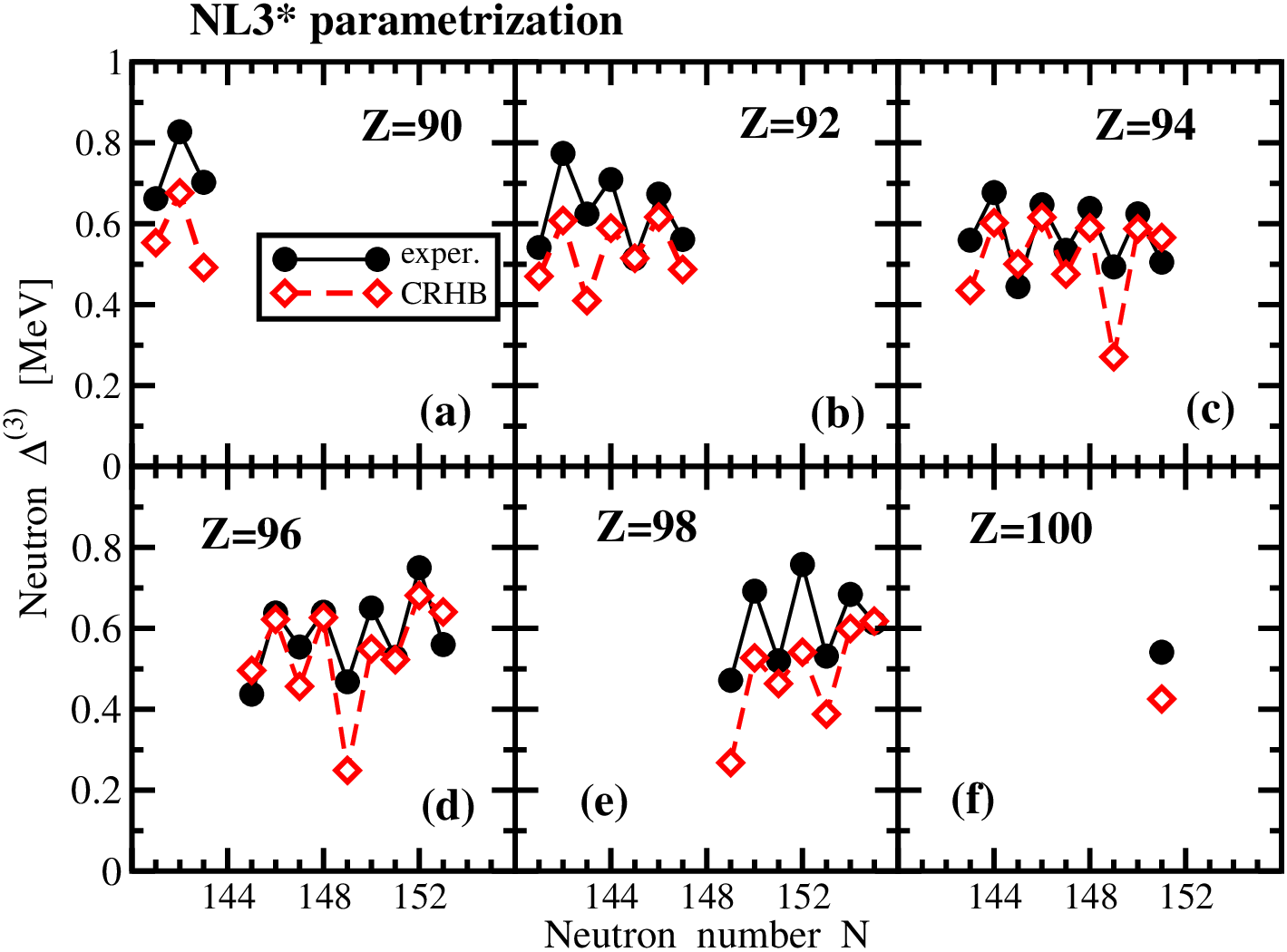}
\caption{(Color online) The same as Fig.\ \ref{Delta3N-NL1} but for the NL3* 
parametrization. Theoretical values have been obtained
within the CRHB framework.}
\label{Delta3N-NL3}
\end{figure}

\begin{figure}[h]
\centering
\includegraphics[width=8.6cm]{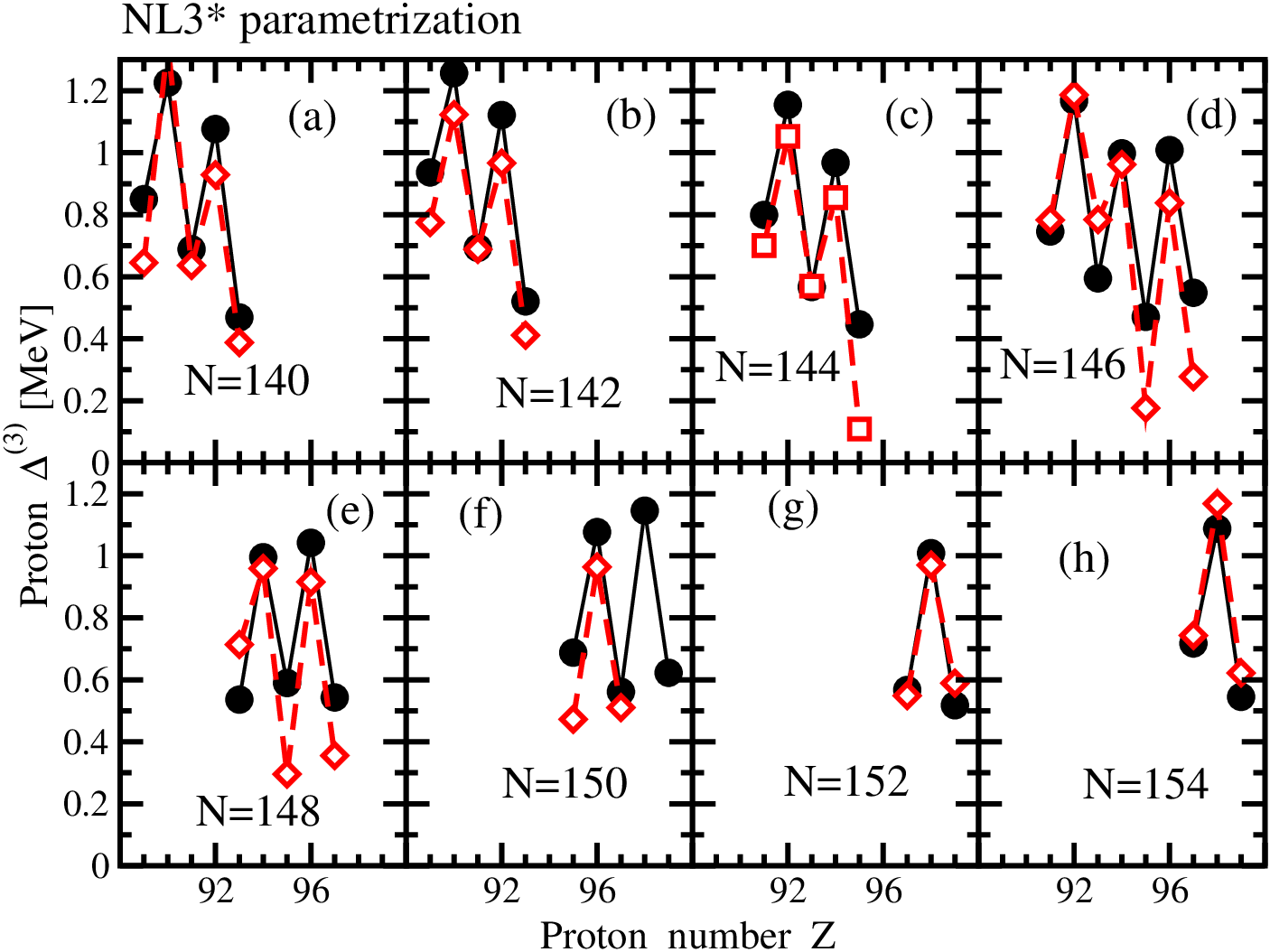}
\caption{(Color online) The same as Fig.\ \ref{Delta3P-NL1} but for the NL3* 
parametrization. Theoretical values have been obtained
within the CRHB framework.}
\label{Delta3P-NL3}
\end{figure}

\subsection{Rotational and deformation properties at low spin.}
\label{Rof-def-scal}

The kinematic moments of inertia at low spin obtained with average 
scaling factors $f_{av}$ are shown in Fig.\ \ref{J1-low}. One can see 
that the moments of inertia are described with an accuracy better
than 10\%. However, the use of average scaling factors leaves some 
unresolved trends as a function of particle number in both 
parametrizations. Figs.\ \ref{Scaling-f} and \ref{J1-low} show that 
the pairing has to be slightly weaker (stronger) in the nuclei with 
high $Z$ and high $N$ (low $Z$ and low $N$) relative to the calculations 
performed with $f_{av}$. Similar deviations from experiment exists at 
low spin also in the CRHB+LN calculations in rare-earth region (see 
Fig.\ 1 in Ref.\ \cite{J1Rare}).

 Direct experimental information on the deformations of nuclei 
from Coulomb excitation and lifetime measurements is quite limited 
\cite{RMMNS.87}. An alternative method is to derive a quadrupole 
moment from the $2^+ \rightarrow 0^+$ transition energy by employing 
the relation given by Grodzins \cite{G.62} or by later refinements
\cite{No252}. The prescription of Ref.\ \cite{No252} has an 
accuracy of about 10\%.  From the calculated and experimental charge 
quadrupole moments $Q$, the deformation parameters $\beta_2$ are 
derived by the relation
\begin{eqnarray} 
Q=\sqrt{ \frac{16\pi}{5}} \frac{3}{4\pi} Z R_0^2 \beta_2,
\,\,\,\,\,\,\,{\rm where}\,\,\,\,\,\,\, R_0=1.2 A^{1/3}.
\end{eqnarray}
The simple linear expression is used to maintain consistency with 
earlier papers \cite{RMMNS.87}. It is sufficient for comparison
between calculations and experiment because the same relation is
used. Including higher powers of $\beta_2$, e.\ g. as in Ref.\ 
\cite{NZ-def}, yields values of $\beta_2$ that are $\approx 10\%$ 
lower.


 Experimental quadrupole deformations of the nuclei under study are 
rather well reproduced in the CRHB+LN calculations (Fig.\ \ref{beta2-low}).
Thus, they do not represent a major source of the deviations between 
theory and experiment for kinematic moments of inertia. Considering 
typical uncertainties of the extraction of quadrupole deformation 
either in direct \cite{RMMNS.87} or indirect \cite{No252}  methods, 
it is difficult to give a preference to either NL1 or NL3* 
parametrization based on this observable.

\subsection{Three-point indicators $\Delta^{(3)}$.}
\label{Sub-indicat}

  The strength of pairing correlations can also be accessed via the 
three-point indicator \cite{DMNSS.01}
\begin{eqnarray}
\Delta ^{(3)}_{\nu}(N) = \frac{\pi_N}{2} \left[ B(N-1) + B(N+1) - 2 B(N) \right], \nonumber \\ 
\label{neut-OES}
\end{eqnarray} 
which is frequently used to quantify the odd-even staggering (OES) of 
binding energies. Here $\pi_N=(-1)^N$ is the number parity and $B(N)$ is 
the (negative) binding energy of a system with $N$ particles. In Eq.\ 
(\ref{neut-OES}), the number of protons $Z$ is fixed, and $N$ denotes 
the number of neutrons, i.e. this indicator gives the neutron OES. The 
factor depending on the number parity $\pi_N$ is chosen so that the 
OES centered on even and odd neutron number $N$ will both be positive. 
An analogous proton OES indicator $\Delta ^{(3)}(Z)$ is obtained by 
fixing the neutron number $N$ and replacing $N$ by $Z$ in Eq.\ 
(\ref{neut-OES}). The impact of time-odd mean fields on this quantity
has been discussed in detail in Ref.\ \cite{AA.10}.

  In order to extract the  ground state in odd-mass nucleus, the binding 
energies are calculated for a number of the 1-qp configurations based 
on the orbitals active in the vicinity of the Fermi level and then 
the lowest in energy state is assigned to the ground state. 
Such calculations are very 
complicated and time-consuming. As a result, they were done only in 
a few cases (see Refs.\ \cite{AS.11} and references quoted therein) on 
the H(F)B level of the DFT framework. To our knowledge, the detailed 
analysis of the $\Delta^{(3)}$ indicators has not been performed so 
far in the RHB framework because of the complexity of the 
definition of the ground states in odd-mass nuclei. Thus, this 
manuscript represents a first attempt of the systematic analysis of 
pairing correlations via fully self-consistent calculations of the
$\Delta^{(3)}$ indicators in the RHB framework.

  The $\Delta^{(3)}$ indicators are analysed in the CRHB and CRHB+LN 
frameworks. There are several reasons for a such comparative study.
First, the HFB [RHB] calculations without approximate particle number 
projection by means of the LN method are still used in the study of 
rotational bands \cite{ER.00,DGGL.06}, one-quasiparticle states 
\cite{AS.11}, fission barriers \cite{DGGL.06,KALR.10} and fission 
half-lives \cite{WE.12} of actinides and superheavy nuclei in the 
methods which employ the Brink-Booker part of finite range Gogny 
force
in the pairing channel. Second, 
the calculations with the LN method are more time-consuming and 
frequently less numerically stable than the ones without it. As a 
consequence, it is important to understand the similarities and 
differences between the CRHB and CRHB+LN results related to pairing.

  The CRHB calculations were performed with original 
strength of the Brink-Booker part of the Gogny D1S force
(scaling factor $f=1.0$) which according to Ref.\ \cite{A250} 
provides good description of the moments of inertia in $^{254}$No
over experimentally measured spin range. These calculations also 
reasonably well describe the moments of inertia at low spin 
(Fig.\ \ref{J1-low-CRHB}). Systematic non-relativistic investigations 
within cranked HFB approach based on Gogny D1S force also give 
reasonable description of the moments of inertia in actinides 
\cite{ER.00,DGGL.06,HG.07}. For example, the results obtained in 
Ref.\ \cite{HG.07} are close to the CRHB(NL3*) ones.
Considering the similarities of these two 
approaches (CRHB and Gogny HFB) \cite{J1Rare}, this is not surprising.

 Figs.\ \ref{Delta3N-NL1}, \ref{Delta3P-NL1}, \ref{Delta3N-NL3} 
and \ref{Delta3P-NL3} show the $\Delta^{(3)}$ indicators obtained in 
the CRHB calculations using the results of the calculations of odd-mass
nuclei of Ref.\ \cite{AS.11}. In average, they are close to experimental 
data. For proton subsystem, the rms deviations from experimental
$\Delta_{\nu}^{(3)}$ indicators are 0.22 and 0.15 MeV in the CRHB
calculations with the NL1 and NL3* parametrizations, respectively. 
For neutron subsystem, these deviations are 0.10 and 0.125 MeV, respectively. 
This compares favorably with global fits of pairing to the $\Delta^{(3)}$ 
indicators in the Skyrme DFT calculations of Refs.\ \cite{BBNSS.09,MAB.11} 
in which an RMS accuracy of about 0.25 MeV has been obtained for 
$\Delta^{(3)}$.
   
  The inclusion of the LN method into the calculations leads to the
decrease of the scaling factor by approximately 10\% (to $f_{av}=0.9147$ and
$f_{av}=0.899$ in the NL1 and NL3* parametrizations, respectively [see Sect.\ 
\ref{Sel-f}]). The experimental moments of inertia at low spin are 
described rather well with these scaling factors, see discussion in 
Sect.\ \ref{Rof-def-scal}. Figs.\ \ref{Delta3N-NL1} and \ref{Delta3P-NL1} 
show the $\Delta^{(3)}$ indicators calculated in the CRHB+LN approach. 
However, the convergence problems in the calculations of one-quasiparticle
states in odd-mass nuclei, emerging from the interaction of the 
blocked orbital with others, appear more frequently when approximate 
particle number projection by means of the Lipkin-Nogami method is 
employed. This is most likely due to additional nonlinearities of
the LN method. Note that such convergence problems are typical for 
the methods employing iterative diagonalization schemes for the solution of
the mean field equations and appear both in the CRHB and CRHB+LN calculations.
As a consequence, it was not possible to obtain the 
$\Delta^{(3)}$ indicators in the CRHB+LN calculations in a significant 
number of the cases since no reliable definition of the ground state
in some odd-mass nucleus is possible. This is despite the fact that the 
CRHB+LN calculations of low-energy spectra in odd-mass nuclei were 
restricted to three lowest in energy one-quasiparticle configurations 
obtained in the CRHB calculations of Ref.\ \cite{AS.11}. Such a 
simplified procedure [as compared with the one used in the CRHB 
calculations of Ref.\ \cite{AS.11}] can be used  since the quasiparticle 
spectra calculated within CRHB (with original $(f=1.0)$ strength of the
Brink-Booker part of the Gogny D1S force)
and CRHB+LN 
(with attenuated strength of the Brink-Booker part)
are very similar; the difference in the 
energies of three lowest in energy one-quasiparticle configurations
is typically less than 100 keV and the configuration ordering is 
the same (see, for example, Fig.\ 23 in Ref.\  \cite{A250}).
However, the lack of convergence in either of these three configurations 
disqualifies odd-mass nucleus from consideration for the calculations of
the $\Delta^{(3)}$ indicator.

  Because of these convergence problems and time-consuming nature of
the CRHB+LN calculations for one-quasiparticle configurations, the systematic
analysis of the $\Delta^{(3)}$ indicators shown in Figs.\ \ref{Delta3N-NL1} 
and \ref{Delta3P-NL1} has been performed only for the NL1 parametrization. 
The CRHB+LN calculations rather well describe these observables; the rms 
deviations from experiment are 0.11 and 0.084 MeV for proton and neutron 
subsystems, respectively. For the same [as in CRHB+LN calculations] set 
of nuclei, these deviations are 0.18 and 0.077 MeV in the CRHB calculations. 
As a result, for neutron subsystem the results of both calculations are 
similar, while proton $\Delta^{(3)}$ indicators are better described 
in the CRHB+LN calculations. The fit of the strength of pairing force 
to experimental moments of inertia (see Sect.\ \ref{Rof-def-scal}) and 
the fact that the LN method leads to a better (in average) description 
of the  $\Delta^{(3)}$ indicators \cite{BBNSS.09} may be responsible 
for observed differences in the CRHB+LN and CRHB results.

  The comparison of calculated moments of inertia, three-point 
indicators $\Delta^{(3)}$ and individual scaling factors $f_i$ 
allows to make a number of important conclusions. First, there
is a strong correlation between the definitions of pairing strengths
by means of the moments of inertia and three-point indicators. 
For example, the calculations for both of these physical observables 
show that pairing has to be slightly stronger at low values of 
neutron number $N$. The definitions of pairing strength via these
two observables are complimentary. This is because (i) it is 
difficult to disentangle proton and neutron contributions to
pairing when considering the moments of inertia and (ii) the 
$\Delta^{(3)}$ indicators are affected by particle-vibration 
coupling and depend on correct reproduction of the ground 
states in odd-mass nuclei (see Sec.\ \ref{dev-delta} for 
details).

  Second, approximate particle number projection by means of the LN
method is important for a better description of particle number 
dependencies of the moments of inertia.  Although the average 
description of the moments of inertia in the CRHB calculations seen 
in Fig.\ \ref{J1-low-CRHB} can be improved by an increase of the 
strength of pairing by few \%, this increase will not resolve wrong 
particle number dependencies for calculated $J^{(1)}$ and will not 
lead to the same level of accuracy of the description of $J^{(1)}$  
as seen in the CRHB+LN calculations (Fig.\ \ref{J1-low}).

  Third, obtained results clearly show that the strength of pairing
in the CRHB calculations has to be by 10-15\% larger 
than the one in the CRHB+LN calculations in order to reproduce
the experimental observables sensitive to pairing with comparable
level of accuracy. This clarifies the problem with different 
pairing strengths employed previously in the CRHB and CRHB+LN 
approaches which was discussed in the introduction.
 Considering 
weak dependence of the results of the CDFT parametrization, the 
results presented in the current manuscript and the ones obtained in 
Refs.\ \cite{J1Rare,AF.05} suggest that the scaling factors $f$ of
the  Brink-Booker part of the Gogny D1S force
$\sim 0.9$ and $\sim 1.0$ ($\sim 1.0$ and 
$\sim 1.10$) have to be used in the actinides and rare-earth/lighter 
nuclei in the CRHB+LN (CRHB) calculations, respectively. Although
this weak dependence on the CDFT parametrization has been verified
here only for the NL1 and NL3* parametrizations, we believe that it 
will hold also for other modern CDFT parametrizations. The pairing 
properties depend on single-particle level densities which in turn 
are defined by the Lorentz effective mass $m*(k_F)/m$ of nucleons at 
the Fermi surface. However, these effective masses are very similar 
for all successful CDFT parametrizations 
\cite{AS.11,NL1,NL3*,DD-ME2,NVR.08,PC-PK1,TP.05}.
%
 

\subsection{The sources of the deviations between theory and experiment
for the $\Delta^{(3)}$ indicators}
\label{dev-delta}

   The accuracy of the description of the $\Delta^{(3)}$ indicators
depend on a number of factors some of which were investigated in 
Refs.\ \cite{DMNSS.01,BBNSS.09,MAB.11}. Here, we will briefly 
discuss two factors which have been ignored (and even not mentioned) 
in the absolute  majority of the studies of pairing based on 
odd-even staggerings of binding energies. These are the 
correctness of the reproduction of the ground states in odd-mass 
nuclei and the impact of particle-vibration coupling (PVC).
They clearly affect the $\Delta^{(3)}$ indicators and limit the 
accuracy with which the experimental data can be described in 
model calculations.

 The structure of the ground state in odd-mass 
nucleus is not always correctly reproduced in model calculations 
and this can have an impact on the calculated $\Delta^{(3)}$ value. 
Indeed, the calculations within the RHB theory with
the NL1 and NL3* parametrizations \cite{AS.11}, the 
Hartree-Fock+BCS approach with different parametrizations
of the Skyrme forces \cite{BQM.07} as well as the FRDM model 
employing phenomenological folded-Yukawa potential 
\cite{BQM.07} show that only approximately 40\% of the ground 
states in odd-mass deformed nuclei are correctly reproduced. 

However, different single-particle states have different 
polarization effects for quadrupole and hexadecapole moments 
(this is clearly seen in the statistical analysis presented in 
Fig.\ 4 of Ref.\ \cite{AS.11}) and for time-odd mean fields 
(see Table IV  in Ref.\ \cite{A250}). These polarization
effects will impact the binding energies of odd-mass
nuclei. If the structure of calculated ground state differs 
from experimental one, the difference in polarization effects 
of these two states 
contributes into the discrepancy between calculated and experimental 
$\Delta^{(3)}$ values. This effect is expected to be minimal 
(maximal) when these two states have similar (significantly different) 
deformation-driving properties. The analysis of a number of the 
cases suggests that  wrong ground state in odd-mass nucleus can 
sometimes modify the $\Delta^{(3)}$ indicator by as much as
150 keV.

Additional binding  due to time-odd mean fields in odd-mass nuclei
is rather small in actinides and shows weak dependence on the 
blocked orbital (see Ref.\ \cite{AA.10} and  Table IV  in Ref.\ 
\cite{A250}). Thus, even if the ground state in odd-mass
nucleus is wrong in model calculations, the difference in 
polarization effects due to time-odd mean fields of the
wrong and correct states will only marginally (by less than 
20-30 keV) affect the $\Delta^{(3)}$ indicators. 

It is well known from the studies of spherical odd-mass nuclei 
that particle-vibration coupling affects the binding energies 
(see Refs.\ \cite{LA.11,CSB.10}). So far, no similar 
studies are available in deformed nuclei in the PVC models 
based on relativistic or non-relativistic DFT because of 
the complexity of the problem. However, the calculations
within the quasiparticle-phonon model based on phenomenological 
Woods-Saxon potential indicate that the lowest states 
of odd-mass actinides have mainly quasiparticle nature 
\cite{GIMS.71,IKMS.73} and that the corrections to
their energies due to PVC are typically less than 150 keV 
\cite{M.01}. These corrections will definitely have impact 
on the $\Delta^{(3)}$ indicators.

  These two effects can be a source of deviations between 
theory and experiment seen in Figs.\ \ref{Delta3N-NL1}, 
\ref{Delta3P-NL1}, \ref{Delta3N-NL3} 
and \ref{Delta3P-NL3}. However, the fact that on average the 
moments of inertia and the $\Delta^{(3)}$ indicators are 
reasonably well described (especially, in the CRHB+LN calculations) 
with the same strength of pairing suggests that apart from
some combinations of proton and neutron numbers these two 
effects do not contribute significantly. Note that the 
moments of inertia of even-even nuclei are significantly 
less affected by these two effects. So, in some sense they 
are more robust measure of pairing correlations.

\begin{figure}[ht]
\centering
\includegraphics[width=8.6cm,angle=0]{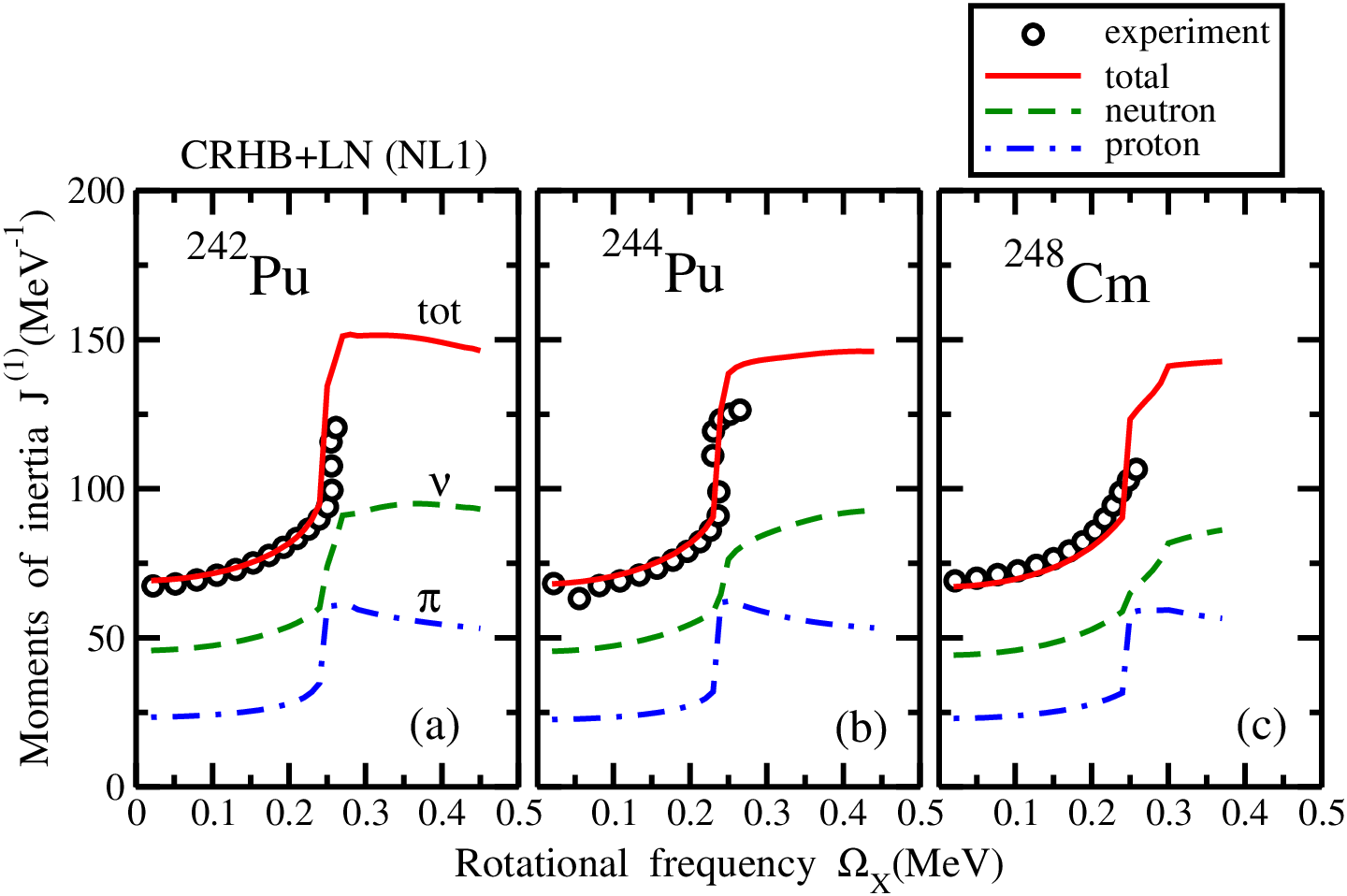}
\caption{(Color online) The experimental and calculated kinematic 
moments of inertia $J^{(1)}$ of ground state rotational bands in 
$^{242,244}$Pu and $^{248}$Cm as a function of rotational frequency 
$\Omega_x$. The calculations are performed with the NL1 parametrization.} 
\label{backbending_NL1}
\end{figure}

\begin{figure}[h]
\includegraphics[width=8.6cm,angle=0]{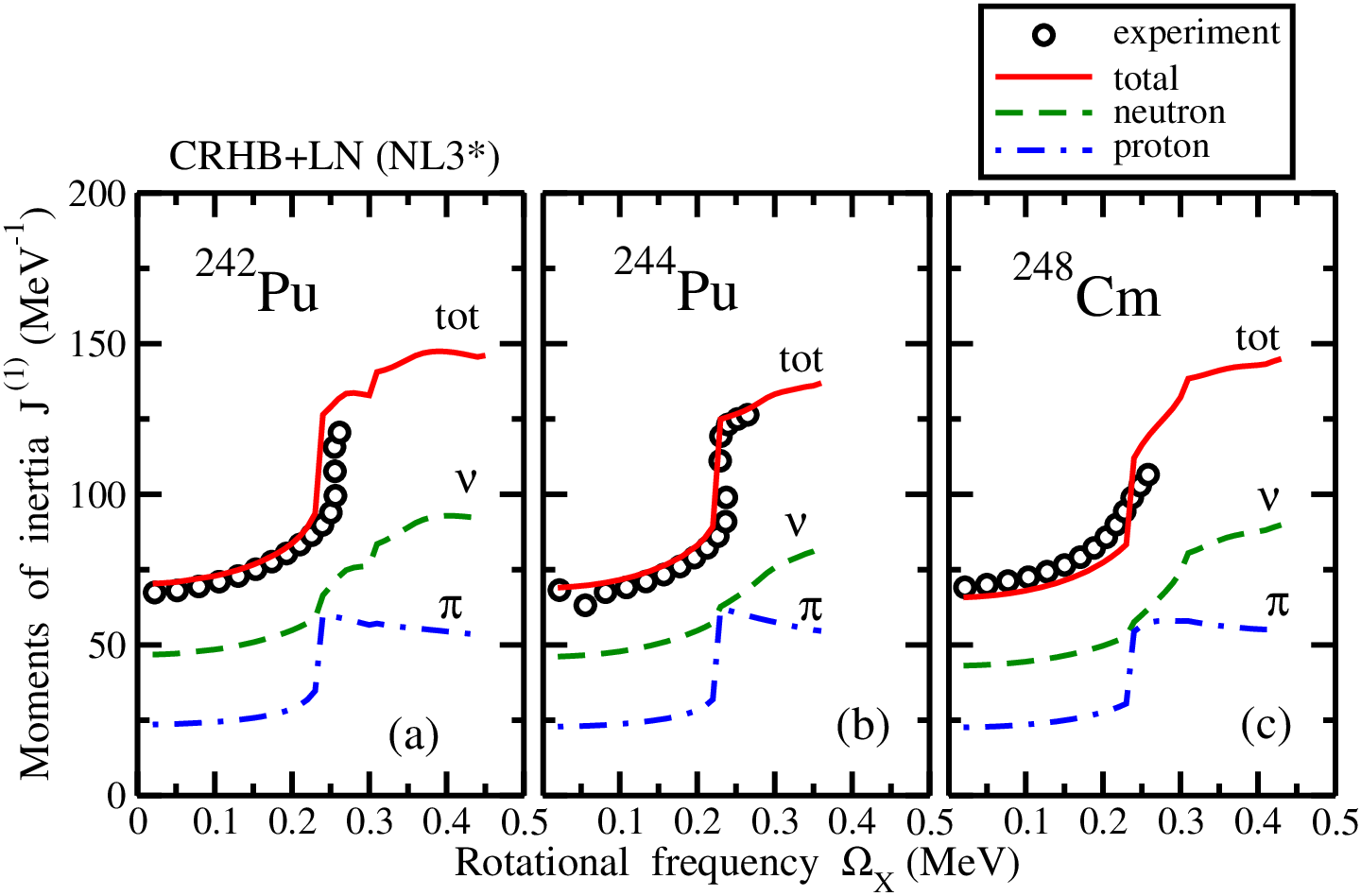}
\caption{(Color online) The same as Fig.\ \ref{backbending_NL1}
but for the results obtained with the NL3* parametrization.} 
\label{backbending_NL3*}
\end{figure}

\section{Rotational properties of even-even nuclei}
\label{Rot-even-even}

\begin{figure*}[ht]
\includegraphics[width=18.9cm,angle=90]{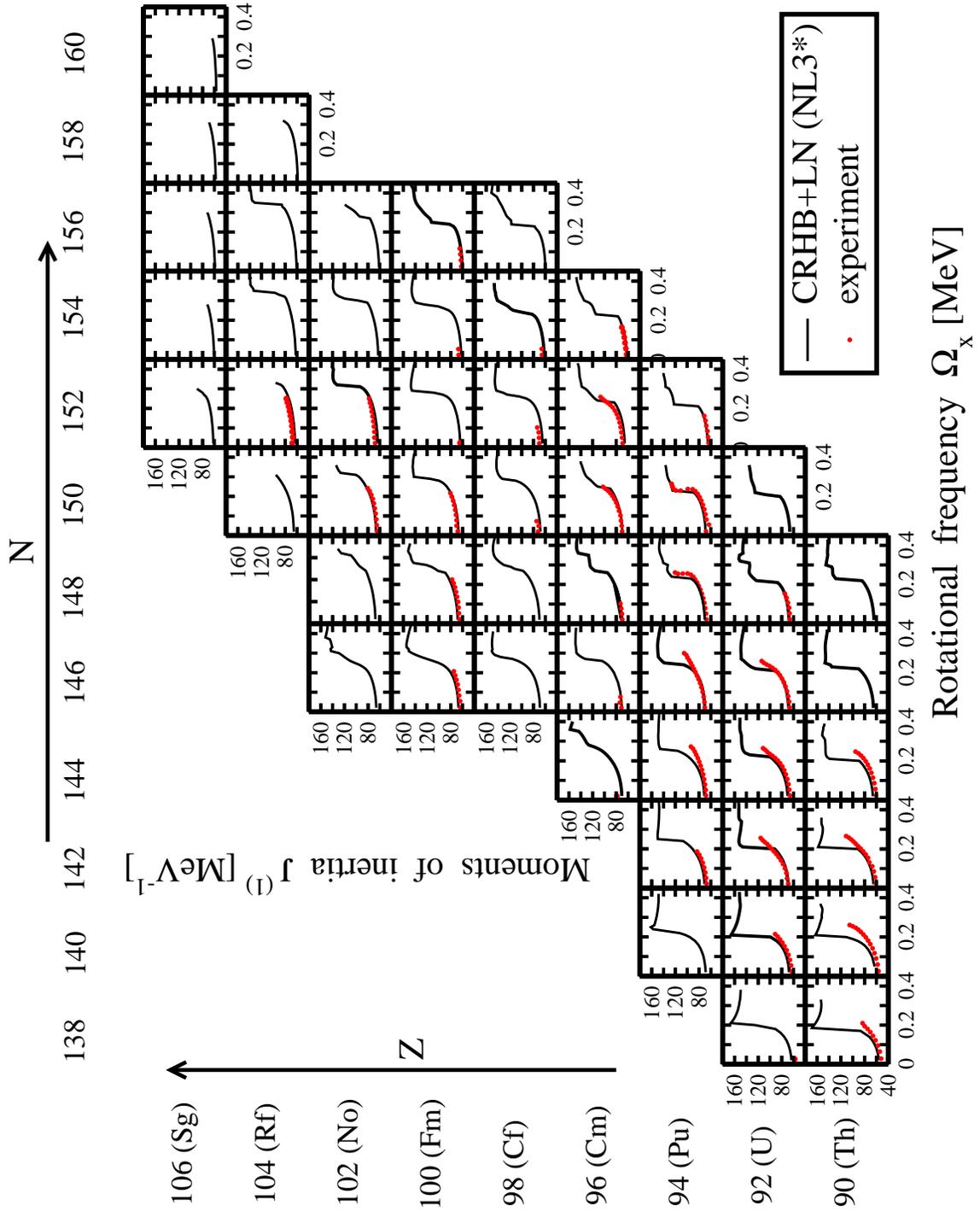}
\caption{(Color online) The experimental and calculated moments 
of inertia $J^{(1)}$ as a function of rotational frequency $\Omega_x$.
The calculations are performed with the NL3* parametrization of CDFT.
Calculated results and experimental data are shown by black lines 
and red small solid circles, respectively. Although some calculations 
suggest that $^{228}$Th is octupole soft (see Fig.\ 7 in Ref.\ 
\cite{BN.96}), its moment of inertia is rather well described in our 
calculations with no octupole deformation.} 
\label{sys-J1-NL3s}
\end{figure*}

\begin{figure*}[ht]
\includegraphics[width=20.0cm,angle=90]{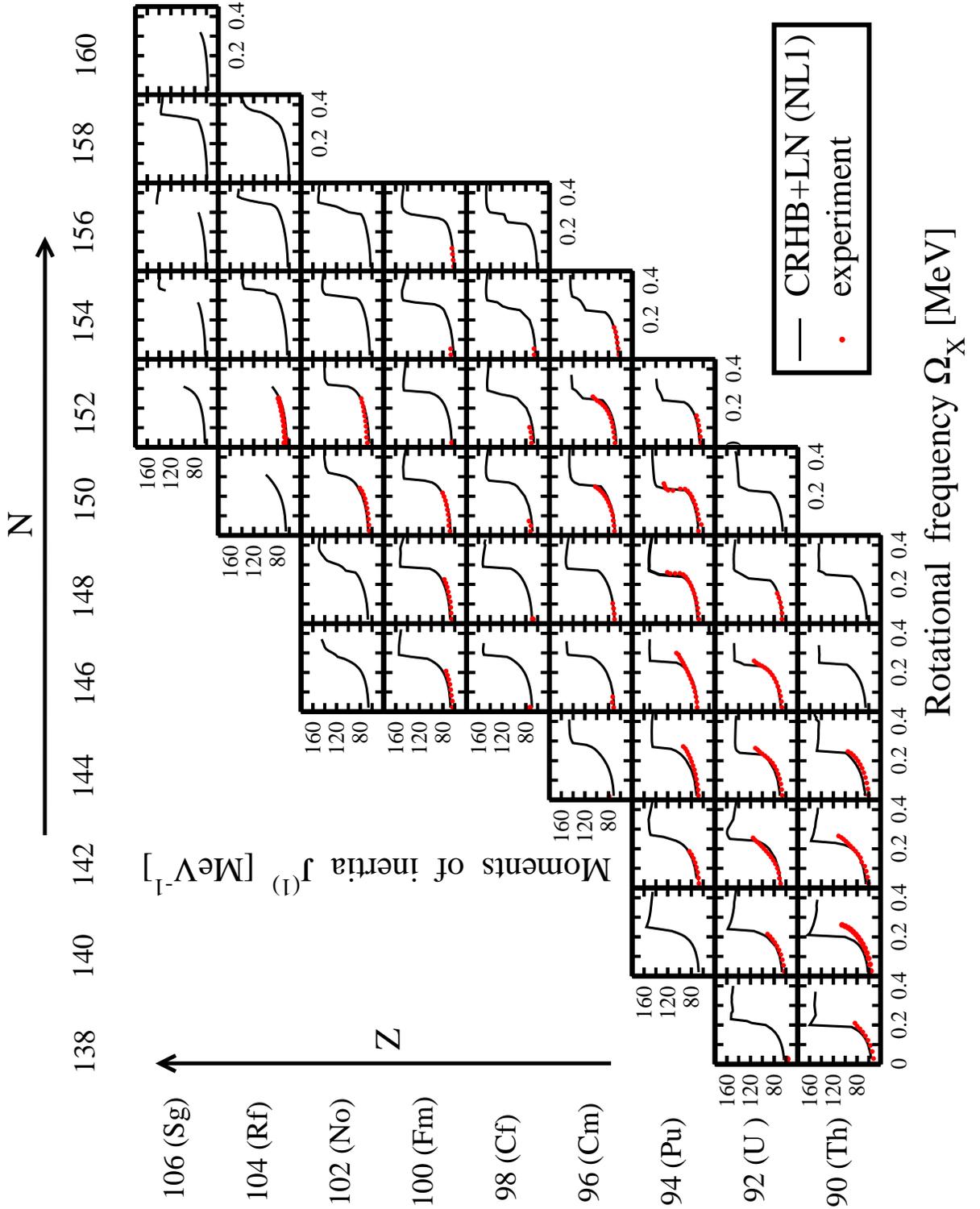}
\caption{(Color online) The same as Fig.\ \ref{sys-J1-NL1} but for
 the calculations with the NL1 parametrization.}
\label{sys-J1-NL1}
\end{figure*}

\begin{figure*}[ht]
\centering
\includegraphics[width=19.0cm,angle=90]{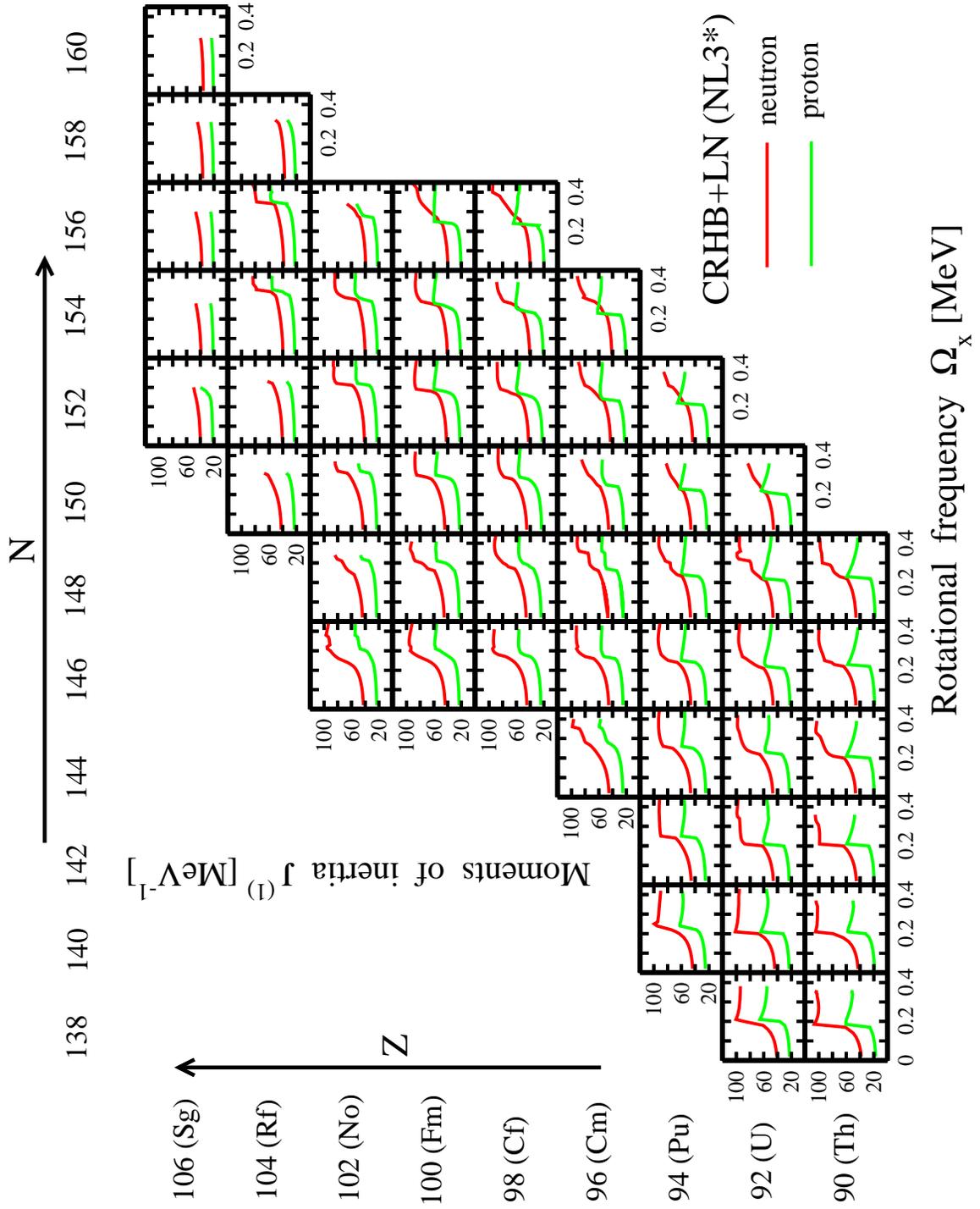}
\caption{(Color online) Calculated proton and neutron contribution
to kinematic moments of inertia $J^{(1)}$ as a function of rotational 
frequency $\Omega_x$. The calculations are performed with the NL3* 
parametrization of CDFT.} 
\label{J1_p_n_NL3s}
\end{figure*}

  Figs.\ \ref{sys-J1-NL3s} and \ref{sys-J1-NL1} show the results
of systematic calculations for the kinematic moments of inertia of
the ground state rotational bands in even-even actinides. Either 
sharp or more gradual increase of the kinematic moments of
inertia is observed at $\Omega_x \approx 0.2-0.30$ MeV. Fig.\ 
\ref{J1_p_n_NL3s} shows the proton and neutron contributions to 
the kinematic moments of inertia. These increases in $J^{(1)}$
are due to the alignments of the neutron $j_{15/2}$ and proton 
$i_{13/2}$ orbitals which in many cases take place at similar 
rotational frequencies (see Fig.\ \ref{J1_p_n_NL3s}). It is clear 
that the situation in actinides is more complicated than in the
rare-earth region in which the $h_{11/2}$ protons align 
substantially later than the $i_{13/2}$ neutrons. This 
simultaneous alignment of proton and neutron orbitals is also 
present in a number of  theoretical models discussed below.

  One can see that the CRHB+LN calculations rather well describe
the experimental kinematic moments of inertia and their evolution
with rotational frequency. The results obtained with the NL1 and 
NL3* parametrizations are rather similar in the majority of the
cases (compare Figs.\ \ref{sys-J1-NL1} and \ref{sys-J1-NL3s}).
Only in the $^{238,240}$Pu, $^{236,238}$U and $^{230,232}$Th 
nuclei, the CRHB+LN(NL1) calculations predict sharp upbending 
in $J^{(1)}$ at $\Omega_x \sim 0.2$ MeV which is not present 
in experimental data. The same problem exists also in the 
CRHB+LN(NL3*) calculations, but in addition to above mentioned 
set of nuclei an earlier alignment (as compared with experiment) 
is seen also in the calculations for the $^{234}$U and $^{234}$Th  
nuclei. This indicates that the calculated details of the band 
crossings depend more on the CDFT parametrization than the 
kinematic moments of inertia before band crossings.

  A number of theoretical calculations based on cranking model 
discussed below also do not reproduce the rotational properties
of these nuclei at the highest spins. This suggests that some 
effect not included in the model framework plays a role 
at the spins at which sharp band crossing takes place in model 
calculations. As discussed below, the stabilization of octupole
deformation at high spin is most likely candidate for this 
effect.

  So far, the observation of sharp upbending has been reported 
in journal publication only in $^{242,244}$Pu\footnote{Similar 
sharp upbendings have also been observed in ground state rotational 
bands of $^{246,250}$Cm and $^{250}$Cf \cite{HC.13}. Their 
properties are well described in the CRHB+LN calculations
with the NL1 and NL3* parametrizations.}. Figs.\ \ref{backbending_NL1} 
and \ref{backbending_NL3*} compare experimental data with model 
calculations. The backbending is complete in $^{244}$Pu 
and the CRHB+LN(NL3*) calculations rather well describe it 
(Fig.\ref{backbending_NL3*}b); the sharp alignment of the 
proton $i_{13/2}$ orbitals is a source of this backbending and
the neutron $j_{15/2}$ alignment proceeds gradually over
extended frequency range. On the contrary, sharp alignments
of the proton and neutron pairs take place at the same
frequency in the CRHB+LN(NL1) calculations (Fig.\ \ref{backbending_NL1}b) 
and they somewhat overestimate the kinematic moment of 
inertia above the band crossing. 
The same situation with the alignments of the proton 
$i_{13/2}$ and neutron $j_{15/2}$ pairs exists also in 
the CRHB+LN(NL1) and CRHB+LN(NL3*) calculations for  
$^{242}$Pu. They accurately reproduce the 
evolution of kinematic moments of inertia with frequency 
and the frequency of the paired band crossing (Figs.\ 
\ref{backbending_NL1}a and \ref{backbending_NL3*}a). However, 
since upbending is not complete in experiment it is impossible 
to judge whether the simultaneous sharp alignments of proton 
and neutron pairs really take place  in nature. 
Smooth upbending takes place in $^{248}$Cm (Fig.\ 
\ref{backbending_NL1}c). The CRHB+LN(NL1) [Fig.\ \ref{backbending_NL1}c] 
and CRHB+LN(NL3*) [Fig.\ \ref{backbending_NL3*}c] calculations 
suggests that this upbending is predominantly due to the proton 
$i_{13/2}$ alignment. However, the interaction between the g- 
and S-bands in the band crossing region is too week in the proton 
subsystem which leads to sharp upbending in model calculations.

 The calculated kinematic moments of inertia at the frequencies
below and above band crossing only weakly depend on the CDFT 
parametrization  (compare Figs.\ \ref{sys-J1-NL3s} and 
\ref{sys-J1-NL1}). On the contrary, the rate of the increase of 
the kinematic moment of inertia in the band crossing region of 
respective subsystem (proton or neutron) depends more sensitively 
on employed parametrization. Proton and neutron contributions to the 
kinematic moments of inertia obtained in the CRHB+LN(NL3*) 
calculations are shown in Fig.\ \ref{J1_p_n_NL3s}. In the CRHB+LN(NL1) 
calculations, the alignments of the proton and neutron pairs are 
similar to the ones of the CRHB+LN(NL3*) calculations in the majority 
of the cases; this is a reason why no figure similar to Fig.\ 
\ref{J1_p_n_NL3s} is presented for the CRHB+LN(NL1) results. 
The largest dependence on the parametrization is seen in the calculated 
alignments of the neutron $j_{15/2}$ pairs for which the increase of 
the neutron $J^{(1)}$ values in the band crossing region is either 
sharper or more gradual in the CRHB+LN(NL1) calculations
as compared with the CRHB+LN(NL3*) ones in $^{236,238}$Th, 
$^{238,240,242}$U, $^{236,240,242,244}$Pu, $^{240,244}$Cm,
$^{246,254}$Cf, $^{248,256}$Fm, and $^{258}$No. On the contrary, 
such differences are seen in proton subsystem only in $^{240,244}$Cm, 
$^{246}$Cf, and $^{246,248}$Fm. These differences between the 
CRHB+LN(NL3*) and CRHB+LN(NL1) results are in part due to 
the differences in the single-particle structure obtained in the 
NL1 and NL3* parametrizations \cite{AS.11}.

  Rotational properties of actinides have been in the focus 
of extensive studies within cranked shell model (CSM) 
\cite{ER.82,CF.83,247-249Cm-249Cf,ZZZZ.11,ZHZZZ.12}, rotating 
shell model \cite{ER.84}, and cranked Hartree-Fock-Bogolibov (CHFB) 
\cite{PF.82,LXW.12} approach  based on the phenomenological 
potentials (Woods-Saxon \cite{247-249Cm-249Cf,LXW.12} and Nilsson 
\cite{ER.84,CF.83,ZZZZ.11,ZHZZZ.12}) or quadrupole-quadrupole force 
Hamiltonian \cite{ER.82,PF.82}). Similar to our calculations, 
the simultaneous or near-simultaneous alignments of neutron 
$j_{15/2}$ and proton $i_{13/2}$ orbitals define rotational 
and band crossing properties. However, these calculations 
suffer from a number of simplifications such as fixed deformations 
\cite{CF.83,247-249Cm-249Cf,PF.82,ER.84,ZHZZZ.12}, reduced or/and fixed 
pairing gaps \cite{247-249Cm-249Cf,ER.84}, the absence of particle 
number projection \cite{ER.82,247-249Cm-249Cf,PF.82}, the restriction 
to axial symmetry \cite{CF.83,LXW.12,ZZZZ.11,ZHZZZ.12} or extensive 
local fit of model parameters to experimental data \cite{ER.84,ZHZZZ.12}. 
The cranked HFB calculations based on the Gogny D1S force 
\cite{ER.00,DGGL.06} have also been performed without particle 
number projection. Such simplifications are avoided in the current
CRHB+LN calculations.

  Let illustrate a typical situation by an example of extensive 
cranking calculations employing the universal parametrization of 
the Woods-Saxon (WS) potential \cite{247-249Cm-249Cf}. 
In these calculations, for each nucleus the deformation was  fixed 
at the value calculated for the ground state. It was found that 
the pairing gaps equal to 80\% of the value defined 
from five-point odd-even mass difference have to be used to better 
explain observed properties. However, required quenching of pairing 
gap has not been explained. The alignments 
of neutron $j_{15/2}$ and proton $i_{13/2}$ orbitals define rotational
and band crossing properties. However, there is no consistent 
explanation for the absence of experimental neutron $j_{15/2}$ band 
crossings in a number of nuclei. It was also concluded that based on 
available data it is difficult to determine how accurately the WS 
calculations predict the crossing frequencies
and interaction strengths.

  While  the CRHB+LN calculations well describe the sharp 
$i_{13/2}$ proton alignment observed in the rotational sequences 
of the $^{242-244}$Pu nuclei (see Sec.\ \ref{odd-neu-nuc} for $^{243}$Pu 
results), they fail to reproduce the absence of such alignments in 
$^{238,240}$Pu, $^{234,236,238}$U and 
$^{230,232,234}$Th (see Figs.\ \ref{sys-J1-NL1} and \ref{sys-J1-NL3s}).
Such problem exists in all cranking calculations, see references
in this section quoted above for details. Although the cranking 
calculations may not be completely adequate for the band-crossing 
region \cite{H.76}, a reasonable description of band crossings in 
$^{242-244}$Pu suggests that this is not main source of the deviations
between theory and experiment.

 It is quite likely that this problem is related to the stabilization 
of octupole deformation at high spin which is not taken into account in 
model calculations. Stable octupole deformation has been shown to 
delay alignment processes  \cite{FP.84} and this may explain the
differences between theory and experiment. Indeed, the analysis of 
the spectra of the ground state positive parity and lowest negative 
parity bands of $^{232}$Th, $^{238}$U and $^{240}$Pu indicates a second 
order phase transition from reflection-symmetric to reflection-asymmetric 
shapes in these bands (Ref.\ \cite{JBJ.12}). This phase transition takes 
place at spins $I\approx 12-15\hbar$. This analysis is based on the 
mathematical techniques of supersymmetric quantum mechanics, two-center 
octupole wave functions ansatz, and the Landau theory of phase transitions. 

  It was also suggested that strong octupole correlations in 
rotational bands of some actinides may be interpreted as the 
rotation-induced condensation
of octupole phonons having their angular momentum aligned with the 
rotational axis \cite{F.08}. When the rotation of the condensate and 
the quadrupole shape of the nucleus synchronize, the collective 
motion becomes the familiar rotation of a static octupole shape.
The experimental data on $^{238,240}$Pu \cite{Pu240} agrees with
such an interpretation and the experimental data in $^{238}$U shows 
the indications of this process \cite{238U}. Indeed, at the highest 
spins the yrast and the octupole bands in $^{238,240}$Pu appear to 
merge into a single sequence of levels with alternating spin and 
parity, and large intrinsic dipole moments were inferred from the 
measured B(E1)/B(E2) ratios \cite{Pu240}.  In addition, there
are indications of the formation of parity-doublets at high spin 
in $^{239}$Pu \cite{237U-239Pu}.  All that suggests the 
stabilization of octupole deformation at the highest spins in these nuclei. 
Furthermore, the systematics of the lowered energies of the $1^-$ states 
and the lowered hindrance factors in $\alpha$-decay populating these 
$1^-$ states suggest an increased octupole correlations for Pu and U 
nuclei with 144 and 146 neutrons \cite{SR.00}.

 New experimental data on $^{230}$Th shows the signatures of the 
stabilization of octupole deformation \cite{J-priv.12}. On the other 
hand, the extension of ground state and especially octupole vibrational 
rotational bands up to higher spin is needed in order to see whether 
this is also a case in $^{234,236}$U and $^{232,234}$Th nuclei.

 In the context of the study of rotational properties of actinides, 
it is interesting to mention that the same $j_{15/2}$ neutron and 
$i_{13/2}$ orbitals lie at the Fermi surface in the superdeformed 
nuclei of the $A\sim 190$ mass region. Remarkably, most superdeformed 
nuclei of this region exhibit a surprisingly smooth and 
gradual increase of their moments of inertia with frequency emerging
from the alignment of these orbitals and
this process is very well described in the CRHB+LN(NL1) calculations
\cite{CRHB}.

\section{Rotational properties of odd-mass nuclei}
\label{Rot-odd}

\begin{figure}[ht]
\includegraphics[width=8.6cm,angle=0]{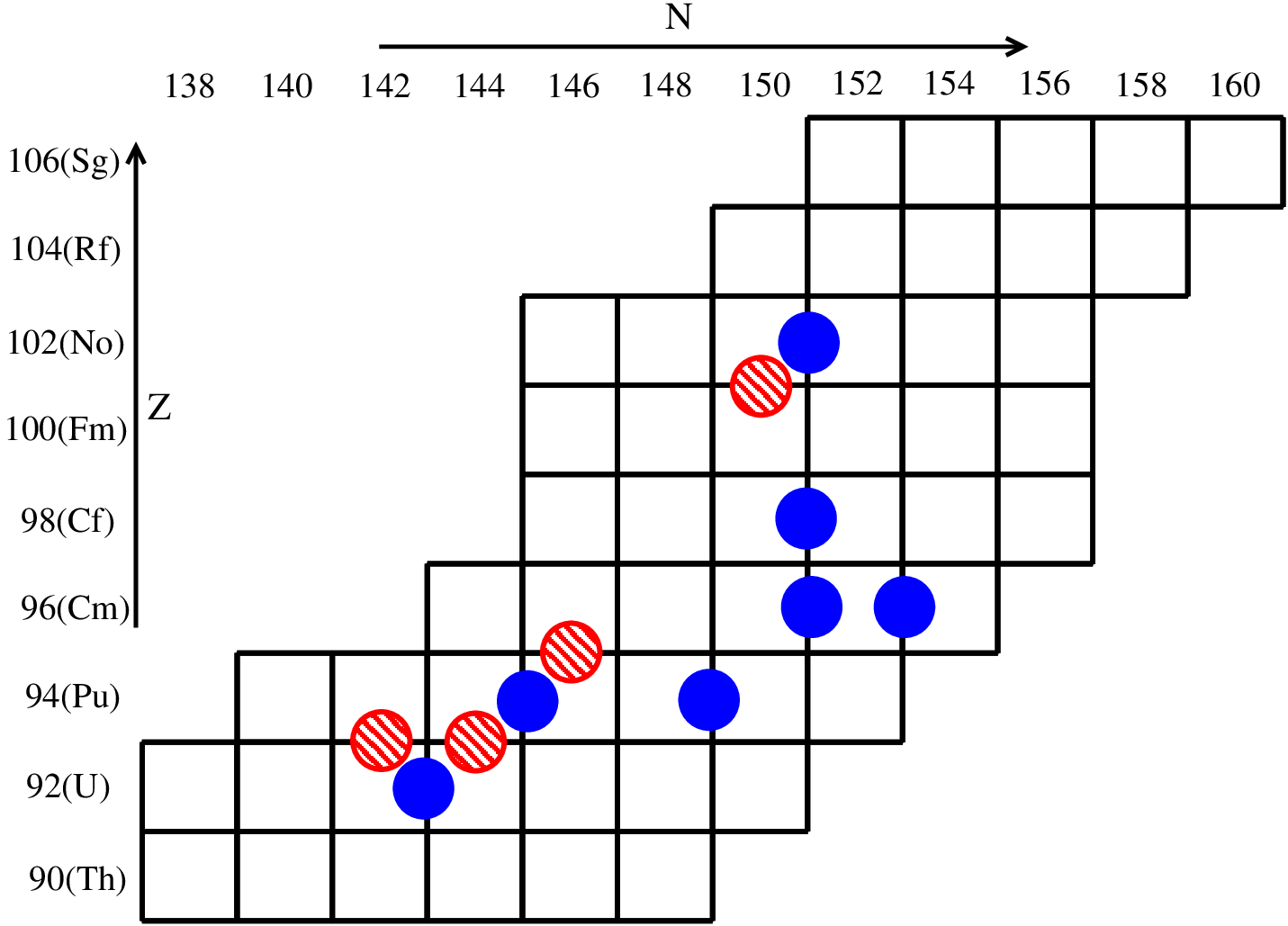}
\caption{(Color online) The chart of nuclei investigated in the current work. 
Solid blue and red shaded circles indicate studied odd-neutron
and odd-proton nuclei, respectively. The nucleus has $Z+1$ protons 
and $N$ neutrons if its circle is located between the $(Z,N)$ and 
$(Z+2,N)$ boxes. Alternatively, it has $Z$ protons  and $N+1$ neutrons 
if its circle is located between the $(Z,N)$ and $(Z,N+2)$ boxes.}
\label{chart-odd}
\end{figure}

\begin{figure*}[ht]
\includegraphics[width=8.6cm,angle=0]{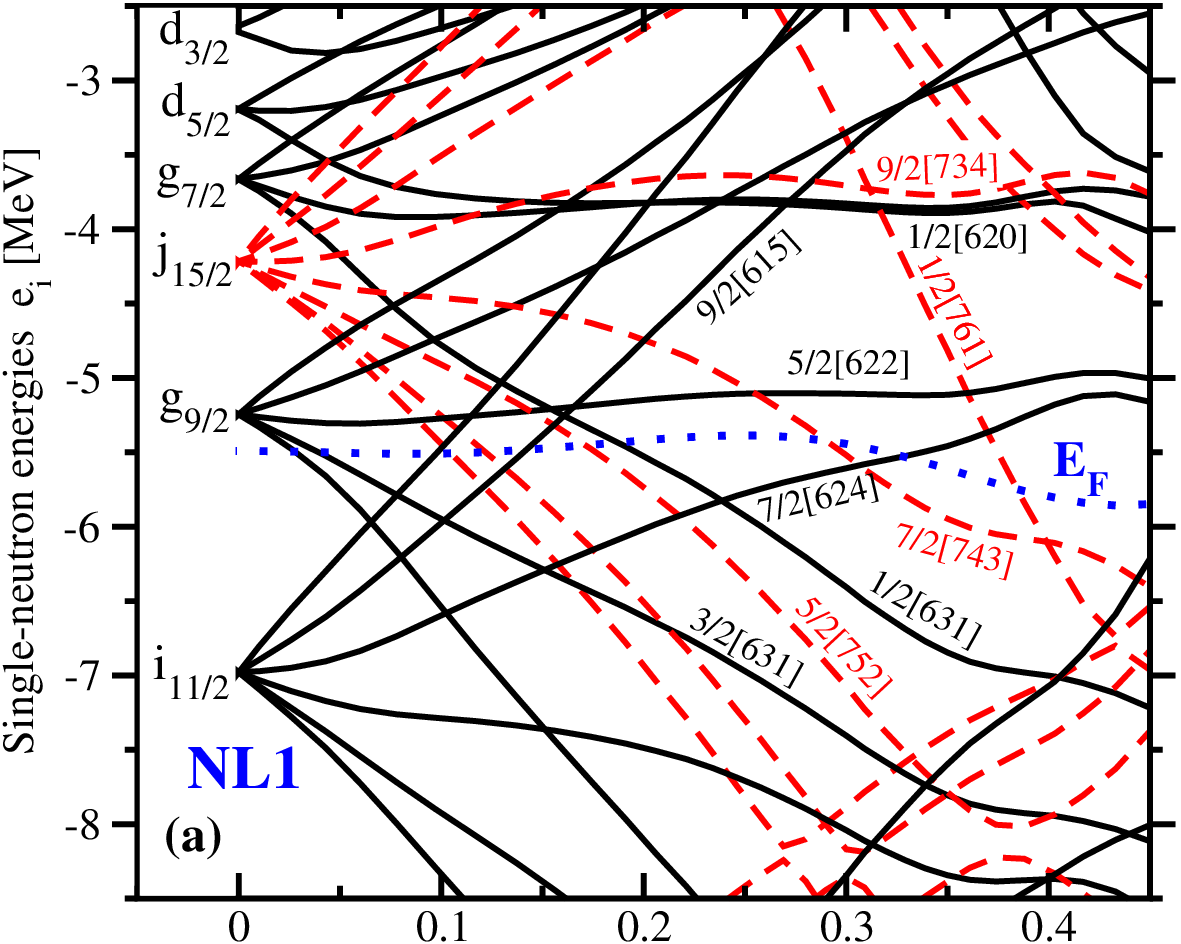}
\includegraphics[width=8.6cm,angle=0]{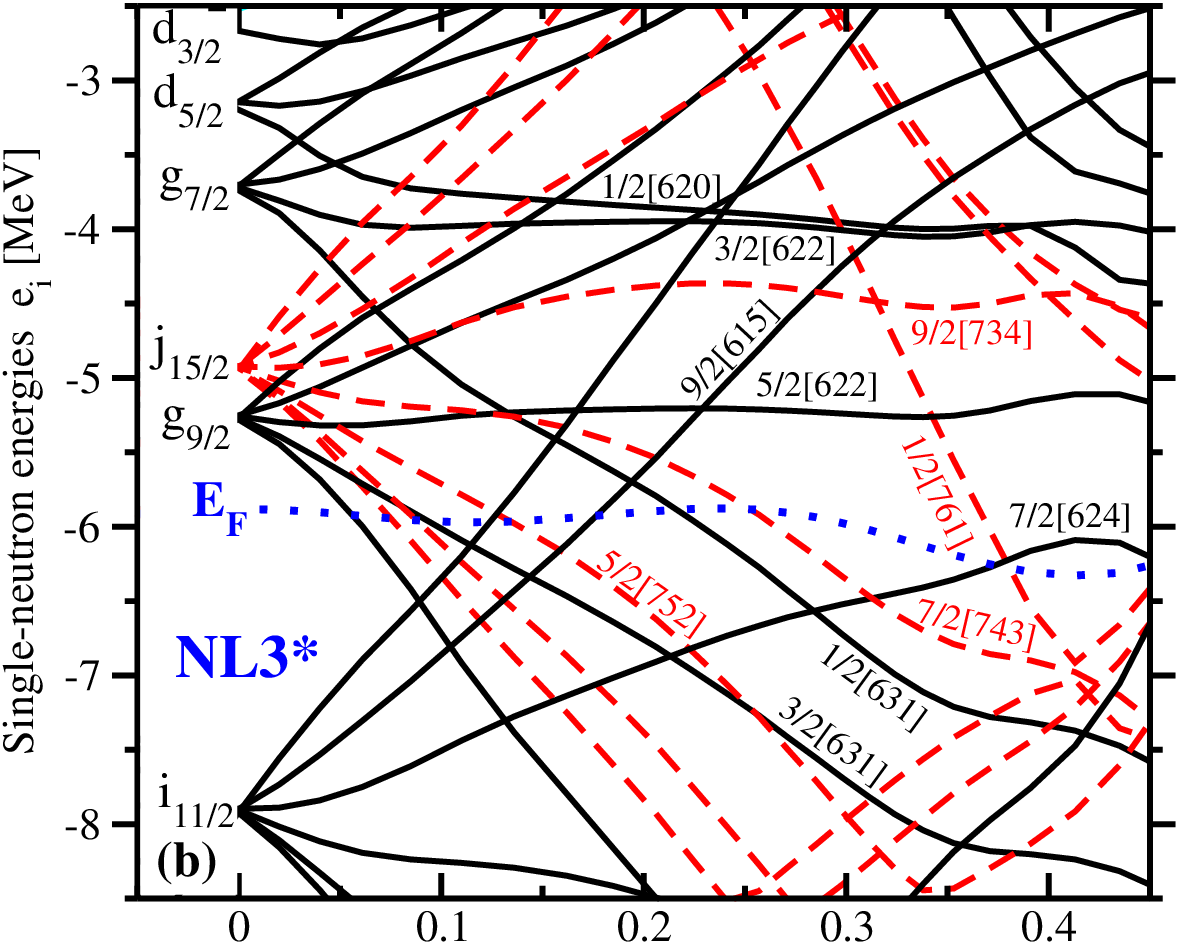}
\includegraphics[width=8.6cm,angle=0]{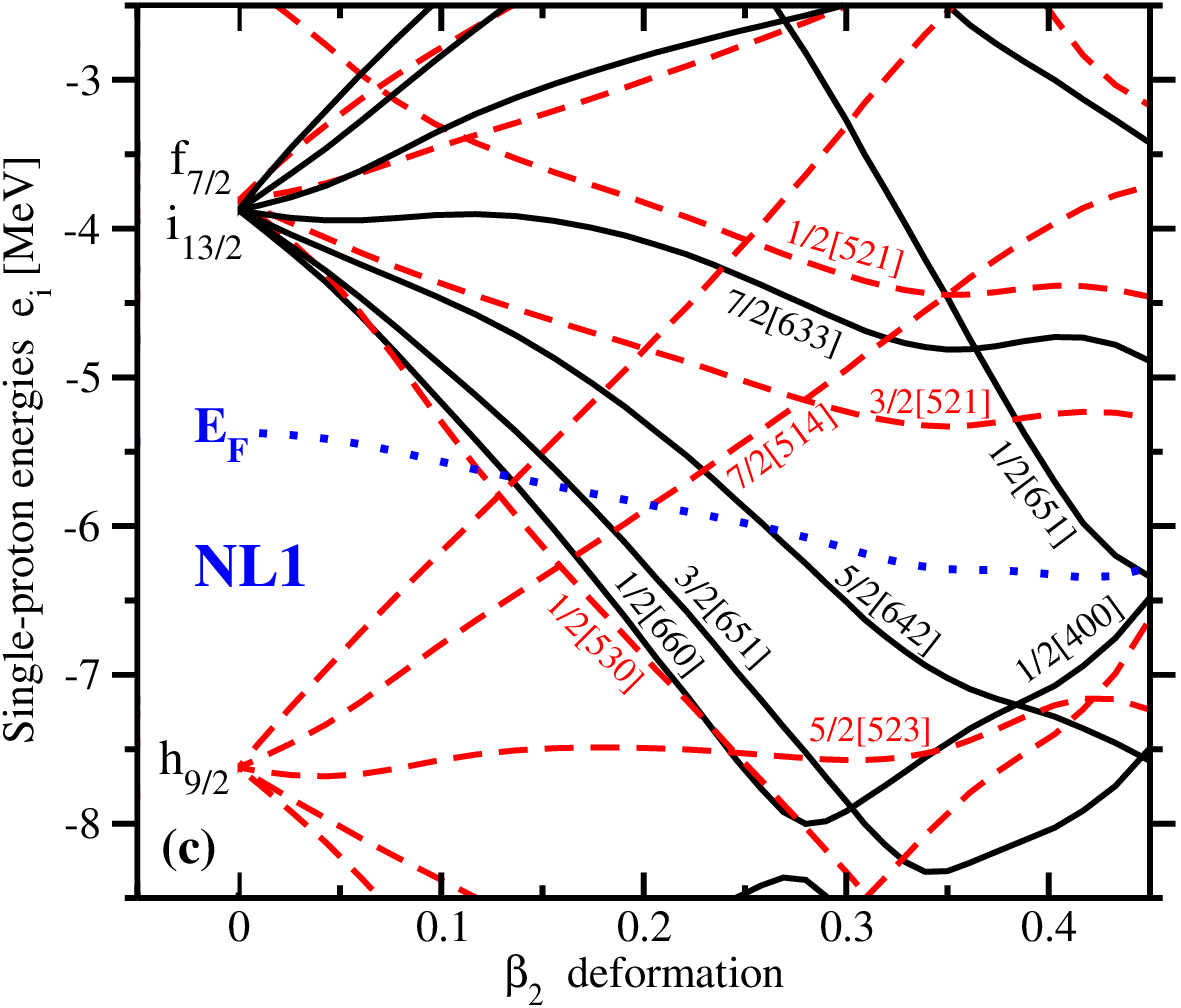}
\includegraphics[width=8.6cm,angle=0]{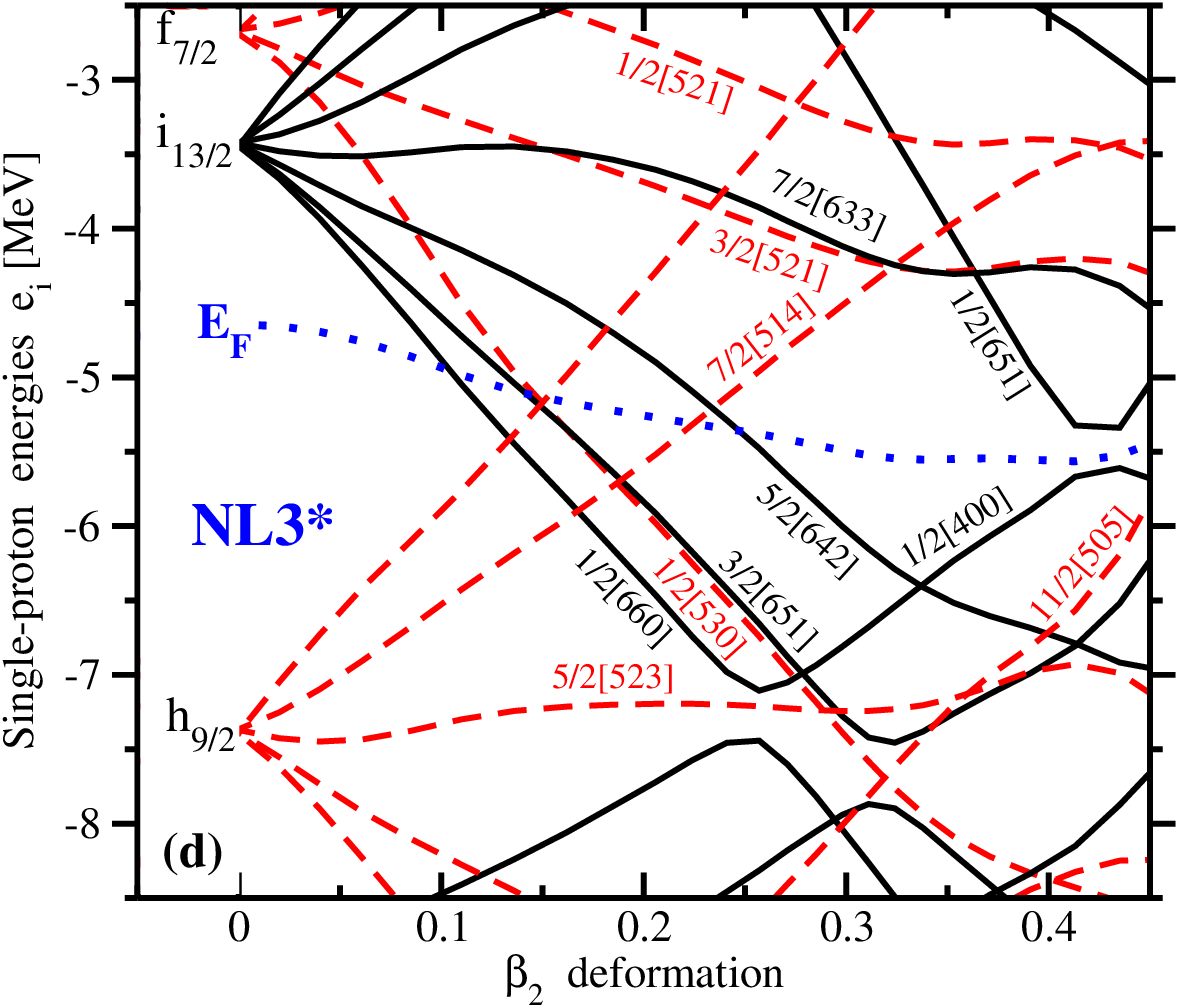}
\caption{(Color online) Proton and neutron single-particle energies
in $^{244}$Cm as a function of quadrupole deformation $\beta_2$
obtained in the calculations with the NL1 and NL3* parametrizations. Solid 
black and red dashed lines are used for positive and negative parity 
states, respectively. The energy of the Fermi level is shown by blue 
dotted line. Deformed single-particle orbitals of interest are labelled 
by the Nilsson quantum numbers $\Omega[Nn_z\Lambda]$.} 

\label{Nilsson-diagrams}
\end{figure*}

\begin{figure*}[ht]
\includegraphics[width=12.0cm,angle=0]{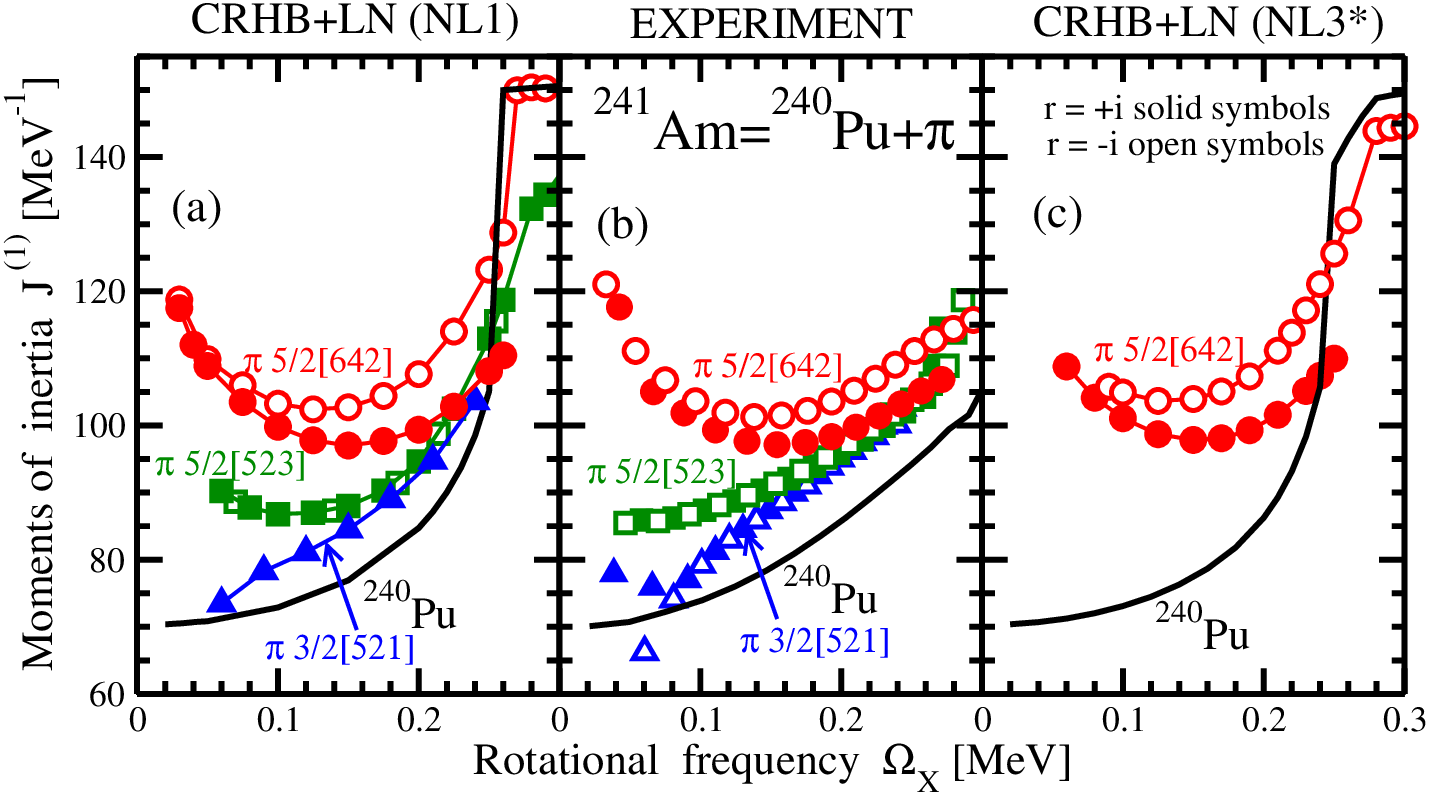}
\caption{(Color online) Calculated and experimental kinematic moments
of inertia $J^{(1)}$ of the indicated one-quasiproton configurations
in the $^{241}$Am nucleus and ground state rotational band in
reference even-even $^{240}$Pu nucleus. Experimental data are shown in
the middle panel, while the results of the CRHB+LN calculations with the
NL1 and NL3* parametrizations in the left and right panels, respectively. 
The same symbols/lines are used for the same theoretical and experimental 
configurations. The symbols are used only for the configurations in odd-mass 
nucleus; the ground state rotational band in reference even-even nucleus 
is shown by solid black line. The label with the following structure
``Odd nucleus = reference even+even nucleus + proton($\pi$)/neutron($\nu$)''
is used in order to indicate the reference even-even nucleus and the type 
of the particle (proton or neutron) active in odd-mass nucleus. The 
experimental data are from Refs.\ \cite{Pu240,241Am-237Np}.} 
\label{J1-241Am}
\end{figure*}

 Rotational properties of one-quasiparticle configurations provide 
an important information on the impact of odd particle/hole on alignment 
and pairing properties. They can also provide an additional constraint 
on the structure of single-particle states which is especially important 
for the light superheavy nuclei at the edge of the region where spectroscopic 
studies are still feasible (the nuclei with masses $A\sim 255$  and proton 
number $Z\geq 102$). This is because alternative methods of configuration 
assignment either provide the results with a low level of confidence or
are not possible \cite{INPC2010}. Unfortunately, our knowledge of the 
accuracy of the description of rotational properties of one-quasiparticle 
configurations in the DFT frameworks is extremely limited since no 
systematic investigation of such properties has been performed so far. 
Across the nuclear chart they were studied only in $^{251}$Md \cite{INPC2010}, 
$^{241}$Am \cite{INPC2010}, $^{253}$No \cite{No253} and $^{255}$Lr 
\cite{Lr255} within the CRHB+LN approach, in $^{253}$No 
\cite{DBH.01,No253,BBDH.03}, $^{255}$No, $^{251}$Md, and
$^{255}$Lr \cite{BBDH.03} as well as in superdeformed rotational bands
of $^{193}$Pb and $^{193}$Hg \cite{TFHB.97} in the cranked HFB approach 
based on Skyrme forces. To our knowledge, such studies have not been
performed in the cranked HFB approach based on the Gogny forces. It is 
also surprising that no systematic  investigation of rotational structures 
in odd-mass actinides is available in the MM approach; the occasional 
cranked shell model calculations characterized by a number of parameters 
adjusted to experimental data should not be considered as a replacement 
for fully fledged MM calculations.         

In order to fill this gap in our knowledge, a systematic investigation
of rotational properties of odd-mass actinides is performed in this 
manuscript. Even with present computational facilities, it is still 
non-trivial problem because of three reasons discussed below. 

  First, a proper description of odd nuclei implies the loss of  
time-reversal symmetry of the mean field, which is broken both by 
the unpaired nucleon \cite{AA.10} and by the rotation \cite{TO-rot}. 
As a consequence, time-odd mean fields and nucleonic currents, which 
cause the {\it nuclear magnetism} \cite{KR.89} have to be taken 
into account.

 Second, the effects of blocking due to odd particle have to be 
included in a fully self-consistent way. This is done in the CRHB+LN 
code according to Refs.\ \cite{RBM.70,EMR.80,RS.80}. The blocked 
orbital can be specified by different fingerprints such as 
\begin{itemize}
\item
  dominant main oscillator quantum number $N$ of the wave
  function, 

\item
  the dominant $\Omega$ quantum number ($\Omega$ is the projection 
  of the angular momentum on the symmetry axis) of the wave 
  function,

\item
  the particle or hole nature of the blocked orbital,

\item
  the position of the state within specific 
  parity/signature/dominant $N$/dominant $\Omega$ block,

\end{itemize} 
or their combination. For a given configuration, possible 
combinations of the blocked orbital fingerprints were defined 
from the analysis of calculated quasiparticle spectra in 
neighboring even-even nuclei and the occupation probabilities
of the single-particle orbitals of interest in these nuclei.

  Third,  variational solutions with blocked orbital(s) are numerically
less stable than the ones for the ground state bands in even-even nuclei. 
This is because at each iteration of the variational procedure blocked 
orbital has to be properly identified. This identification is complicated 
by the fact that $\Omega$ is not conserved quantum number in the CRHB+LN 
code. As a consequence, closely lying orbitals within a given parity/signature
block can interact and exchange a character. The convergence problems, 
emerging from the interaction of the blocked orbital with others, 
appear quite frequently. The interaction strength of these orbitals 
is one ingredient affecting the convergence. Another is the relative energies 
of interacting orbitals. Different CDFT parametrizations are characterized
by different single-particle specta \cite{A250} (see also Nilsson diagrams
presented in Fig.\ \ref{Nilsson-diagrams}). As a result, the 
convergence problems for specific blocked solution can show up in one 
parametrization but will not affect the solution in another 
parametrization. The structure of wave function of blocked orbital and 
the energy of this orbital with respect of other orbitals change as a 
function of rotational frequency. While converging in some frequency range 
the solution for a given blocked orbital may face the convergence problems 
outside this range. 
The convergence also depends on the initial conditions; for some
configurations the solution at 
given frequency converges if we start 
from self-consistent solution of the neighboring frequency point but 
does not converge if we start from the fields generated by the 
Woods-Saxon potential and diagonal $\Delta$-matrix. This feature has 
been used in the calculations. The employed combination of blocked 
orbital fingerprints also affects the numerical convergence; the 
solution can converge for one combination but face the convergence 
problems for another one. Thus, for a number of configurations several 
combinations of blocked orbital fingerprints have been used.


  The results of systematic calculations are presented in Figs.\ 
\ref{J1-241Am}, \ref{J1-np237},  \ref{J1-Md251}, 
\ref{J1-np235}, \ref{J1-247Cm},
\ref{J1-cf249}, \ref{J1-No253}, \ref{J1-u237},  \ref{J1-pu239},
\ref{J1-249Cm}, and \ref{J1-pu243}. All odd-mass nuclei with long 
rotational sequences are considered in this investigation; the
only exception is the $^{255}$Lr nucleus since the configuration
assignment for observed rotational structure is still under 
debate \cite{Lr255}. These nuclei are shown by circles in 
Fig.\ \ref{chart-odd}. Other odd-mass nuclei, the rotational 
sequences of which contain only few low-spin states, are 
ignored in this investigation since we are interested in the
evolution of rotational properties with spin. 

 Fig.\ \ref{Nilsson-diagrams} shows the Nilsson diagrams obtained 
for $^{244}$Cm in the calculations with the NL1 and NL3* parametrizations. 
This nucleus is located in the center of the region of odd-mass 
nuclei under study (Fig.\ \ref{chart-odd}). The single-particle
orbitals which can be observed in odd-mass nuclei of the region
under study are labelled by the Nilsson labels. Moreover, long rotational 
sequences built on some of these orbitals have been experimentally 
observed (see discussion in Secs.\ \ref{Odd-proton} and \ref{odd-neu-nuc}). 
Although the general structure of the Nilsson diagrams is the same in 
the NL1 and NL3* parametrizations, the relative energies of different
single-particle orbitals and their energies with respect of the Fermi 
level at the deformation $\beta_2 \sim 0.3$ typical for nuclei under 
study (see Fig.\ \ref{beta2-low}) depend strongly on parametrization.
For example, the $\pi 3/2[521]$ and $\pi 7/2[633]$ orbitals are nearly 
degenerate in the NL3* parametrization (Fig.\ \ref{Nilsson-diagrams}d). 
However, they are separated by 0.5 MeV gap in the NL1 parametrization 
(Fig.\ \ref{Nilsson-diagrams}c). Such differences in the energies of 
deformed states can be traced back to the differences in the 
single-particle energies at spherical shape \cite{A250}.

\begin{figure}[h]
\includegraphics[width=8.6cm,angle=0]{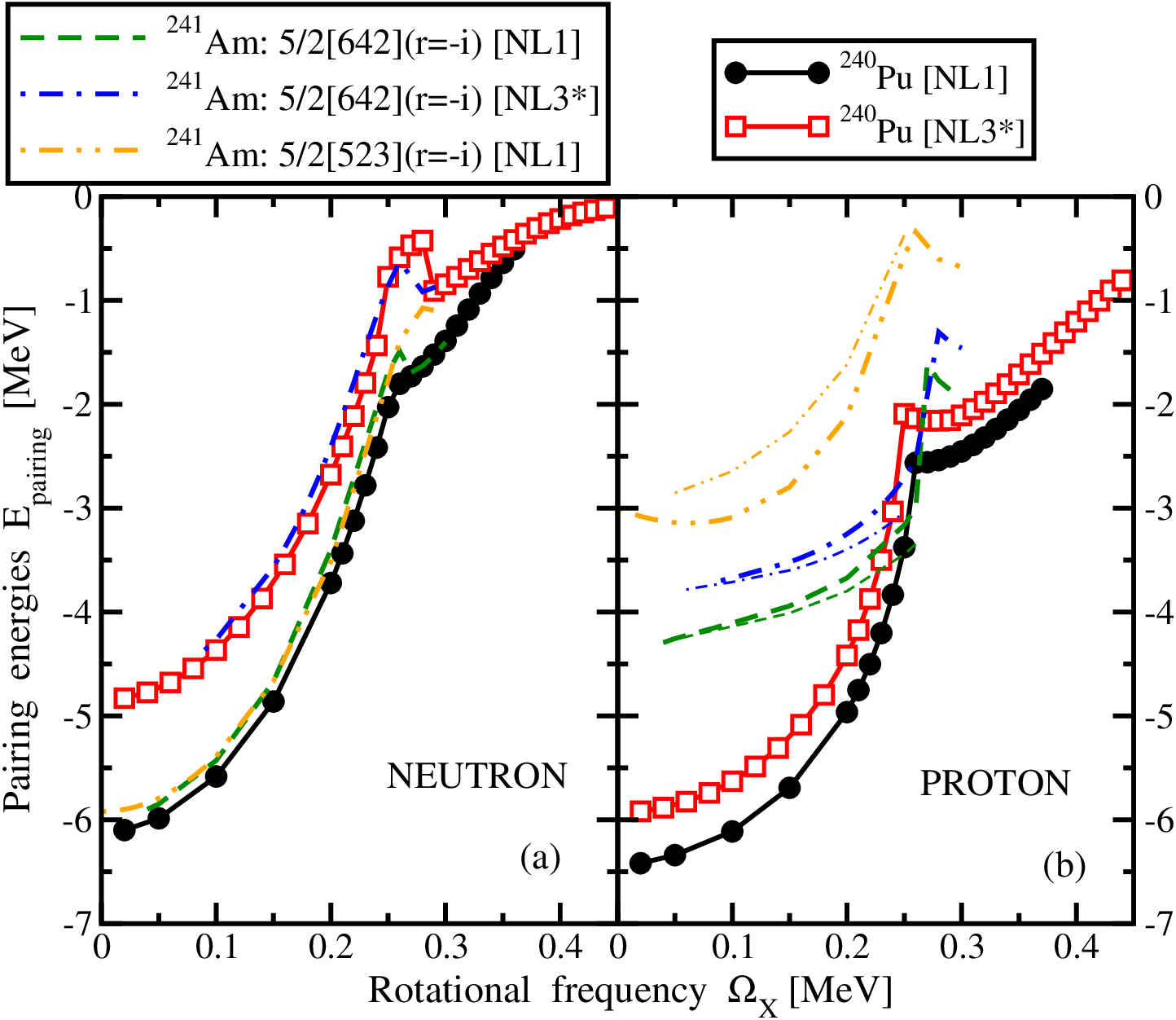}
\caption{(Color online) Calculated proton and neutron pairing 
energies in ground state rotational band of $^{240}$Pu and
one-quasiparticle rotational bands of $^{241}$Am. Thick and 
thin lines are used for the $(r=-i)$ and $(r=+i)$ branches
of one-quasiparticle configurations, respectively. Note that
neutron pairing almost does not depend on the signature of 
blocked proton orbital. As a result, only the $(r=-i)$ branches 
are shown in panel (a).} 
\label{pairing}
\end{figure}

\begin{figure*}[ht]
\includegraphics[width=12.0cm,angle=0]{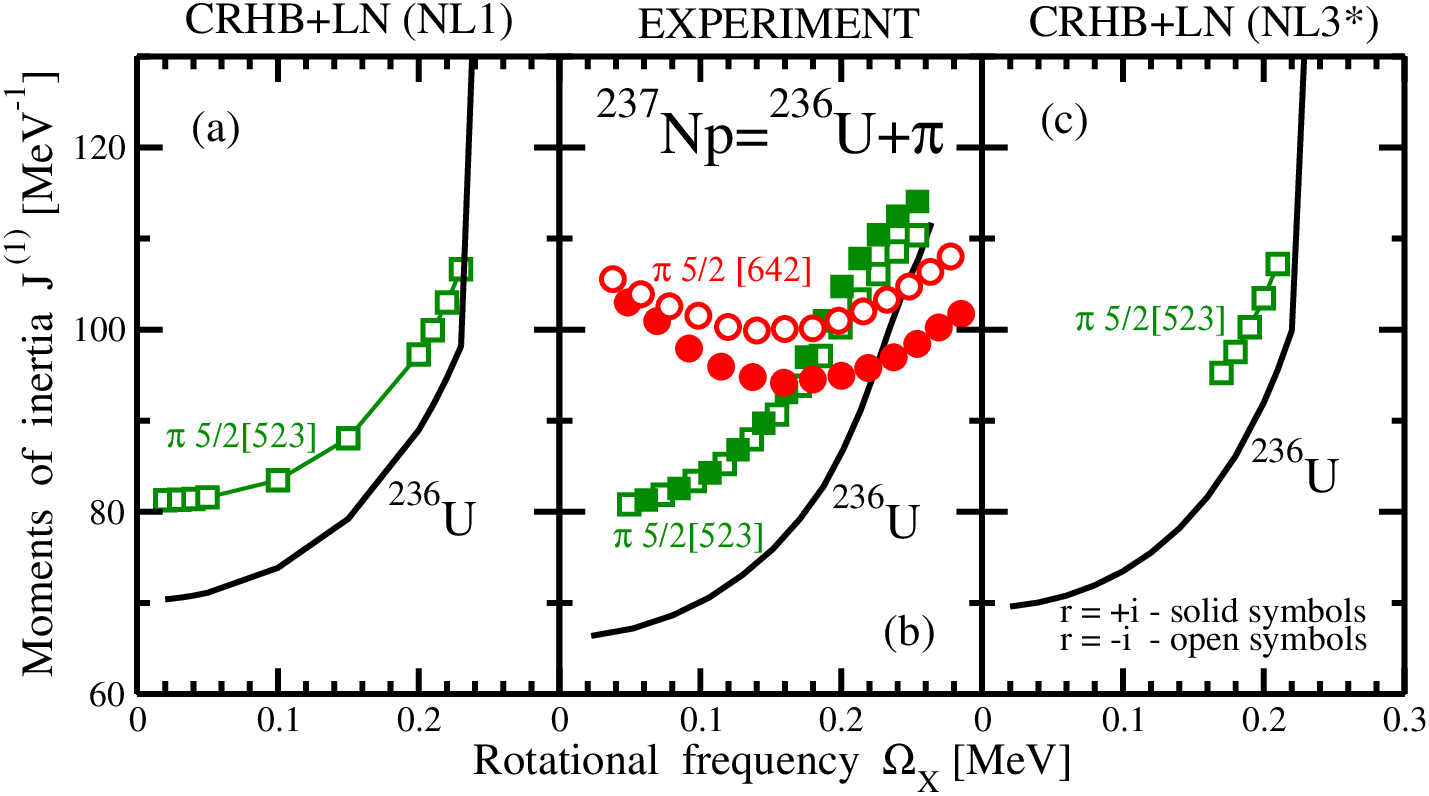}
\caption{(Color online)  The same as Fig.\ \ref{J1-241Am} but 
for $^{237}$Np. The experimental data are from Ref.\ \cite{241Am-237Np}.} 
\label{J1-np237}
\end{figure*}

  We compare experimental and calculated kinematic 
moments of inertia $J^{(1)}$ of  one-quasiparticle configurations
in odd-mass nuclei and ground state rotational band in reference
even-even mass nucleus. Figure \ref{J1-241Am} is an example of such
a comparison and the figures for other nuclei follow its
pattern. Two parametrizations, namely, NL1 and NL3*, are used
in the calculations in order to see how the results depend on the
parametrization. We drop the panel with specific parametrization 
when it was not possible to obtain the converged solution for it. 
The calculations were attempted  for all experimentally observed 
configurations of odd-mass nuclei indicated in Fig.\ 
\ref{chart-odd}; the absence of calculated curve for specific 
configuration indicates
that no convergence has been obtained for it.

\subsection{Odd-proton nuclei}
\label{Odd-proton}

 Long rotational bands based on different single-particle 
orbitals have been observed in odd-proton $^{241}$Am, $^{235,237}$Np 
and $^{251}$Md nuclei. We discuss them below separately, 
nucleus by nucleus.

 {\bf The $^{241}$Am nucleus.} The rotational bands based 
on the Nilsson orbitals $\pi 5/2[642]$ (from the $i_{13/2}$ 
spherical subshell), $\pi 5/2[523]$ (from the $h_{9/2}$ subshell) 
and  $\pi 3/2[521]$ (from the $f_{7/2}$ subshell) have been 
observed in this nucleus. As can be seen in Fig.\ \ref{J1-241Am}b, 
at low frequencies they have distinctly different kinematic moments 
of inertia $J^{(1)}$. Theoretical calculations (Fig.\ 
\ref{J1-241Am}a,c) describe well the absolute values of the 
kinematic moments of inertia of different configurations and their 
evolution with rotational frequency. In particular, the splitting 
of two signatures of the $\pi 5/2[642]$ configuration is  
well described in the model calculations. The results of the 
CRHB+LN(NL1) and CRHB+LN(NL3*) calculations for this configuration 
are similar.

  On the contrary, the $\pi 5/2[523]$ and $\pi 3/2[521]$ bands show 
(with exception of very low frequencies in the case of the 
$\pi 3/2[521]$ band) no signature splitting. The CRHB+LN(NL1) 
calculations for the two signatures of the $\pi 5/2[523]$ configuration 
show explicitly this feature. Unfortunately, it was not possible to get 
a convergence in the case of the $\pi 3/2[521](r=+i)$ configuration.
However, the analysis of the quasiparticle routhian diagram confirms 
that the $\pi 3/2[521](r=\pm i)$ configurations have to be degenerate in 
energy up to rotational frequency $\Omega_x \sim 0.16$ MeV in 
agreement with experimental observations. At higher frequencies, 
small signature separation is expected in the calculations.

 In addition to the above mentioned features, the relative properties
of different bands both with respect of each other and with respect to 
ground state band in reference nucleus $^{240}$Pu are well described 
in the model calculations. The increase of the kinematic moment of inertia
in the bands of $^{241}$Am as compared with the one of ground state band in
$^{240}$Pu is caused by the blocking effect which results in a
decreased proton pairing (see Fig.\ \ref{pairing}).

\begin{figure*}[ht]
\includegraphics[width=12.0cm,angle=0]{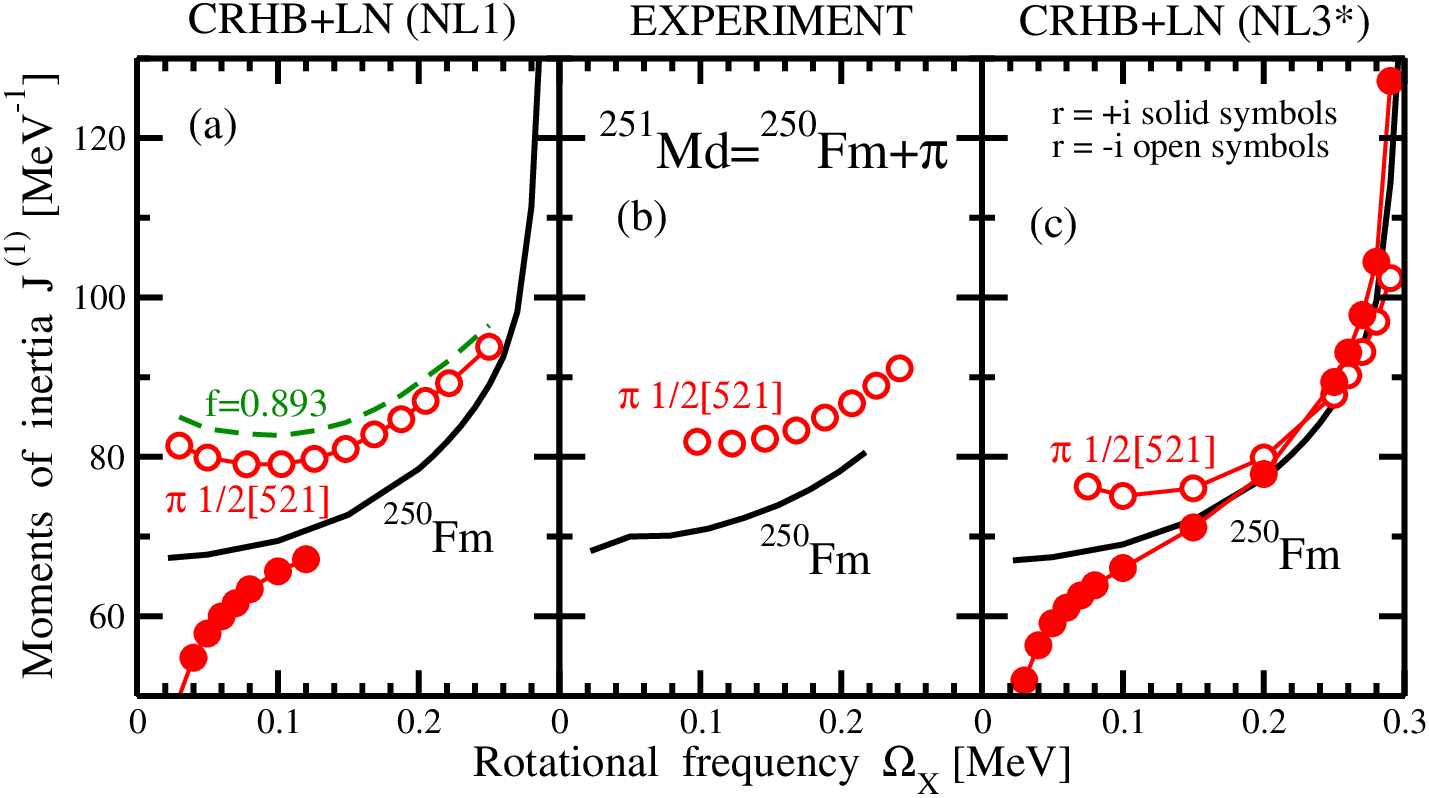}
\caption{(Color online) The same as Fig.\ \ref{J1-241Am} but 
for $^{251}$Md. Experimental data are taken from Refs.\ 
\cite{Fm250,Md251}.} 
\label{J1-Md251}
\end{figure*}

\begin{figure*}[ht]
\includegraphics[width=12.0cm,angle=0]{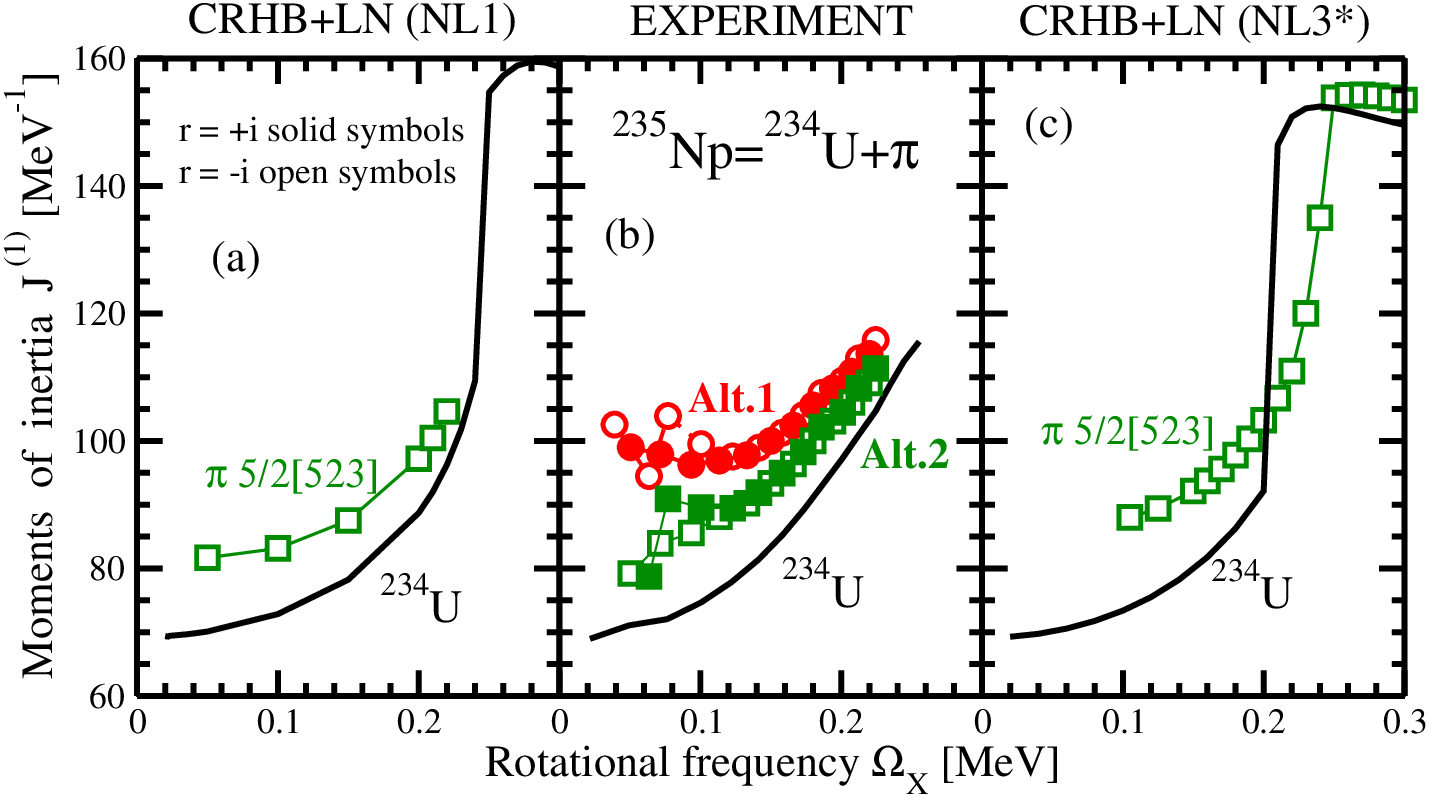}
\caption{(Color online) The same as Fig.\ \ref{J1-241Am} but 
for $^{235}$Np. Experimental data are taken from Ref.\ 
\cite{235Np}. See text for details.} 
\label{J1-np235}
\end{figure*}

 {\bf The $^{237}$Np nucleus.} Similar to $^{241}$Am, the $\pi 5/2[642]$ 
and $\pi 5/2[523]$ rotational bands have been observed in this nucleus. 
Unfortunately, it was possible to obtain a convergent solution only for 
the $\pi 5/2[523](r=-i)$ configuration (Fig.\ \ref{J1-np237}).  This 
configuration rather well describes both its experimental counterpart 
and relative properties of the $\pi 5/2[523](r=-i)$ band in $^{237}$Np 
and ground state band in $^{236}$U. The CRHB+LN(NL1) and CRHB+LN(NL3*)
results are very similar for this configuration. 
In the quasiparticle routhian diagram, the $\pi 5/2[523](r=\pm i)$ 
orbitals are degenerate in energy up to rotational frequency 
$\Omega_x \sim 0.15$ MeV with small signature separation developing 
at higher frequencies.  These features agree with experimental 
observations.

\begin{figure}[h]
\includegraphics[width=8.6cm,angle=0]{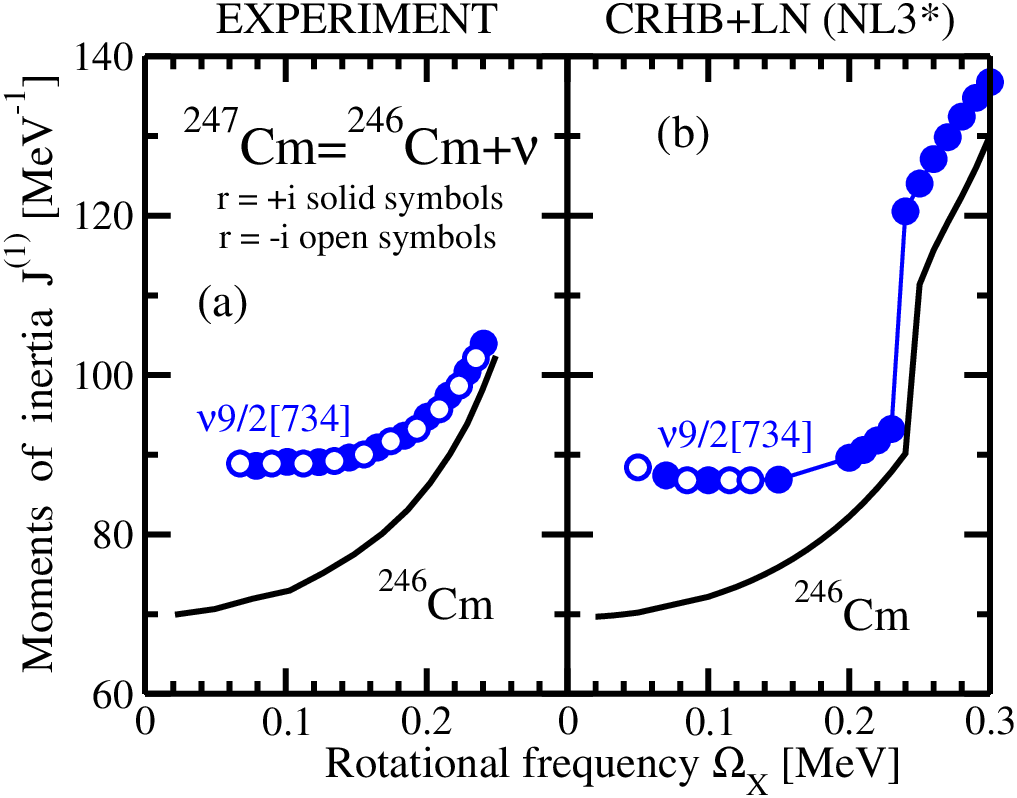}
\caption{(Color online)  The same as Fig.\ \ref{J1-241Am} but 
for $^{247}$Cm. Experimental data are taken from Ref.\ 
\cite{247-249Cm-249Cf}.} 
\label{J1-247Cm}
\end{figure}

{\bf The $^{251}$Md nucleus.} 
The $\pi 1/2[521](r=-i)$ configuration has been assigned to
single decoupled band observed recently in the odd-proton nucleus 
$^{251}$Md \cite{Md251}. In the CRHB+LN(NL1) calculations, this 
configuration accurately describes both its experimental counterpart 
and the relative properties of the $\pi 1/2[521](r=-i)$ band in 
$^{251}$Md and ground state band in $^{250}$Fm (Fig.\ 
\ref{J1-Md251}a,b). On the contrary, it somewhat underestimates 
experimental $J^{(1)}$ values and the difference between the
$J^{(1)}$ values of the $\pi 1/2[521](r=-i)$ band in $^{251}$Md 
and the ground state band in $^{250}$Fm (Fig.\ \ref{J1-Md251}b,c)
in the CRHB+LN(NL3*) calculations.

\begin{figure}[h]
\includegraphics[width=8.6cm,angle=0]{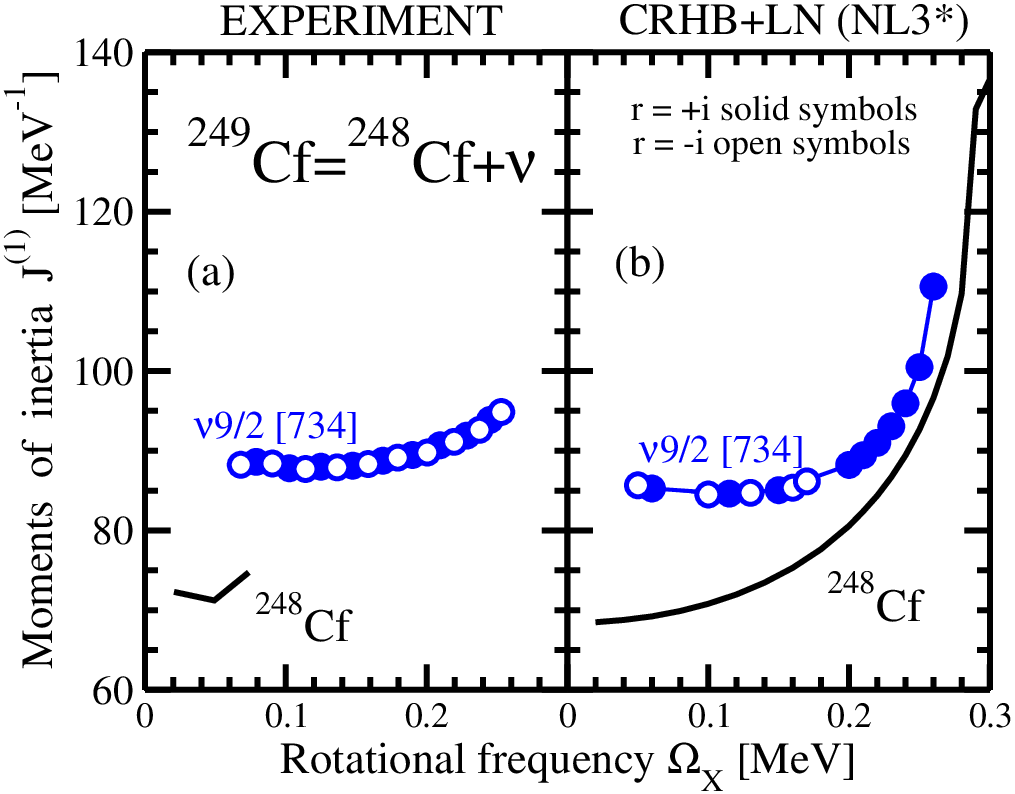}
\caption{(Color online) The same as Fig.\ \ref{J1-241Am} but 
for $^{249}$Cf. Experimental data are taken from Ref.\ 
\cite{247-249Cm-249Cf}.} 
\label{J1-cf249}
\end{figure}

The results of the CRHB+LN(NL1) calculations for this configuration 
obtained using an average scaling factor $f_{av}=0.9147$ defined in 
the current manuscript (open red circles) are compared with the ones 
(dashed green line) obtained employing scaling factor $f=0.893$ of 
Ref.\ \cite{A250} in Fig.\ \ref{J1-Md251}a. One can see that the 
results of the calculations are very sensitive to actual value of 
$f$; the modification of scaling factor by 2\% leads to visible 
changes in the calculated $J^{(1)}$ values but do not change much 
the slope of calculated $J^{(1)}$ curve.

\begin{figure*}[ht]
\includegraphics[width=12.0cm,angle=0]{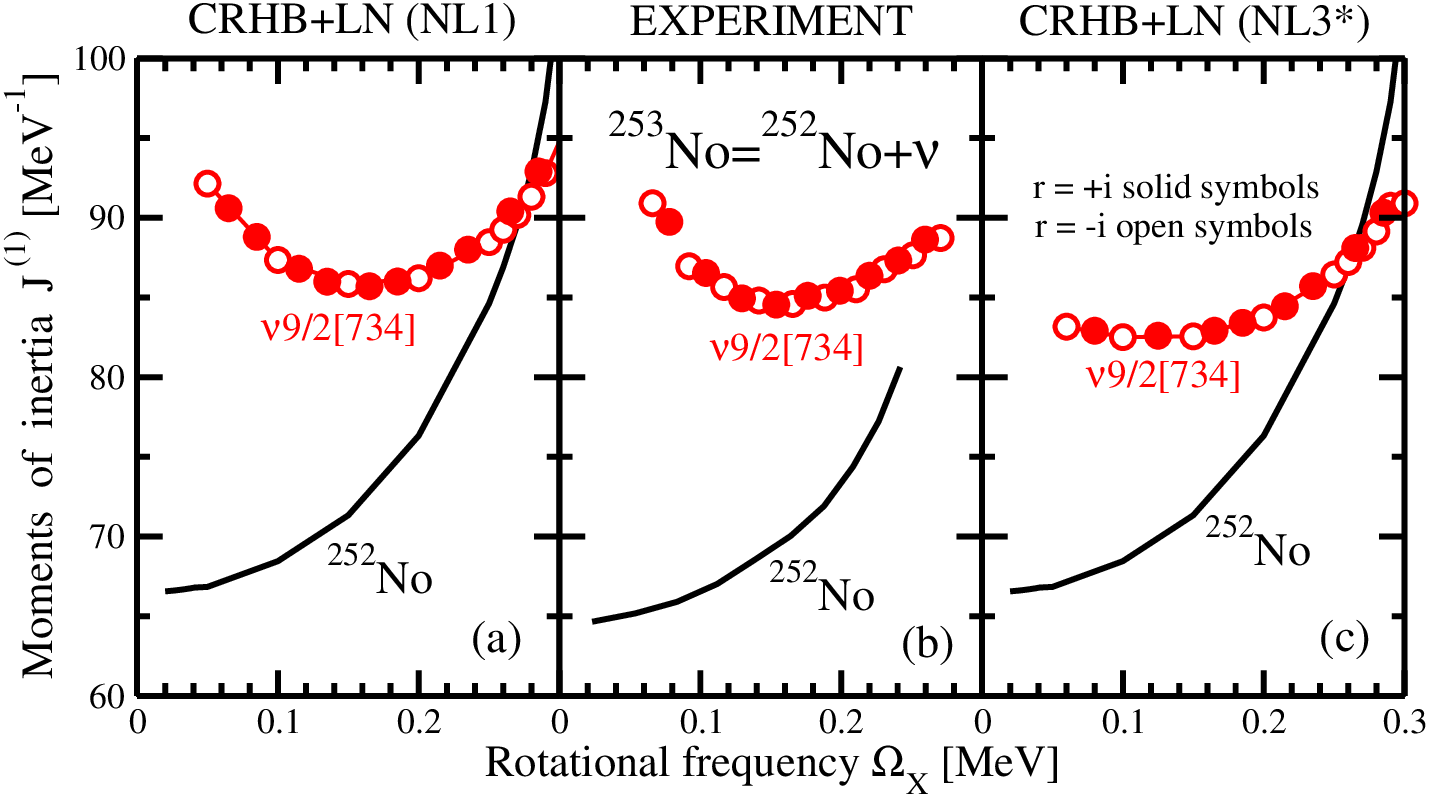}
\caption{(Color online) The same as Fig.\ \ref{J1-241Am} but 
for $^{253}$No. Experimental data are taken from Refs.\ 
\cite{No252,No253}.} 
\label{J1-No253}
\end{figure*}

 Note that the CRHB+LN(NL1) calculations of Ref.\  \cite{INPC2010} 
for $^{251}$Md have been performed with $f=0.893$. These calculations 
showed that the $\pi 7/2[633]$, $\pi 3/2[521]$, $\pi 9/2[624]$ and 
$\pi 9/2[505]$ configurations cannot be theoretical counterparts of 
observed decoupled band because they lead either to signature degenerate 
bands or to the bands with small signature splitting. As a consequence, 
both signatures are expected to be observed in experiment. 


 {\bf The $^{235}$Np nucleus.}  Two rotational sequences, presumably
the two signatures of the ground state band, have been observed
in Ref.\ \cite{235Np}. The authors of this reference proposed the 
$\pi 5/2[624]$ configuration for this band. Fig.\ \ref{J1-np235}b shows 
the kinematic moments of inertia of these two sequences by  
solid and open red circles under spin/parity assignments of 
Ref.\ \cite{235Np}; these curves are labeled as 'Alt. 1'. The origin 
of the disturbances visible at low frequency in the $J^{(1)}$ 
values of the $r=-i$ sequence is not clear. However, these two 
sequences are signature degenerate above $\Omega_x \geq 0.1$ 
MeV. This is in clear contrast with the behavior of the
$\pi 5/2[624]$ bands in $^{237}$Np (Fig.\ \ref{J1-np237}b) and $^{241}$Am
(Fig.\ \ref{J1-241Am}b), the two signatures of which show substantial
signature splitting. The signature splitting is usually a
robust fingerprint of the configuration. In addition, the removal
of two neutrons from $^{237}$Np should not change the signature
splitting in the proton $\pi 5/2[642]$ band since this process will not 
change the deformation substantially. Because of these two 
reasons, we believe that the $\pi 5/2[642]$ configuration 
assignment is not well justified.

 The $\pi 5/2[523]$ configuration has also been mentioned as a
possible (but less likely) candidate for observed band in Ref.\ 
\cite{235Np}. The $\pi 5/2[523]$ band is signature degenerate
in $^{241}$Am (Fig.\ \ref{J1-241Am}b), but it develops small signature
splitting at high spin in $^{237}$Np (Fig.\ \ref{J1-np237}b). From 
our point of view, the $\pi 5/2[523]$ configuration assignment for 
observed band in $^{235}$Np is more likely than the assignment of the 
$\pi 5/2[642]$ configuration because of the reasons discussed below.
However, such reassignment would require the modification of the level scheme 
in $^{235}$Np which is not prohibited since there is neither firm
evidence for the lowest member of each rotational sequence
nor firm parity assignment \cite{235Np}.
Thus, we suggest the following modifications.
The (78) keV transition linking the $9/2^+$ 
and $5/2^+$ states in the sequence labeled as 1 in Fig.\ 5 
of Ref.\ \cite{235Np} as well as the $5/2^+$ state
have to be dropped from the level scheme and the
spins of observed states have to be lowered by $1\hbar$ (so the 
sequence 1 runs from $I^{\pi}=7/2^-$ up to $I^{\pi}=51/2^-$).
The spins of the states in sequence 2 have also to be lowered
by $1\hbar$, so this sequence runs from $I^{\pi}=5/2^-$ up 
to $I^{\pi}=49/2^-$. With these modifications this band looks
very similar to the $\pi 5/2[523]$ band in $^{237}$Np shown in Fig.\ 1
of Ref.\ \cite{241Am-237Np}.

  The kinematic moments of inertia of observed sequences under 
these spin/parity changes are shown by open and closed green squares 
in Fig.\ \ref{J1-np235}b; these curves are labeled as 'Alt.2'. 
One can see that this alternative is rather well 
described by the $\pi 5/2[523](r=-i)$ configuration both in terms of 
absolute $J^{(1)}$ values and their evolution with spins
(Fig.\ \ref{J1-np235}). In 
addition, the relative properties of the bands in $^{235}$Np and 
$^{234}$U are rather well reproduced in model calculations. 
Unfortunately, it was not possible to obtain opposite signature 
configuration in model calculations. However, the analysis of the 
routhian diagrams in $^{235}$Np and the results of the calculations for the 
$\pi 5/2[523](r=\pm i)$  configurations in 
$^{241}$Am (Fig.\ \ref{J1-241Am}) suggest that the latter
configurations should either be signature degenerate of have small 
signature splitting in $^{235}$Np.

\subsection{Odd-neutron nuclei}
\label{odd-neu-nuc}

  Long rotational bands based on different single-particle 
orbitals have been observed in odd-neutron $^{237}$U, 
$^{239,243}$Pu, $^{247,249}$Cm, $^{249}$Cf and $^{253}$No nuclei. 
Considering that the experimental systematics for odd-neutron 
systems is larger than for odd-proton ones, the discussion of 
former systems is performed here on the 'band by band' basis.

\begin{figure*}[ht]
\includegraphics[width=12.0cm,angle=0]{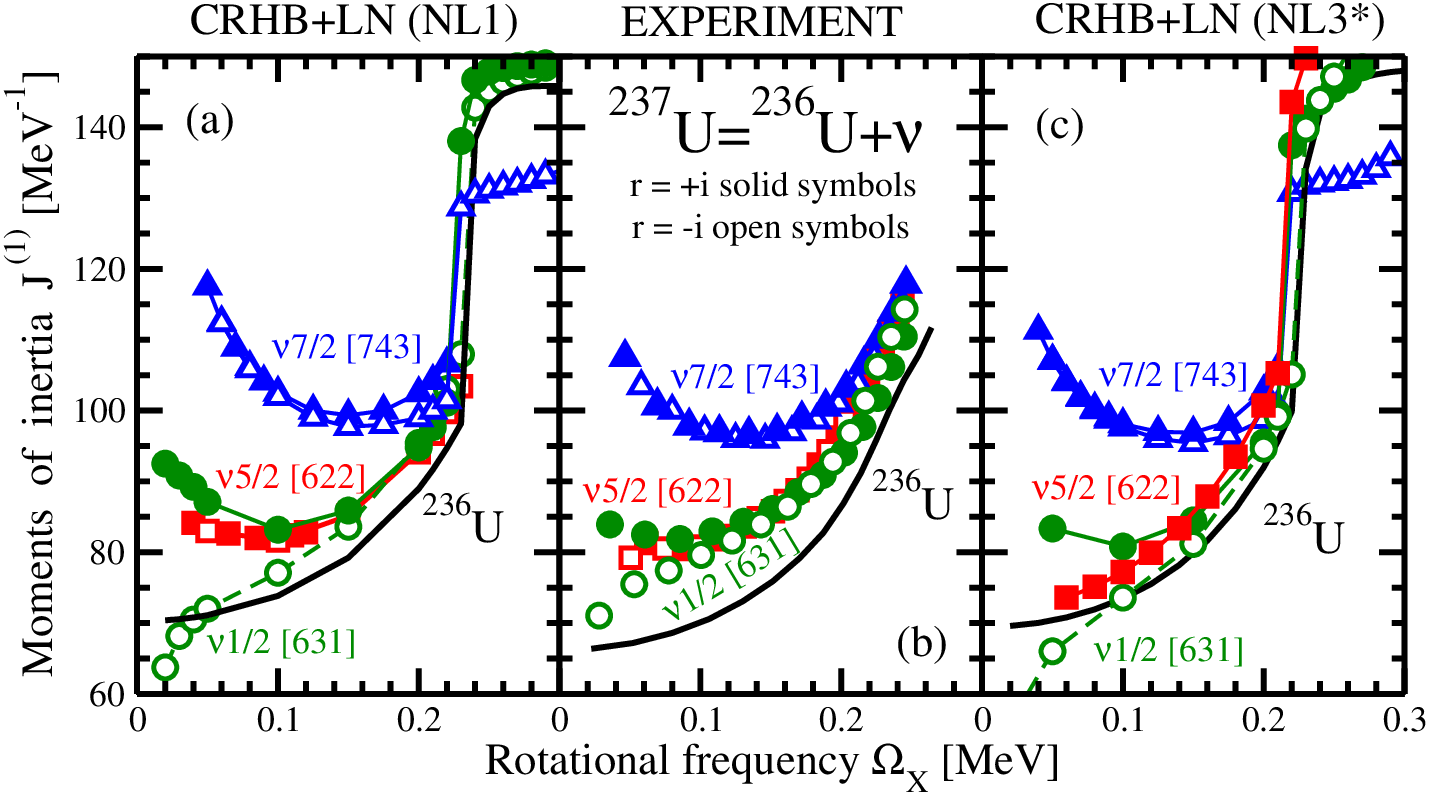}
\caption{(Color online) The same as Fig.\ \ref{J1-241Am} but 
for $^{237}$U.} 
\label{J1-u237}
\end{figure*}

{\bf The $\nu 9/2[734]$ rotational band.} This band has been 
observed in $^{247}$Cm, $^{249}$Cf and $^{253}$No. Figs.\ 
\ref{J1-247Cm}, \ref{J1-cf249} and \ref{J1-No253} show the comparison 
between theory  and experiment for it. In all three nuclei,
the signature degeneracy of observed band is well reproduced.

 In $^{247}$Cm, the CRHB+LN(NL3*) calculations accurately reproduce 
the relative properties of the $\nu 9/2[734]$ band and the reference 
band in $^{246}$Cm. The absolute  $J^{(1)}$ values of experimental
band  are well described up to $\Omega_x \sim 0.15$ MeV (Fig.\ 
\ref{J1-247Cm}).  However, the calculations underestimate the 
increase of $J^{(1)}$ seen at higher frequencies. This is due to
the fact that the increase of $J^{(1)}$ with rotational frequency
is underestimated in the reference $^{246}$Cm ground state band.
The CRHB+LN(NL3*) calculations accurately reproduce the absolute 
$J^{(1)}$ values of the $\nu 9/2[734]$ band in $^{249}$Cf and 
their evolution with frequency as well as its relative properties 
(at low frequency) with respect of the reference band in 
$^{248}$Cf (Fig.\ \ref{J1-cf249}).

 The CRHB+LN(NL1) calculations reproduce very well the $\nu 9/2[734]$ 
band in $^{253}$No and its relative properties with respect of the 
ground state band in $^{252}$No (Fig.\ \ref{J1-No253}a,b). Similar 
accuracy of the reproduction of the  $\nu 9/2[734]$ band in 
$^{253}$No is achieved in the CRHB+LN(NL3*) calculations at 
$\Omega_x\geq 0.12$ MeV (Fig.\ \ref{J1-No253}b,c). However, at 
lower frequencies these calculations do not reproduce the increase 
of the $J^{(1)}$ moments with decreasing $\Omega_x$.

\begin{figure*}[ht]
\includegraphics[width=12.0cm,angle=0]{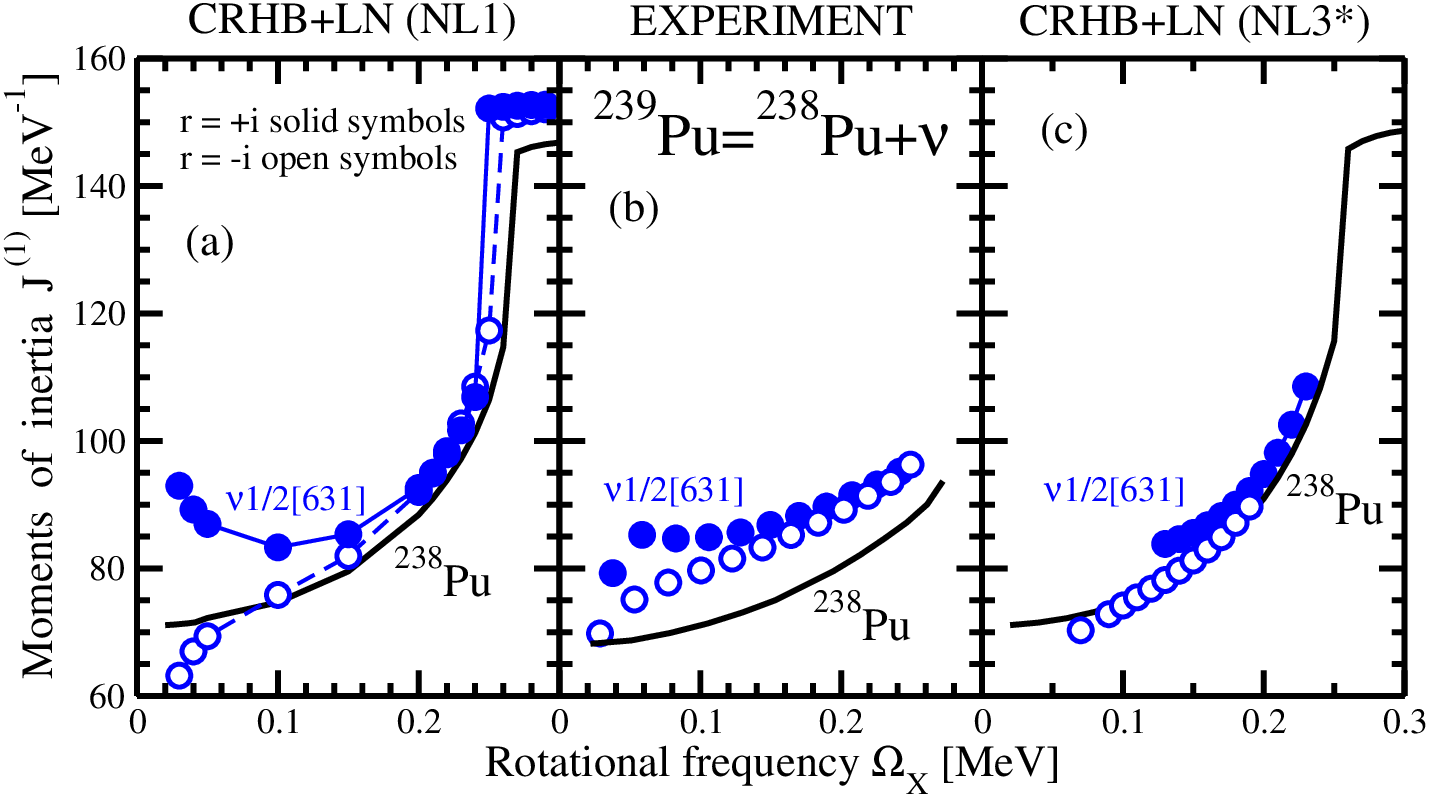}
\caption{(Color online) The same as Fig.\ \ref{J1-241Am} but 
for $^{239}$Pu. } 
\label{J1-pu239}
\end{figure*}

{\bf The $\nu 1/2[631]$ rotational band}. This band has been observed
in $^{237}$U and $^{239}$Pu (Figs.\ \ref{J1-u237} and \ref{J1-pu239}). 
There is large separation between the $J^{(1)}$ values corresponding 
to the $(r=\pm i)$ branches of the $\nu 1/2[631]$ band at low frequency 
which gradually decreases and finally vanishes at high frequency. This 
feature and the fact that the $(r=-i)$ branch has lower values of 
$J^{(1)}$ at low frequency are rather well reproduced in model calculations. 
In experiment, the $J^{(1)}$ values in odd-mass nucleus are higher
than the ones in reference band of even-even nucleus. However, this 
difference is underestimated in model calculations (see Figs.\ 
\ref{J1-u237} and \ref{J1-pu239}). 

{\bf The $\nu 7/2[743]$ rotational band}. This band has been observed
in $^{237}$U (Fig.\ \ref{J1-u237}). It is signature degenerate
at low rotational frequencies. A small separation between the $J^{(1)}$
values of the $(r=\pm i)$ branches is seen at medium and high frequencies;
at these frequencies, the $(r=+i)$ branch has larger $J^{(1)}$ values. 
These features are well reproduced in the model calculations with
both CDFT parametrizations. The model calculations also reproduce the
absolute $J^{(1)}$ values and their evolution with frequency
as well as their relative properties with respect of reference band
in $^{236}$U. However, the increase of $J^{(1)}$ in the band crossing 
region is sharp in model calculations but more gradual in experiment. 
This is similar to the situation seen in the reference $^{236}$U nucleus. 
As discussed in Sec.\ \ref{Rot-even-even}, this discrepancy between 
theory and experiment in $^{236}$U may be due to the stabilization of 
octupole deformation at high spin which leads to the delay of the 
alignment of the  proton $\pi i_{13/2}$  and neutron $\nu j_{15/2}$ orbitals. 
If that is a case in nature, similar situation can be expected also
in $^{237}$U.

{\bf The $\nu 5/2[622]$ rotational band}. This band has been observed
in $^{237}$U (Fig.\ \ref{J1-u237}). It is signature degenerate and this 
feature is reproduced in the CRHB+LN(NL1) calculations. Only the $r=+i$ 
branch of this band has been obtained in the CRHB+LN(NL3*) calculations. 
However, the $\pi 5/2[523](r=\pm i)$ orbitals are signature degenerate
in the frequency range of interest in the quasiparticle routhian 
diagram obtained with the NL3* parametrization. The absolute values 
of $J^{(1)}$ and their evolution with frequency are reproduced in model 
calculations. The NL1 parametrization somewhat better reproduces the 
properties of this band with respect of reference band in $^{236}$U than 
the NL3* parametrization which underestimates the increase of the 
$J^{(1)}$ values due to blocking of the $\nu 5/2[622](r=\pm i)$ 
orbitals.

{\bf The $\nu 7/2[624]$ rotational band}. This band has been observed
in $^{243}$Pu (Fig.\ \ref{J1-pu243}) and it is the only odd-mass nucleus 
band in the whole actinide region in which sharp upbend is observed.
The model calculations extremely well reproduce the properties of this
band including the absolute $J^{(1)}$ values and their evolution with
rotational frequency, signature degeneracy of the $(r=\pm i)$ branches,
the relative properties of this band and the reference band in $^{242}$Pu 
and the properties in the band crossing region. In experiment, the 
band crossing  in the $\nu 7/2[624]$ band takes place earlier (by 
0.01 MeV) than the one in the ground state rotational band of the 
reference $^{242}$Pu nucleus (Fig.\ \ref{J1-pu243}b). However, in 
the CRHB+LN(NL1) calculations both crossings take place at the same 
frequency (Fig.\ \ref{J1-pu243}a).

\begin{figure}[h]
\includegraphics[width=8.6cm,angle=0]{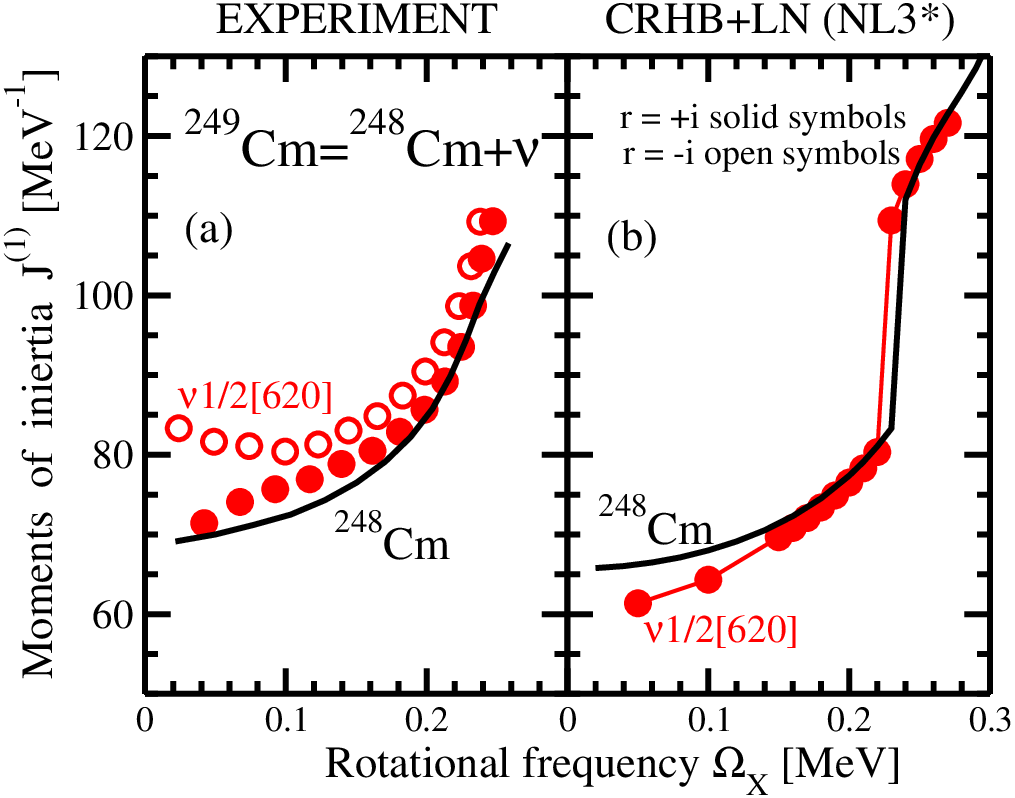}
\caption{(Color online)  The same as Fig.\ \ref{J1-241Am} but 
for $^{249}$Cm. Experimental data are taken from Ref.\ 
\cite{247-249Cm-249Cf}.} 
\label{J1-249Cm}
\end{figure}

\begin{figure}[ht]
\includegraphics[width=8.6cm,angle=0]{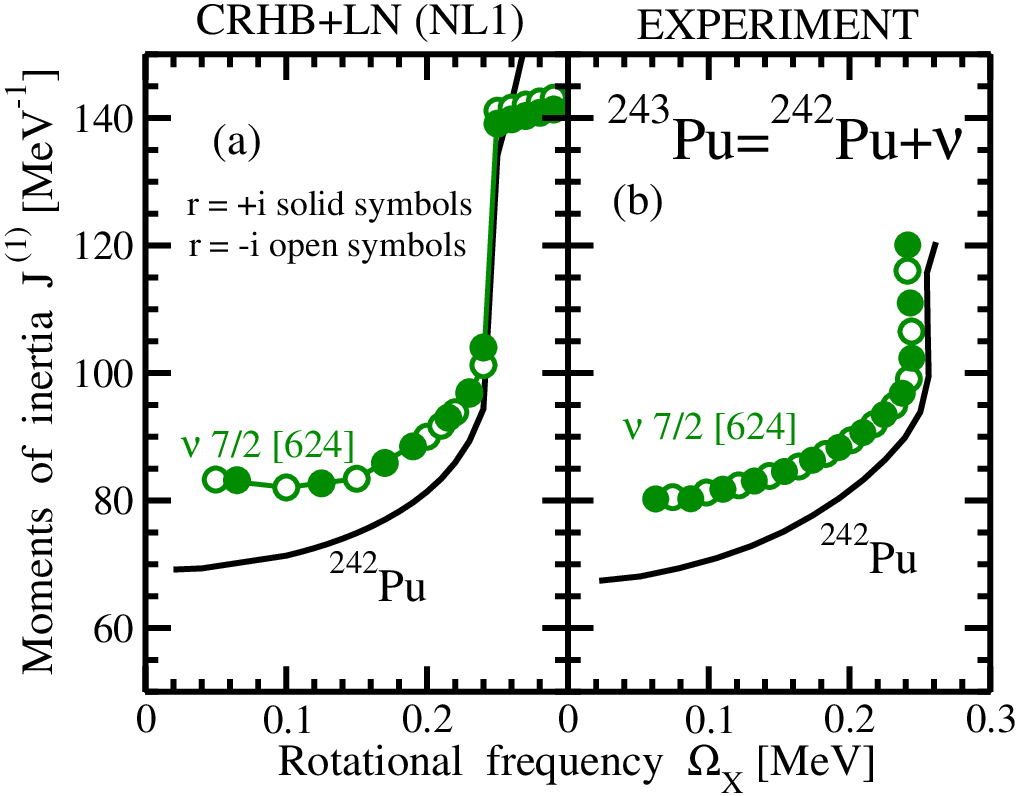}
\caption{(Color online) The same as Fig.\ \ref{J1-241Am} but 
for $^{243}$Pu.} 
\label{J1-pu243}
\end{figure}

{\bf The $\nu 1/2[620]$ rotational band}.  This band has been 
observed in $^{249}$Cm (Fig.\ \ref{J1-249Cm}). It was possible to obtain only
the $(r=+i)$ branch of the $\nu 1/2[620]$ configuration and only in
the CRHB+LN(NL3*) calculations. The slope of experimental $J^{(1)}$ 
curve as a function of frequency is reasonably well reproduced before
the band crossing. However, the relative properties of this band and 
the ground state band in the reference $^{248}$Cm nucleus are not 
completely reproduced. Dependent of frequency the $J^{(1)}$ values 
of the former band are somewhat larger (or similar) than that (to that) 
of the later band in experiment. However, opposite situation is seen 
in the calculations. The band crossing region is reproduced in general 
in the calculations. However, in experiment the band approaches the band 
crossing point in a more gradual way than in the calculations.

\subsection{Rotational properties as a tool of configuration
assignment}
 
  On-going experimental investigations of odd-mass light superheavy
nuclei at the edge of the region where spectroscopic studies 
are still feasible (the nuclei with masses $A\sim 255$ and 
proton  number $Z\geq 102$)  \cite{INPC2010} require reliable
theoretical tools for the assignment of one-quasiparticle
configurations. This is due to inherent restrictions of the
studies at the limits of experimental capabilities. Rotational
properties have been occasionally used for that purpose. However,
only with the completion of this systematic study of odd-mass
nuclei it becomes possible to reliably estimate theoretical 
errors of the description of rotational properties of such
nuclei and the robustness of configuration assignment based 
on such properties.

   Indeed, rotational properties of one-quasiparticle configurations 
give an important information on their underlying structure, thus 
providing an extra tool for a configuration assignment.  The 
rotational properties reflected through the following fingerprints
\begin{itemize}
\item the presence or absence of signature splitting,

\item the relative properties of different configurations with 
respect of each other and/or with respect to the ground state
band in  reference  even-even nucleus,

\item the absolute values of the kinematic moments of inertia 
(especially at low rotational frequencies) and their evolution 
with rotational frequency

\end{itemize}
provide useful tools for quasiparticle configuration assignments.
Our systematic investigation shows that with few exceptions these 
features of rotational bands are well described in model calculations. 
 The presence or absence of signature separation and its magnitude is 
the  most reliable fingerprint which is reproduced in model calculations 
with good accuracy. The moments of inertia and their evolution with
frequency are generally well described in model calculations. As a
consequence, the relative properties of different configurations 
with respect of each other and/or with respect to the ground state band 
in reference even-even nucleus provide a reasonably reliable 
fingerprint of configuration. This fingerprint is especially useful at 
low frequencies where the largest difference between the configurations
is observed. Only in the case of the $\nu 1/2[631]$ and $\nu 1/2[620]$ 
configurations, the calculations fail to describe their relative 
properties with respect of reference band in even-even nucleus.

 However, it is necessary to recognize that the configuration 
assignment based on rotational properties has to be complemented 
by other independent methods and has to rely on sufficient experimental
data. This is because such method of configuration assignment not 
always leads to a unique candidate configuration due to theoretical 
inaccuracies in the description of the moments of inertia. The 
interpretation of the rotational band in $^{253}$No is quite 
illustrative in this respect. Initially, it was interpreted as based 
on the $\nu 7/2[624]$ configuration 
\cite{No253-PRL}. However, improved experiments allowed to identify 
the M1 transitions between opposite signatures of the observed band 
\cite{No253} which led to the $\nu 9/2[734]$ configuration assignment. 
The kinematic moments of inertia of the observed band under two 
configuration assignments are described within a typical theoretical 
uncertainty and, as a result, the configuration assignment based only 
on rotational properties cannot be fully reliable. The branching 
ratios of observed M1 and E2 transitions have to be used in order 
to distinguish different configuration assignments \cite{No253}.


\section{Rotational and deformation properties of fission isomers}
\label{Fis-def-rot}

\begin{figure}[h]
\includegraphics[width=8.6cm,angle=0]{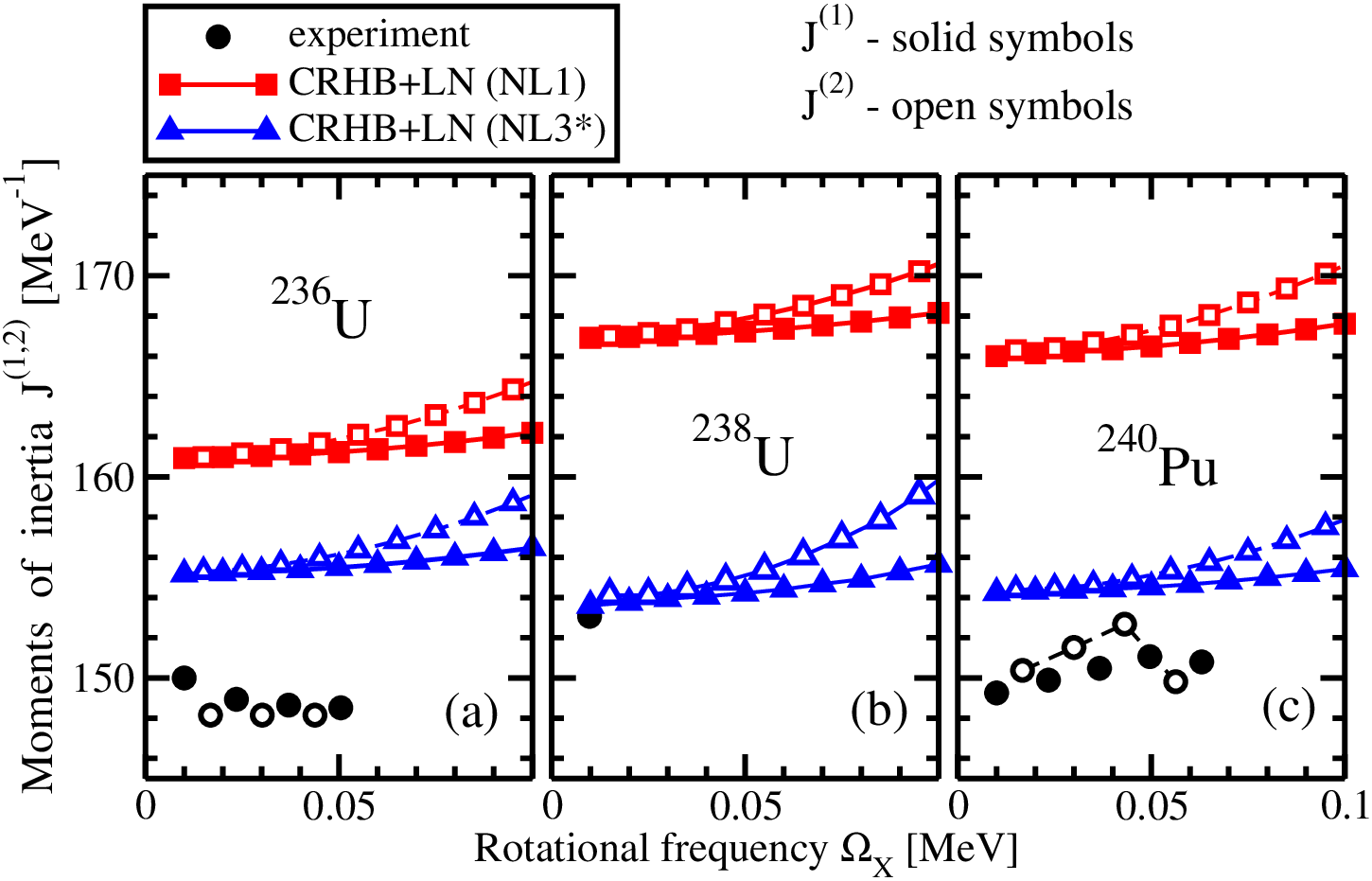}
\caption{(Color online) Experimental and calculated kinematic 
$(J^{(1)})$ and dynamic $(J^{(2)})$ moments of inertia of SD rotational
bands in $^{236,238}$U and $^{240}$Pu. The notation of the lines 
and symbols is given in the figure.} 
\label{J1-SD}
\end{figure}
\begin{table*}[ht] 
\caption{Experimental and theoretical charge quadrupole moments $Q$ 
of SD fission isomers. The results of the CRHB+LN calculations with 
the NL1 and NL3* parametrizations are presented. Experimental data 
for the U and Pu isotopes are 
taken from Ref.\ \protect\cite{H.89}, while the one for $^{242}$Am from
Ref.\ \protect\cite{Am242}.}
\begin{tabular}{|c|c|c|c|c|c|c|} \hline
                   & $^{236}$U & $^{238}$U & $^{236}$Pu & $^{239}$Pu &
                   $^{240}$Pu & $^{242}$Am \\ \hline
$Q^{exp}$ ($e$b) & $32\pm 5$ & $29\pm 3$ & $37\pm 10$ & $36\pm 4$ &
                   & $35.5\pm 1.0_{st}\pm 1.2_{mod}$ \\
$Q^{\rm NL1}$ ($e$b)  &  35.8 & 37.3 & 36.1 & & 38.2 & \\
$Q^{\rm NL3*}$ ($e$b) &  33.9 & 33.7 & 34.8 & & 34.9 & \\ \hline 
\end{tabular}
\label{QT-SD}
\end{table*}

 The investigation of fission isomers provides a
valuable information on rotational and pairing properties in
the superdeformed (SD) minimum of actinides. The latter is important 
for an understanding of fission barriers which sensitively depend on  
pairing properties (see Ref.\ \cite{KALR.10} and references quoted
therein). Although 
some attempts were made in 70ies to extract the information on 
pairing properties at fission saddles \cite{BH.74}, 
they did not lead to reliable estimates. Thus, fission isomers
provide only available tool to estimate the evolution of pairing
with deformation in actinides. Such an estimate is available only 
through the study of rotational properties of $^{236,238}$U and 
$^{240}$Pu nuclei; these are only nuclei for which SD rotational 
bands were experimentally measured\footnote{The information on
pairing cannot be extracted from odd-even mass staggerings in the
SD minimum, since the inaccuracies of the measuruments of 
excitation energies of fission isomers in odd-mass nuclei are 
at least 200 keV but can reach 400 keV \cite{SZF.02}.}. 
Although fission isomers in 
actinides have been observed more than 50 years ago their rotational 
and single-particle properties 
are significantly less known experimentally than in other regions of 
superdeformation. For example, no reliable experimental data on 
single-particle states in odd-mass actinides exist.

  The experimental and calculated kinematic and dynamic moments of inertia 
of the SD rotational bands in $^{236,238}$U and $^{240}$Pu are shown in
Fig.\ \ref{J1-SD}. The calculated kinematic and dynamic moments of inertia 
increase with increasing rotational frequency $\Omega_x$ (Fig.\ 1). In 
addition, the difference between these moments grows with the increase 
of $\Omega_x$ since $J^{(2)}$ raises faster than $J^{(1)}$. In the 
calculations, these features are predominantly 
due to gradual alignment of the $N=8$ neutrons and $N=7$ protons and a 
smooth decrease of pairing correlations with increasing $\Omega_x$. They 
are similar to the ones observed in the $A\sim 190$ region of 
superdeformation, see Ref.\ \cite{CRHB} and references quoted therein.

  The experimental data in $^{240}$Pu shows such features for $J^{(1)}$
and $J^{(2)}$. However, the highest $J^{(2)}$ point deviates from this 
trend most likely due to the fact that the energy of the $10^+ \rightarrow 8^+$ 
transition has been measured with lower accuracy than that of other 
transitions within the SD band \cite{ENSDF}. On the other hand, such features
are not seen in $^{236}$U. This is again can be related to insufficient
accuracy of the measurements of the $\gamma-$transitions energies in
the SD band of $^{236}$U; these energies in the SD bands of $^{236}$U and 
$^{240}$Pu are measured with typical accuracy of 1.0 keV and 0.1 keV, 
respectively \cite{ENSDF}.

 The experimental kinematic moments of inertia are best described by
the NL3* parametrization; the deviation from experiment does not exceed
3.4\% (Fig.\ \ref{J1-SD}). Note that similar to low spin results in the 
ND minimum (see Sect.\ \ref{Sel-f}), the minor variations in the experimental 
$J^{(1)}$ values with particle number are not reproduced. The fact that 
the moments of inertia of rotational structures in two different minima 
(ND and SD) are accurately described with the same pairing strength 
strongly suggests that the evolution of pairing correlations with 
deformation is properly described in the CRHB+LN(NL3*) 
framework by the Brink-Booker part of the Gogny D1S force.
This is important for the 
investigation of fission barriers, the properties of which sensitively 
depend on employed pairing interaction \cite{KALR.10}.

\begin{figure}[h]
\includegraphics[width=8.6cm,angle=0]{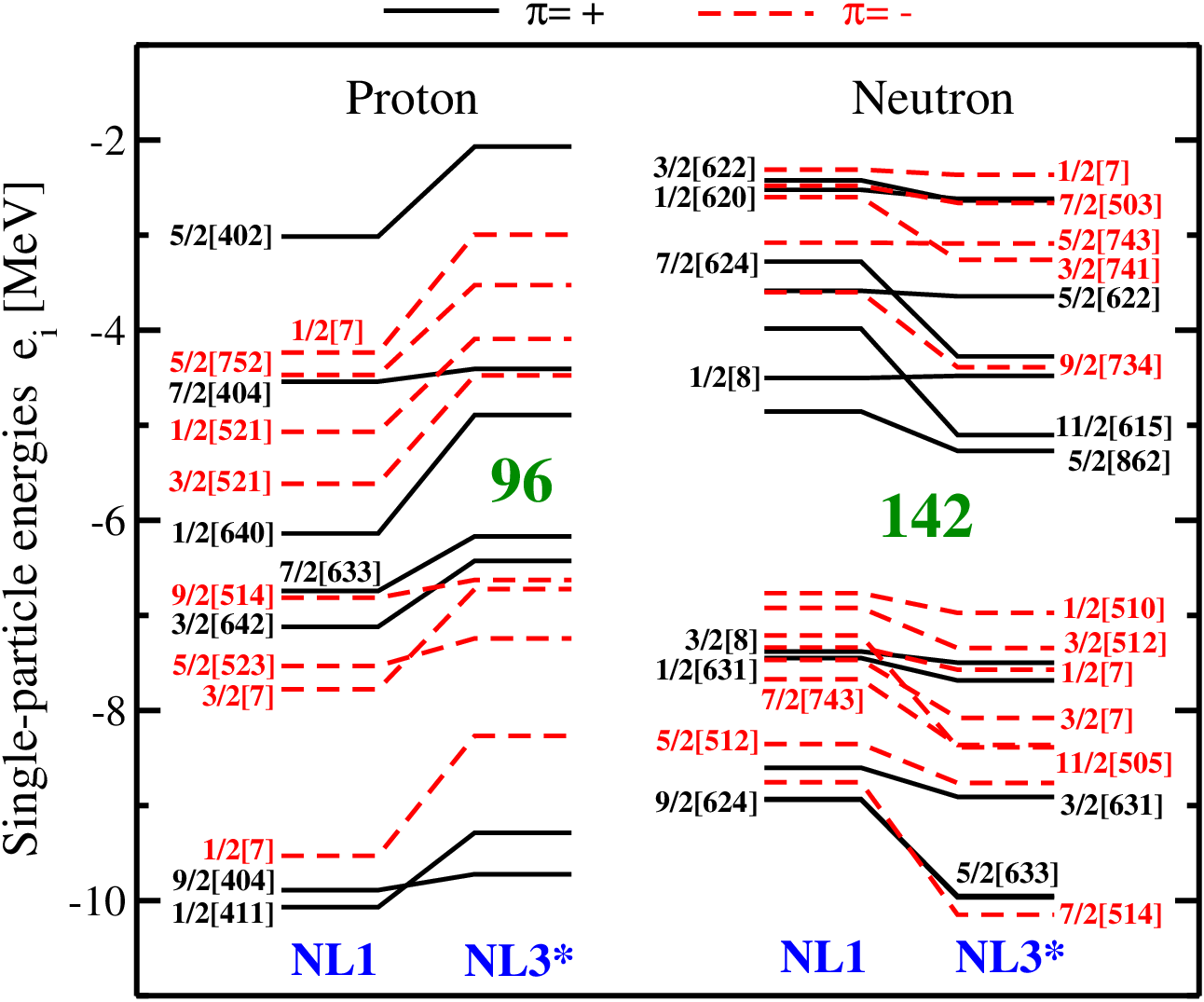}
\caption{(Color online) The single-particle levels in the superdeformed
minimum of $^{236}$U obtained in the calculations with the NL1 and NL3* 
parametrizations. They are labelled by the Nilsson labels 
$\Omega [Nn_z\Lambda]$ when the squared amplitude of the dominant component
of the wave function exceeds 0.5. Otherwise, the states are labelled by 
$\Omega [N]$ where $N$ represents the dominant principal quantum number.} 

\label{SD-shell-str}
\end{figure}

However, this is not always the case since the CRHB+LN(NL1) calculations
substantially overestimate the experimental moments of inertia in the 
SD minimum [for example, by 11.3\% in $^{240}$Pu] (Fig.\ \ref{J1-SD})
while they reproduce the low-spin moments of inertia in the ND minimum 
with the same level of accuracy as the CRHB+LN(NL3*) calculations (see 
Fig.\ \ref{J1-low}). It turns out that reasonable description of the 
moments of inertia at SD can be achieved in the CRHB+LN(NL1) calculations 
only if the original strength of the 
Brink-Booker part of the Gogny D1S force
(scaling factor 
$f=1.0$) is used in the calculations (see Fig.\ 1 in Ref.\ \cite{CRMF-Hung}). 
This clearly indicates that even the pairing force carefully fitted to 
experimental data at normal deformation does not guarantee accurate
description of pairing at SD (and as a consequence also at fission 
saddle). The origin of such behavior is not completely clear but the 
difference between the CRHB+LN(NL3*) and CRHB+LN(NL1*) results for 
$J^{(1)}$ at SD may also partially originate from the differences in 
the single-particle structures at superdeformation  obtained with the 
NL3* and NL1 parametrizations (Fig.\ \ref{SD-shell-str}). While
the large $N=142$ SD shell gap exists in both parametrizations, somewhat
smaller $Z=96$ SD shell gap is seen only in the NL3* parametrization
(Fig.\ \ref{SD-shell-str}). In
addition, the ordering of the single-particle levels is different in
these two parametrizations.


 Table \ref{Pairing-SD-ND} shows that for a specific parametrization 
the calculated pairing energies in the ND and SD minima are comparable 
in a given subsystem (proton or neutron). They are typically within 
0.5 MeV. The only exception is the proton subsystem for which the 
pairing energies in the SD minimum are larger than those in the 
ND one by 1.4 MeV. Note that the pairing is stronger in the NL1 
parametrization. In part, this is a consequence of the fact that 
average scaling factor $f_{av}$ is larger for the NL1 parametrization
(see Sec.\ \ref{Sel-f}).

\begin{table}[ht] 
\caption{Averaged pairing energies $E_{pairing}^{\nu/\pi}$ [in MeV]
(Eq.\ (\ref{Epair}) in the normal- (ND) and superdeformed (SD) minima 
for the NL3* and NL1 parametrizations. These quantities are averaged 
over $^{236,238}$U and $^{240}$Pu nuclei; the individual pairing
energies in each of these nuclei do not deviate from averaged 
ones by more than 0.5 MeV.}
\begin{tabular}{|c|c|c|c|c|} \hline
 Parametrization   & \multicolumn{2}{c|}{Neutron} &
                    \multicolumn{2}{ c|}{Proton} \\ \hline
                   &     ND     &    SD      &   ND &  SD  \\ \hline
   NL3*            &    4.99    &  5.51      & 6.56 & 6.90 \\
   NL1             &    6.26    &  6.52      & 6.97 & 8.35 \\ \hline
\end{tabular}
\label{Pairing-SD-ND}
\end{table}

\begin{figure}[h]
\includegraphics[width=8.6cm,angle=0]{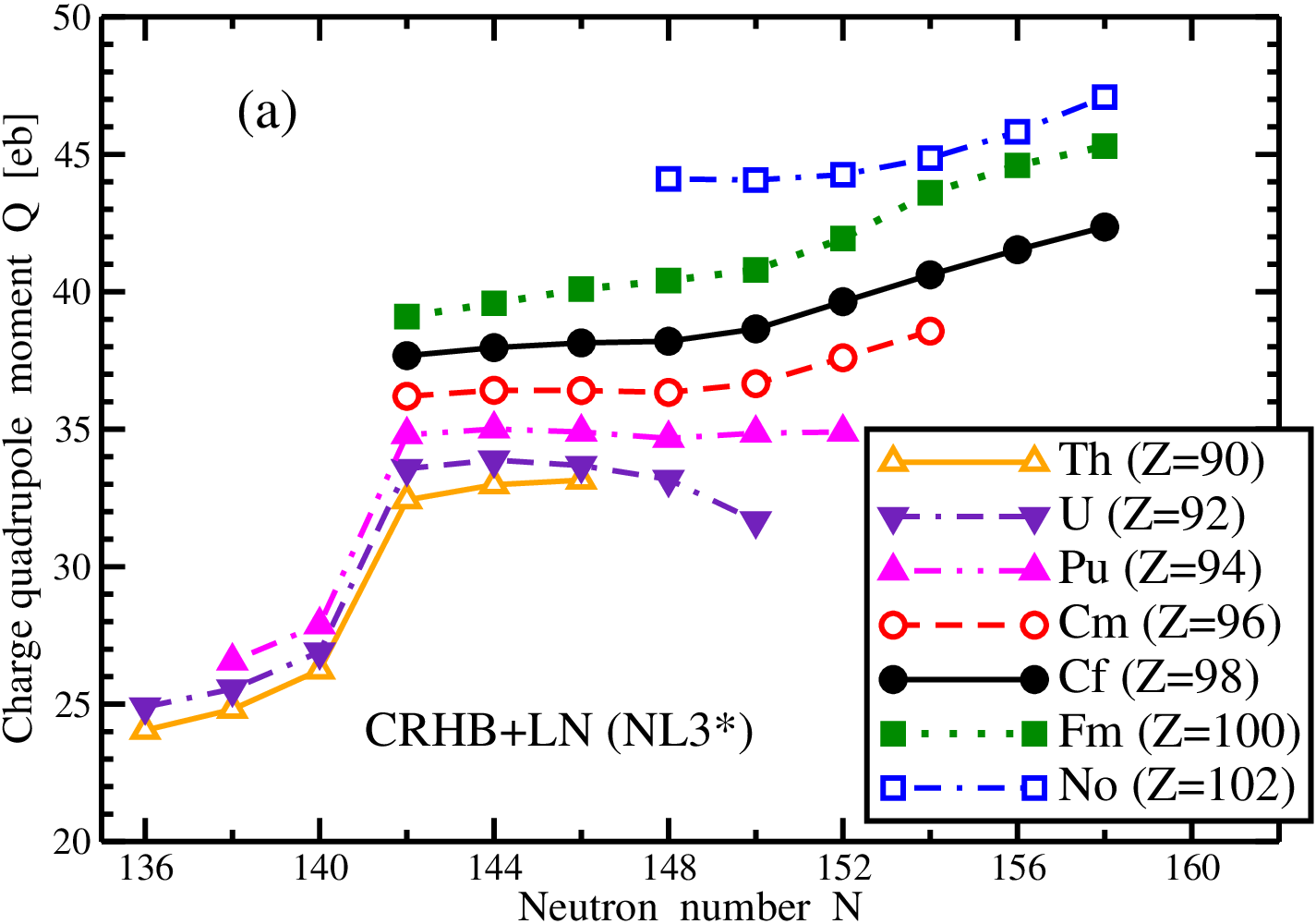}
\includegraphics[width=8.6cm,angle=0]{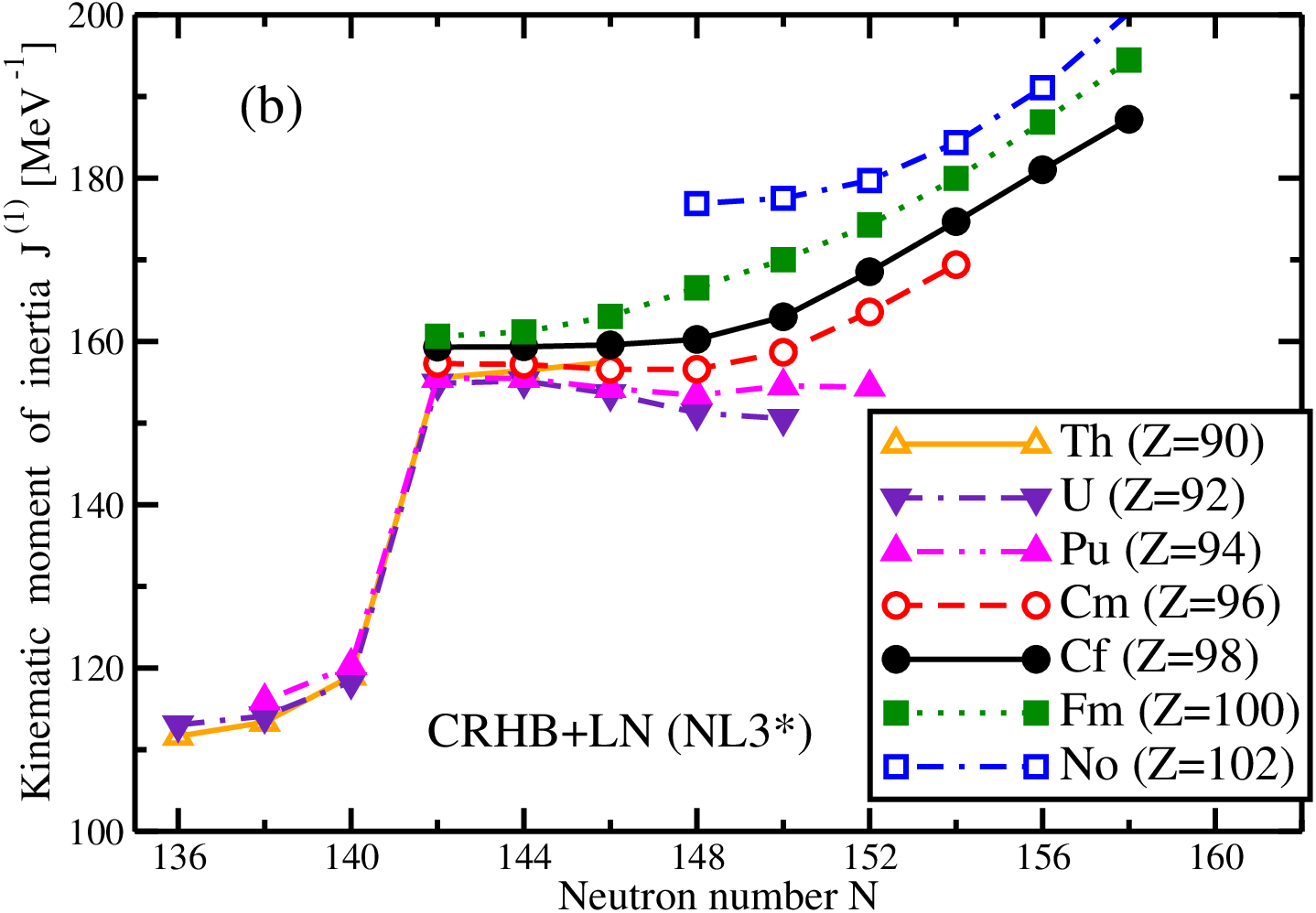}
\caption{ (Color online) Charge quadrupole moments 
$Q$ (panel (a)) and kinematic moments of inertia $J^{(1)}$ 
(panel (b)) of the yrast SD bands as a function of neutron 
number $N$. The results are obtained in the CRHB+LN(NL3*)
 calculations at $\Omega_x=0.01$ MeV.
}
\label{QT-J1-SD-sys}
\end{figure}

The calculated charge quadrupole moments $Q$ are compared with 
available experimental data in Table \ref{QT-SD}. One should note that 
the small error bars on the experimental values of $Q$ given for 
$^{238}$U and $^{240}$Am nuclei should be treated with caution since 
even modern experiments do not provide an accuracy of the absolute 
$Q$ values better than 15\%, see discussion in Ref.\ \cite{CRHB}. In 
addition, when comparing the calculations with experiment one should
take into account that {\bf (i)} the $Q^{exp}$ values have been
obtained with different experimental techniques \cite{H.89}, {\bf
(ii)} it is reasonable to expect that an addition of one neutron to 
$^{239}$Pu will not change considerably the $Q$ value, and, thus, 
$Q^{exp}$($^{239}$Pu) could be used for comparison with the 
calculated $Q$($^{240}$Pu).

 With these considerations in mind, it is clear that the 
CRHB+LN(NL3*) results come reasonably close 
to experiment.  The CRHB+LN(NL1) results are also not far 
away from experimental data but they substantially overestimate 
experimental $Q$ value in $^{238}$U.  The $Q$ values obtained in 
the CRHB+LN(NL1) calculations are always higher than the ones 
for CRHB+LN(NL3*), which also may be a reason why the 
CRHB+LN(NL1) calculations systematically overestimate kinematic 
moments of inertia at SD.

 The systematic analysis of low-spin properties of the yrast SD 
bands presented in Fig.\ \ref{QT-J1-SD-sys} is performed with 
the NL3* parametrization since it describes better available 
experimental data on deformation and rotational properties of 
fission isomers. The $Q$ values generally increase with increasing 
proton number. For a given isotope chain they stay nearly constant 
in the $N=142-150$ range. There is a gradual increase of $Q$ at 
$N>150$ in the Cm, Cf and Fm isotopes. In the Th, U and Pu isotopes,
the $Q$ values drop by 5-7 $e$b on going from $N=142$ to 
$N=140$ (Fig.\ \ref{QT-J1-SD-sys}a). This change in equilibrium 
deformation of second minimum as a function of neutron number is 
clearly visible in the deformation energy curves obtained 
in the RMF+BCS calculations with monopole pairing and the NL3* 
parametrization (Fig.\ 7 in Ref.\ \cite{AAR.10}). It is caused 
by the changes in the underlying shell structure; this is supported 
by the fact that the RMF+BCS calculations of Ref.\ \cite{AAR.10} with 
monopole pairing  and the current CRHB+LN(NL3*) calculations with  
the Brink-Booker part of the Gogny D1S force
in the pairing channel bring similar values 
for equilibrium deformation in second minimum despite different 
treatment of pairing.

  The evolution of the kinematic moments of inertia $J^{(1)}$ of 
the yrast SD bands at low spin as a function of proton and neutron 
numbers closely resembles the one of charge quadrupole moments 
$Q$ (Fig.\ \ref{QT-J1-SD-sys}). This is mostly due to the fact 
that the values of kinematic moments of inertia of the SD bands 
in the limit of no pairing are typically close to the rigid-body 
values \cite{TO-rot}, and, thus, are strongly defined by the deformation 
properties. The pairing lowers the calculated kinematic moments of inertia 
but does not remove this connection.

\section{Deformation and rotational of superheavy nuclei}
\label{SHE-def-rot}

\begin{figure}[h]
\includegraphics[width=8.6cm,angle=0]{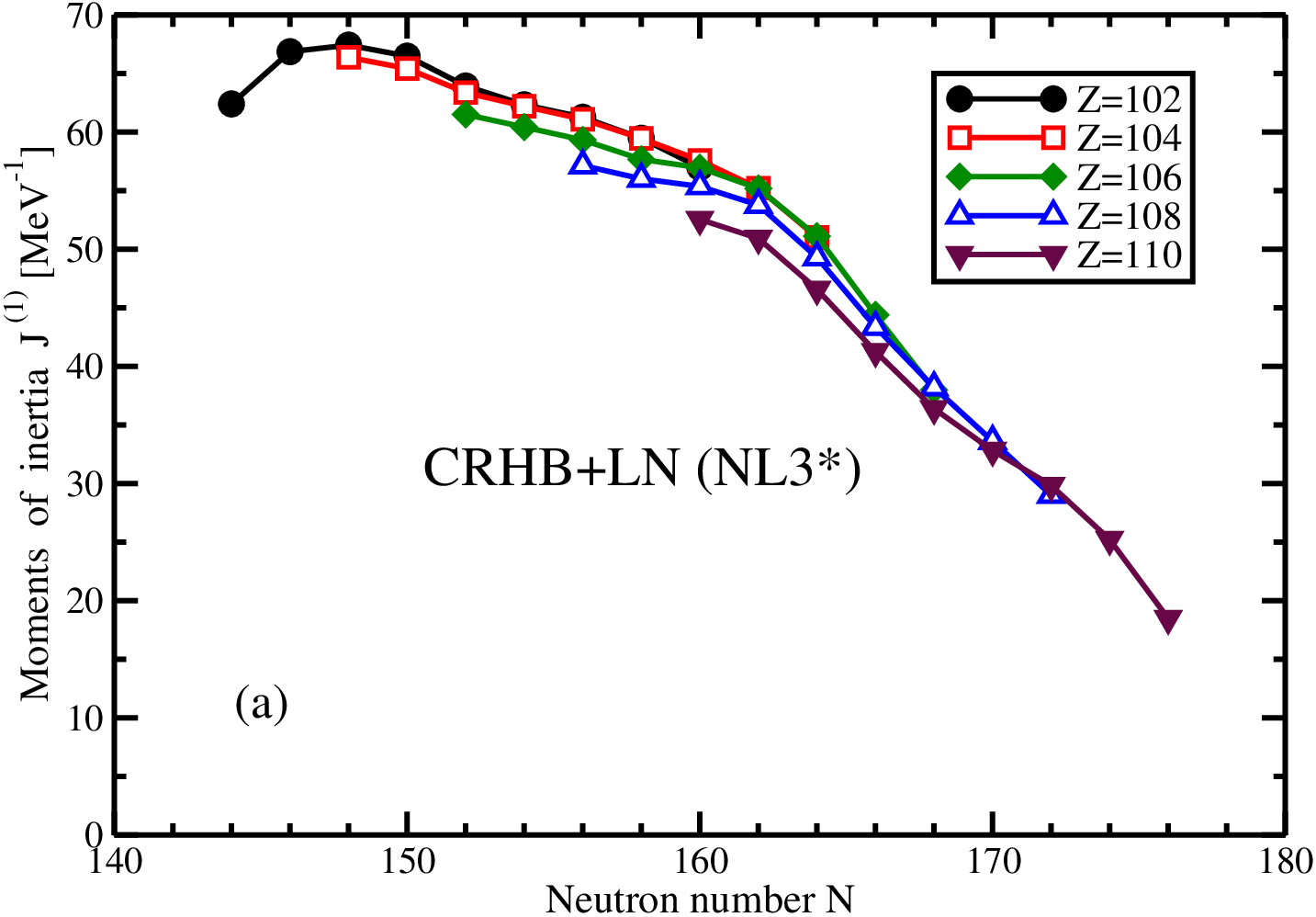}
\includegraphics[width=8.6cm,angle=0]{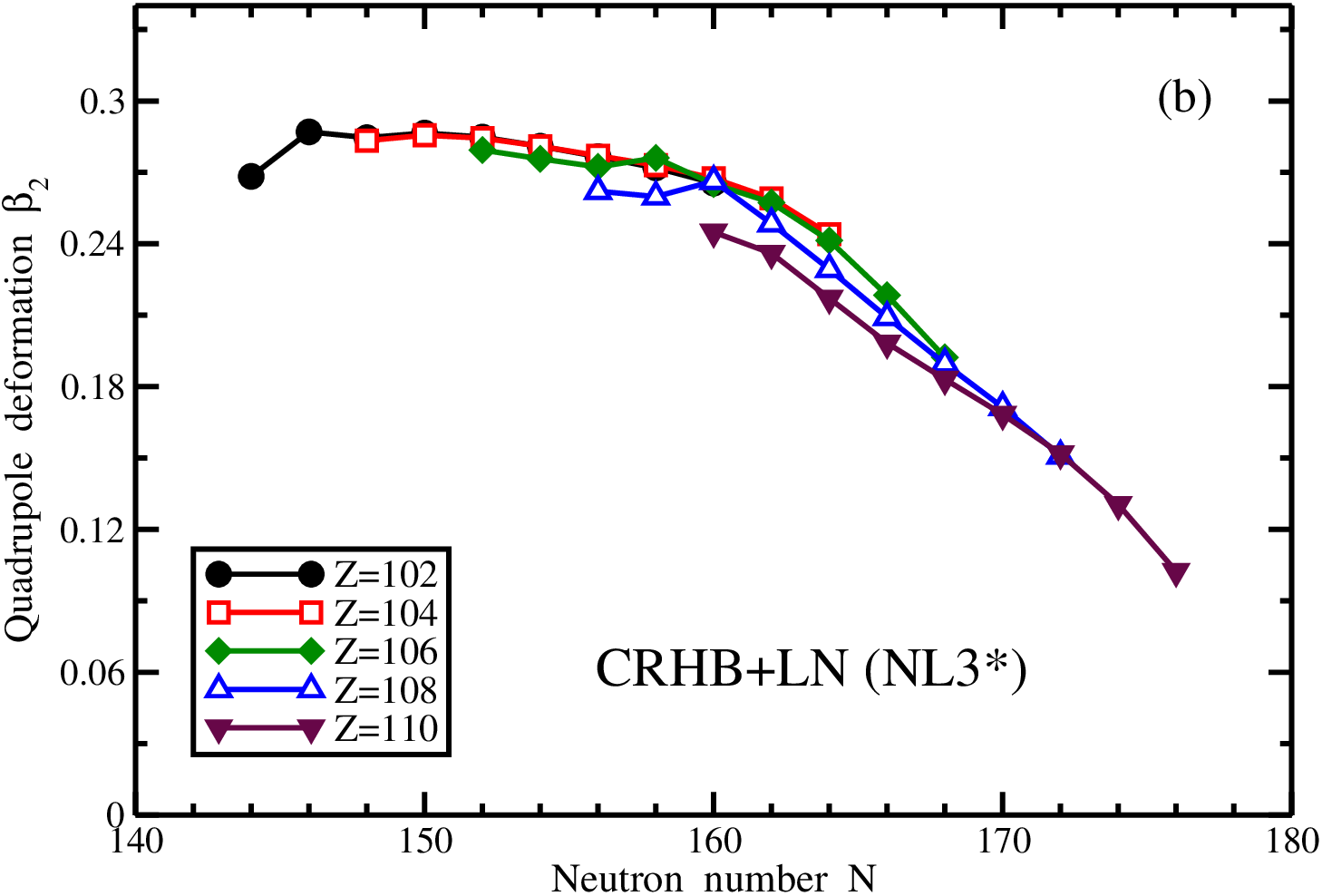}
\caption{(Color online) Kinematic moments of inertia (panel (a)) of 
ground state bands of superheavy nuclei calculated at $\Omega_x=0.02$ MeV 
and their quadrupole deformations $\beta_2$ (panel (b)) as a function of 
neutron number $N$.} 
\label{SHE-sys}
\end{figure}

 Fig.\ \ref{SHE-sys} shows the calculated quadrupole deformations 
$\beta_2$ and kinematic moments of inertia $J^{(1)}$ of even-even 
superheavy nuclei with $Z=102,104,106,108$ and 110 as a function of 
neutron number $N$. The nuclear region selected roughly corresponds to 
the one where superheavy nuclei either have already been measured or 
may be experimentally studied (including rotational properties) within 
the next one or two decades. We do not extend our studies to higher 
$Z$ values since in these nuclei the potential energy surfaces in the 
normal deformed minimum become very soft (see Refs.\ \cite{AAR.12,PNLV.12}) 
so that a description on the mean-field level may not be adequate and 
the methods beyond mean field  \cite{BBH.06,LNVML.09} may be required.

  The CRHB+LN calculations of Fig.\ \ref{SHE-sys} are performed 
with the NL3* parametrization. However, earlier investigations of the 
Fm ($Z=100$) isotope and $N=152$ isotone chains in Ref.\ \cite{A250} 
show that the general trends of the evolution of the  $J^{(1)}$ and 
$\beta_2$ quantities as a function of neutron and proton numbers
only weakly depend on the CDFT parametrization
(see Figs. 13 and 14 in Ref.\ \cite{A250}). Fig.\ \ref{SHE-sys} shows 
that calculated quadrupole deformation $\beta_2$ stays more or less 
constant for the neutron numbers $N\leq 162$. However, at higher $N$ 
it decreases gradually with increasing neutron number. The evolution 
of calculated kinematic moments of inertia correlates strongly with 
the one for the quadrupole deformations. Indeed, with the exception 
of the lightest $Z=102$ nuclei, the $J^{(1)}$ values decrease very 
slowly with increasing neutron number, but above $N=162$ the rate of 
the decrease of $J^{(1)}$ becomes substantially larger.

\section{Conclusions}
\label{Concl}

  The cranked relativistic Hartree-Bogoliubov theory 
has been applied for a study of actinides and light 
superheavy nuclei. The systematic investigation of 
rotational properties of even-even and odd-mass nuclei 
at normal deformation has been performed for the first 
time in the density functional theory framework. In 
addition,  the pairing properties have been systematically 
studied via the $\Delta^{(3)}$ indicators for the first 
time in the CDFT theory. The main results can be 
summarized as follows:

\begin{itemize}
\item
 In order to reproduce the moments of inertia in actinides
and light superheavy nuclei, the strength of the Brink-Booker 
part of the Gogny D1S force
in the particle-particle channel of the CRHB+LN theory 
has to be attenuated by $\approx 10\%$. With 
this attenuation, the moments of inertia below band 
crossings and the $\Delta^{(3)}$ indicators are well
reproduced. In  contrast, the moments of inertia of 
lighter nuclei with $A\leq 200$ are well described  with 
the original strength of the Brink-Booker part of the Gogny 
D1S force
in the CRHB+LN calculations.

\item
The strengths of pairing defined by means of the moments of 
inertia and three-point $\Delta^{(3)}$ indicators strongly 
correlate. This is known result in non-selfconsistent models 
based on phenomenological Woods-Saxon or Nilsson potentials. 
However, this is non-trivial result in the DFT framework since 
time-odd mean fields (absent in phenomenological potentials) 
strongly affect the moments of inertia \cite{TO-rot} and have 
an impact on three-point $\Delta^{(3)}$ indicators \cite{AA.10}.

\item 
 The definitions of pairing strength via these two observables 
are complimentary. This is because (i) it is 
difficult to disentangle proton and neutron contributions to
pairing when considering the moments of inertia and (ii) the 
$\Delta^{(3)}$ indicators are affected by particle-vibration 
coupling and depend on correct reproduction of the ground 
states in odd-mass nuclei (see Sec.\ \ref{dev-delta} for 
details).

\item
The calculations with approximate particle number projection 
by means of the Lipkin-Nogami method provide a better 
description of the absolute values and particle number 
dependencies of the moments of inertia as compared with the 
calculations which do not include it. Similar improvement
is observed for the $\Delta^{(3)}$ indicators. However, more
systematic calculations of the $\Delta^{(3)}$ indicators in the
CRHB+LN and CRHB frameworks are needed to make this observation
conclusive.


\item
  Sharp upbendings observed in a number of rotational bands
of the $A\geq 242$ nuclei are
well described in the model calculations. The calculations 
also predict similar upbendings in lighter nuclei but they 
have not been seen in experiment. The analysis suggests that 
the stabilization of octupole deformation at high spin, not
included in the present CRHB+LN calculations, can be responsible 
for this discrepancy between theory and experiment.

\item
  The proper description of the evolution of pairing with deformation 
implies an accurate reproduction of the moments of inertia of 
rotational structures in normal- and superdeformed minima with the
same strength of pairing. This condition is satisfied only in the 
CRHB+LN(NL3*) calculations. On the contrary, the strength of pairing
in the SD minimum has to be increased by almost 10\% as compared with
the ND minimum in order to reproduce the moments of inertia of the 
SD bands in the CRHB+LN(NL1) calculations. This clearly indicates 
that even the pairing force carefully fitted to experimental data at 
normal deformation does not always guarantee accurate description 
of pairing at SD (and, as a consequence, also at fission saddle). 
The origin of such behavior is not completely clear but partially 
maybe related to the dependence of the shell structure at SD on 
the parametrization.

\item
   It is well known fact that the present generation of the 
density functional theories do not provide the same accuracy 
of the description of the energies of the single-particle states 
as the models based on the phenomenological Woods-Saxon or Nilsson
potentials (see Ref.\ \cite{AS.11} and references quoted therein).
Despite that many aspects of the single-particle motion such as
deformation polarization effects due to particle(s)/hole(s) and 
the impact of the particle(s)/hole(s) on angular momentum  
alignments/moments of inertia  are well described in the DFT 
models in the regime of no or weak pairing \cite{ALR.98,MADLN.07}. 
The current systematic study of rotational bands in odd-mass nuclei 
confirms for the first time this observation also for pairing regime.
This is because with few exceptions  the impact of particle on the 
rotational properties of the bands in odd-mass nuclei is well described 
in model calculations. As a consequence, the absolute and relative 
properties of different configurations/bands in odd-mass nucleus with 
respect of each other and/or with respect to the ground state band 
in reference even-even nucleus provide a reasonably reliable 
fingerprint of underlying one-quasiparticle configuration 
of rotational band in odd-mass nucleus.

\end{itemize}

  The authors would like to thank S.\ Frauendorf,  R.\ Herzberg,
R.\ V.\ F.\ Janssens,  T.\ L.\ Khoo and P.\ Ring for valuable 
discussions and suggestions. This work has been supported by 
the U.S. Department of Energy under the grant DE-FG02-07ER41459.
This research was also supported by an allocation of advanced 
computing resources provided by the National Science Foundation. 
The computations were partially performed on Kraken at the National 
Institute for Computational Sciences (http://www.nics.tennessee.edu/).


\end{document}